%% file: main.tex
\documentclass[reprint, superscriptaddress, aps, twocolumn, prx, floatfix]{revtex4-2}
\input{header.tex}
\usepackage{needspace}
\colorlet{BLUE}{blue}
\colorlet{RED}{red}

\begin{document}
\title{Product Formulae with Exponentially Improved Overhead}
\author{Keisuke Murota}
\affiliation{Department of Physics, The University of Tokyo, Tokyo 113-0033, Japan}
\affiliation{Mathematical Institute, University of Oxford, Woodstock Road, Oxford OX2 6GG, United Kingdom}

\author{Joshua W. Dai}
\affiliation{Mathematical Institute, University of Oxford, Woodstock Road, Oxford OX2 6GG, United Kingdom}
\affiliation{Department of Materials, University of Oxford, Parks Road, Oxford OX2 6UD, United Kingdom}

\author{B\'alint Koczor}
\affiliation{Mathematical Institute, University of Oxford, Woodstock Road, Oxford OX2 6GG, United Kingdom}

\date{\today}
\begin{abstract}
For a Hamiltonian $H=\sum_a H_a$, product formulae approximate the time evolution under $H$ for a time $t$ using exponentials of each term $H_a$. Standard recursive constructions improve accuracy by repeatedly canceling the leading error term, but the number of exponentials per step grows exponentially with the order of recursion. We show how to avoid this exponential overhead by canceling many error terms at once, between approximations to the same evolution with different time steps. Combining approximations at $q$ distinct time steps makes the error exponentially small in $q$ at a cost of only $\widetilde O(q^2)$ exponentials. In particular, on a quantum computer, our product formula implements this evolution to precision $\varepsilon$ with $O(t\log^{4/3+o(1)}(t/\varepsilon))$ exponentials for long times $t$, and with circuit depth $O(t\log^{2+o(1)}(Nt/\varepsilon))$ for local Hamiltonians on $N$ sites. Up to polylogarithmic factors, this matches the scaling of ancilla-assisted algorithms, and it is the best known scaling for a single circuit without ancilla qubits.
\end{abstract}
\maketitle

\section{Introduction}
\label{sec:introduction}

\begin{table*}[t]
\caption{\label{tab:intro-comparison}Simulating $e^{-itH}$ by a unitary circuit to precision $\varepsilon$, for a time-independent Hamiltonian. \emph{Upper block:} simulation costs for $H=\sum_{a=1}^{m}H_a$ with $\norm{H_a}\le1$ and a fixed number $m$ of generators.
The cost is quantified in terms of the number of operations named in the last column, where one exponential is an application of $e^{-i\tau H_a}$ for a single generator $H_a$. \emph{Lower block:} local Hamiltonians on $N$ sites in $d$ dimensions. The cost is specified in terms of circuit depth.
Ancilla qubits are those beyond the system register, $p$ is a fixed order, and $L$ is the number of unitaries in the linear combination that block-encodes $H$. Constants depending on the Hamiltonian, such as on $m$ or $d$, are not explicitly detailed. \Cref{app:related-work} gives more details and also discusses extrapolation~\cite{RWW24,WW25} and randomized methods~\cite{ZSJZ25,WZWZSZGL26,GD24,KK25,MKRST26} that estimate expectation values.}
\small
\setlength{\tabcolsep}{1pt}
\begin{tabular*}{\textwidth}{@{\extracolsep{\fill}}lcll@{}}
\toprule
Method & Ancilla qubits & Cost & Counted in\\
\midrule
\multicolumn{4}{@{}l}{\emph{General Hamiltonians}}\\
QSP and qubitization~\cite{LC17a,LC19,GSLW19} & $O(\log L)$ & $O(t+\log(1/\varepsilon))$ & block-encoding queries\\
Multiproduct formulae~\cite{LKW19,AAT24,MIF26} & $O(\log\log(t/\varepsilon))$ & $O(t\log^{2}(t/\varepsilon))$ & controlled exponentials\\
Suzuki--Trotter, fixed order~\cite{S91,BACS07,CSTWZ21} & 0 & $O(t^{1+1/p}\varepsilon^{-1/p})$ & exponentials\\
Suzuki--Trotter, optimized order~\cite{BACS07,WBHS10} & 0 & $t\,e^{O(\sqrt{\log(t/\varepsilon)})}$ & exponentials\\
\textbf{POLY (This work)} & 0 & $O(t\log^{4/3}(t/\varepsilon)\log^{2}\log(t/\varepsilon))$ & exponentials\\
\midrule
\multicolumn{4}{@{}l}{\emph{Local Hamiltonians on $N$ sites}}\\
Multiproduct formulae~\cite{LKW19,M26} & $O(\log\log(Nt/\varepsilon))$ & $O(N^{1/(p+1)}t\log^{2}(Nt/\varepsilon)\log^{2}\log(Nt/\varepsilon))$ & circuit depth\\
Lieb--Robinson blocks~\cite{HHKL18} & $O(N)$ & $O(t\log^{d}(Nt/\varepsilon))$ & circuit depth\\
Lieb--Robinson, corrected blocks~\cite{SWYYZ26} & $O(N)$ & $O(t\log(Nt/\varepsilon))$ & circuit depth\\
Suzuki--Trotter, fixed order~\cite{CS19,CSTWZ21} & 0 & $O(t(Nt/\varepsilon)^{1/p})$ & circuit depth\\
\textbf{POLY (This work)} & 0 & $O(t\log^{2}(Nt/\varepsilon)\log^{2}\log(Nt/\varepsilon))$ & circuit depth\\
\bottomrule
\end{tabular*}
\end{table*}

Product formulae approximate the exponential of a sum of operators by a product of exponentials of the individual terms. Since the work of Lie and Trotter~\cite{T59}, they have been developed into symmetric and higher-order splittings~\cite{S68,S76,R83,FR90,Y90,S90,S91,S92}, notably Strang's symmetric second-order splitting~\cite{S68} and the compositions of Yoshida and Suzuki, which reach arbitrary even order by composing a second-order formula with itself at different time steps~\cite{Y90,S90,S91}. They are now a mature part of numerical analysis~\cite{MQ02,HLW06,BCM24}. Product formulae are used wherever a complicated evolution is computed from simpler pieces, including quantum Monte Carlo~\cite{S76b,HSSB82}, structure-preserving integrators for molecular dynamics~\cite{V67,TBM92} and celestial mechanics~\cite{WH91}, grid-based solutions of the time-dependent Schr\"odinger equation~\cite{FFS82}, and tensor-network simulations~\cite{V04,WF04}.

We state the problem for the real-time evolution of a quantum system, computed either on a classical computer, as in the grid-based and tensor-network simulations above, or on a quantum computer. This task, called Hamiltonian simulation, is to approximate $U(t)=e^{-itH}$ for a Hamiltonian $H=\sum_{a=1}^{m}H_a$ by an operator $V$ with error $\norm{V-U(t)}\le\varepsilon$, where $\norm{\cdot}$ is the operator norm. We call the terms $H_a$ \emph{generators} and assume that $e^{-i\tau H_a}$ can be applied exactly for every generator and every real time $\tau$. A product formula constructs $V$ as an ordered product of these exponentials. We measure its cost by the number of exponentials, as a function of the evolution time $t$, the target precision $0<\varepsilon<1$, and the system size.

The weakness of product formulae is the cost of accuracy over long times. One step of a Suzuki--Trotter formula $S_p$ of order $p$, with time step $\Delta$, has error $O(\Delta^{p+1})$. Repetition then gives an upper bound of $O(t^{1+1/p}\varepsilon^{-1/p})$ exponentials, which is superlinear in $t$. For local Hamiltonians on $N$ sites, whose terms each act on at most $k$ neighboring sites, the same overhead appears in the spacetime volume, as $(Nt)^{1+1/p}$~\cite{CS19,CSTWZ21}. The standard way to improve the scaling is to raise the order. However, higher order has been regarded as exponentially expensive: Suzuki's recursion multiplies the number of Strang stages by five for each increase of the order by two, so that one step of even order $p$ consists of $5^{p/2-1}$ Strang stages~\cite{S90,S91,CSTWZ21}. Here a \emph{Strang stage} is one application of Strang's second-order formula, and the inverse of a stage is again a stage. Balancing the better scaling against the cost per step, with an order proportional to $\sqrt{\log(t/\varepsilon)}$, gives $t\,e^{O(\sqrt{\log(t/\varepsilon)})}$ exponentials~\cite{BACS07,WBHS10}, an overhead over linear cost that is still superpolylogarithmic.

Since the cost of simulating generic Hamiltonians must grow at least linearly in $t$~\cite{BACS07,HHKL18}, this overhead raises a basic question about the capability of product formulae: \emph{can a deterministic, ordered product of the exponentials $e^{-i\tau H_a}$, acting on the system qubits alone, approximate $e^{-itH}$ to precision $\varepsilon$ with $O(t\,\mathrm{polylog}(t/\varepsilon))$ exponentials?} Intuition suggests that the answer is no. As long as the cost per step grows exponentially with the order, such a cost seems out of reach. Much work has improved product formulae, by optimizing compositions at finite orders~\cite{Y90,MQ02,HS05,BZ20,O23,MCPBSB25,MO26,MO26b}, refining their error analysis~\cite{WBHS10,PZ12,SM23,AFL22,MIF26,BFHJY24,HHBFY25,LWFGC26,XXW26}, and adapting them to more problem structure~\cite{BCDGMS25,BMBSGACGA24,LSTT23,S16,PKB26,HP18}. Despite these improvements, the cost per step of known constructions still grows exponentially with the order, as does the number of order conditions, the equations that the time steps must satisfy to reach a given order~\cite{S92,MQ02,BCM24}. For these reasons, product formulae are used at low order in practice~\cite{CMNRS18}. Yet it has not been proven that this exponential overhead is unavoidable, and known lower bounds~\cite{HHBFY25,KG26,ZAH26} do not exclude such a product formula.

The question is particularly important in quantum computing, because product formulae are among the most widely used quantum algorithms. Simulating quantum dynamics has been one of the primary motivations for building a quantum computer~\cite{F82,L96,AL97}, and early work by Lloyd provided blueprints for a universal quantum simulator based on a product formula~\cite{L96}. Hamiltonian simulation remains one of the most promising applications of quantum computers, both in its own right and as a subroutine of algorithms such as phase estimation and ground-state preparation~\cite{DMBBCGHKKKSWB25,LT22,DLT22}; for recent reviews see~\cite{MRTC21,WYFZ26,O26}. The simplicity of product formulae has made them the method of choice in digital quantum-simulation experiments~\cite{LHNMGZSBRKHZBR11,KEAWVRNWZTK23}. They also underlie many resource estimates in chemistry and condensed-matter physics~\cite{WBCHT14,RWSWT17,KGBWMSJRFANB20,C22,PHWWDT15,BMWAW15,MCLFAJANGM26,BIB25}, where they often compare favorably with asymptotically better methods discussed below (\cref{app:related-work})~\cite{CMNRS18,KGBWMSJRFANB20,C22}, in part because their error is governed by nested commutators of the generators~\cite{CSTWZ21,CS19,L22,SHC21} and observed errors are often far smaller than the worst-case bounds suggest~\cite{TCSCG20,CXZY26,SS21,HSA24,ZZSLC22,CB24,ZZC25,FCZ25,HHZ19,YXZ24}. Removing the exponential overhead of higher order could combine these practical advantages with better asymptotic scaling.

Better scaling was also the motivation for post-Trotter quantum algorithms based on linear combinations of unitaries and quantum signal processing~\cite{BCCKS15,LC17a,LC19,GSLW19} (\cref{app:related-work}). These algorithms reach a cost linear in $t$ up to polylogarithmic factors, and even $O(t+\log(1/\varepsilon))$. However, they access $H$ through a block encoding, a larger unitary that contains a rescaled $H$ as a block, and therefore require ancilla qubits and controlled operations. Product formulae, in contrast, act on the system qubits alone and need no controlled operations~\cite{L96,BACS07}. Furthermore, block encodings 
do not respect the geometric structure of Hamiltonian interactions; the best known algorithms for local Hamiltonians on a lattice instead decompose the evolution into small blocks using Lieb--Robinson bounds, and again simulate each block with algorithms that use ancilla qubits~\cite{HHKL18,SWYYZ26}. These ancilla-assisted techniques are specific to quantum computing, whereas product formulae are used much more broadly in classical simulations as we noted above.

\label{sec:intro-results}

\textbf{In this work} we show that a single product formula, acting on the system qubits alone, can approximate $e^{-itH}$ to precision $\varepsilon$ with $O(t\,\mathrm{polylog}(t/\varepsilon))$ exponentials. \textbf{We thereby answer the basic question above in the affirmative:} the exponential overhead of higher order is not intrinsic to product formulae. We construct a product formula, POLY~\footnote{The name refers to polygon cancellation (\cref{sec:error-cancellation}) and to the polynomial growth of the number of exponentials with the cancellation order.}, whose error decreases exponentially with the cancellation order $q\ge2$, which corresponds to the order $p=2q$ of a Suzuki--Trotter formula, while the number of exponentials it uses grows only polynomially, as $\widetilde{O}(q^2)$. In the long-time regime, the circuit simulates $e^{-itH}$ with $O(t\log^{4/3}(t/\varepsilon)\,\log^2\log(t/\varepsilon))$ exponentials (see \cref{tab:intro-comparison}). For local Hamiltonians on $N$ sites, the circuit depth of our product formula is $O(t\log^2(Nt/\varepsilon)\log^2\log(Nt/\varepsilon))$, with a power of the logarithm independent of the lattice dimension and without ancilla qubits.

\begin{figure*}[t]
 \centering
\includegraphics[width=\textwidth]{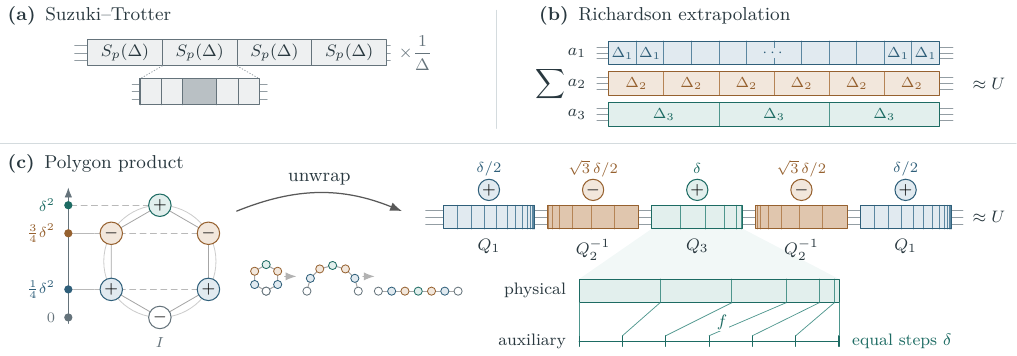}
 \caption{\textbf{Combining approximations additively and multiplicatively.}
 Suppose we want to simulate the time evolution $U=e^{-iHt}$ for time $t=1$. Each bar is a circuit acting on the system qubits (gray wires), divided into Strang stages whose widths are proportional to their physical time steps.
 (a)~Suzuki's recursion. One step $S_p(\Delta)$ of order $p$ is repeated $1/\Delta$ times. Each step consists of $5^{p/2-1}$ Strang stages, shown for $p=4$; the shaded stage has a negative time step.
 (b)~Richardson extrapolation cancels errors by a linear combination, with coefficients $a_j$, of Strang approximations $S_2(\Delta_j)^{1/\Delta_j}$ with uniform time steps $\Delta_j$. Multiproduct formulae implement this linear combination coherently with ancilla qubits. When applied to measured expectation values, it needs no ancilla qubits but does not produce a single unitary.
 (c)~The polygon product instead multiplies approximations $Q_j=Q(\delta_j)$ and their exact inverses (shaded). Left: the vertices of a regular $2q$-gon, here $q=3$, projected onto the vertical axis, give the squared auxiliary time steps $\delta_j^2=\delta^2\sin^2(\pi j/(2q))$, and their signs alternate between $Q_j$ and $Q_j^{-1}$. Cutting the polygon at the vertex $\delta_0=0$, which contributes the identity, and unwrapping it gives the order of the factors. Bottom right: each $Q(\delta)$ places its Strang stages at equal steps $\delta$ in auxiliary time, which the map $f$ sends to unequal physical time steps, so that $\delta$ can take any positive real value (\cref{sec:continuous-family-construction}). Note that no ancilla qubits are used.
}

 \label{fig:construction-overview}
\end{figure*}

Our construction works as follows. Instead of canceling the leading error term repeatedly, as Suzuki's recursion does, we cancel the first $q{-}1$ error terms at once by multiplying approximations of the same unit-time evolution with different time steps (\cref{fig:construction-overview}). Consider the second-order product formula $S_2(\Delta)^{1/\Delta}$, which consists of $1/\Delta$ Strang stages with equal time steps $\Delta$. Its error has an expansion in even powers of $\Delta$ whose coefficients do not depend on $\Delta$. We first show that if $\Delta$ could take any positive real value, multiplying $2q{-}1$ such approximations, with alternating forwards and inverse evolutions,
at the time steps $\Delta\sin(\pi j/(2q))$ would cancel all error terms below order $\Delta^{2q}$ (\cref{fig:construction-overview}(c)), even though the coefficients of the expansion are operators that need not commute. We call this \emph{polygon cancellation} (\cref{sec:error-cancellation}). The time step of $S_2(\Delta)^{1/\Delta}$, however, is restricted to $\Delta=1/r$ for integer $r$, whereas the first cancellation, for example, combines $\Delta$ and $\Delta/\sqrt2$. We remove this restriction by introducing an auxiliary time: the Strang stages are placed at equal steps $\delta$ in auxiliary time and mapped to unequal steps in physical time, with the total physical time kept equal to one (\cref{fig:construction-overview}(c)). The resulting circuits $Q(\delta)$ are defined for every positive real auxiliary time step $\delta$, and we show that their errors follow an even expansion in $\delta$ closely enough for polygon cancellation to apply (\cref{sec:continuous-family-construction}). With $\delta$ chosen so that the error decreases exponentially in $q$, the entire product uses $\widetilde O(q^2)$ Strang stages over unit time, whereas Suzuki's formula of order $p=2q$ uses $5^{q-1}$ Strang stages in each step (\cref{fig:construction-overview}(a)). The advantage is asymptotic, and the low orders drawn in the figure are purely illustrative.

Richardson extrapolation~\cite{R11,RG27} also cancels error terms between approximations with different time steps, but by a linear combination (\cref{fig:construction-overview}(b)). Multiproduct formulae~\cite{LKW19,M26} similarly take linear combinations of product formulae and require additional ancilla qubits for an index register to implement the linear combination.
In contrast, in the present work, we multiply individual product formulae, so the entire product is a single circuit acting only on the system qubits (\cref{fig:construction-overview}(c)). Related cancellation ideas also appear in splitting methods and, most closely, in dynamical decoupling and composite pulses, which we detail in \cref{app:related-work}.

The rest of the paper is organized as follows. \Cref{sec:main-result} fixes the setting and notation, states our main result, \cref{thm:main-informal}, and outlines its proof. The next two sections give the two ingredients of the proof. \Cref{sec:error-cancellation} presents polygon cancellation. \Cref{sec:continuous-family-construction} constructs the circuits $Q(\delta)$, states their error and number of exponentials at fixed $q$ (\cref{thm:order-cost-informal}), and completes the proof of \cref{thm:main-informal} by choosing $q$ and dividing the evolution time. \Cref{sec:discussion} discusses practical relevance, possible improvements of the bounds, and other uses of polygon cancellation.
\section{Setting and main result}
\label{sec:main-result}

\subsection{Setting and notation}
\label{sec:setting}
We present the construction for a fixed number of generators, $H=\sum_{a=1}^m H_a$. The generators $H_a$ are Hermitian operators of finite norm, and we assume $e^{-i\tau H_a}$ can be applied exactly for every $a$ and every real $\tau$.

\begin{definition}[Cost of a product formula]
\label{def:native-exponential-count}
For $H=\sum_a H_a$, the cost of a product formula $V$ is its number of exponentials $e^{-i\tau H_a}$. When $V$ approximates $e^{-itH}$ to precision $\varepsilon$ in operator norm, we write $D(tH,\varepsilon)$ for its cost.
\end{definition}

A Strang stage $S_2(\tau)$ (\cref{eq:strang-step}) consists of $2m-1$ exponentials, so counting Strang stages instead changes the cost only by a constant factor.

\begin{definition}[Local Hamiltonians and their decomposition]
\label{def:local-hamiltonian}
An operator on $N$ sites is \emph{$k$-local} if it can be written as a sum of terms each supported on at most $k$ sites. For simulating local Hamiltonians we assume a given decomposition $H=\sum_{a=1}^m H_a$ into $m$ layers. Each layer $H_a$ is a sum of Hermitian terms of norm at most one, each supported on at most $k$ sites, with pairwise disjoint supports within the layer.
\end{definition}

For such a local decomposition, each exponential $e^{-i\tau H_a}$ is a product of commuting gates that act on disjoint sets of at most $k$ sites and can be applied in parallel, so the circuit depth is $O(D)$ and the number of gates is $O(ND)$.
Product formulae are exact when all the generators commute. Otherwise, their approximation errors directly depend on commutators between the generators~\cite{CSTWZ21}.

\subsection{Main result}
\label{sec:main-theorem}

We construct a product formula that simulates a time-independent Hamiltonian to precision $\varepsilon$ with a number of exponentials linear in $t$ up to polylogarithmic factors. We give the scaling  for a fixed number $m$ of generators $H_a$ and, in the local case, a fixed number $k$ of sites on which each term acts. The implied constants in \cref{thm:main-informal} depend only on $m$ and, in part~(b), on $k$. \Cref{app:proof-map} gives the formal statements of this theorem and supporting results.

\begin{informaltheorem}[Product formulae with polylogarithmic overhead, informal]
\label{thm:main-informal}
For $t\ge1$ and small $\varepsilon$, a polygon product formula approximates $e^{-itH}$ to operator-norm error $\varepsilon$ with the following costs.

\noindent(a) \emph{General Hamiltonians.} Put $\lambda=\sum_a\|H_a\|$ and $\ell=\log(\lambda t/\varepsilon)$. The number of required exponentials is
\[ O\!\left((\lambda t\ell^{4/3}+\ell^2)\log^2\ell\right). \]

\noindent(b) \emph{Local Hamiltonians.} For a layer decomposition on $N$ sites as in \cref{def:local-hamiltonian}, the number of required exponentials is
\[ O\!\left(t\log^2(Nt/\varepsilon)\,\log^2\log(Nt/\varepsilon)\right). \]
\end{informaltheorem}

Both parts repeat the same circuit and differ only in the length of the time interval that one circuit simulates, as a general time $t$ is built up from these small intervals. The $\ell^2$ term in (a) accounts for the cost of a single interval. For sufficiently long times $\lambda t\ge\ell^{2/3}$ the term $\lambda t\ell^{4/3}$ dominates, giving the scaling quoted in \cref{tab:intro-comparison}. For (b), locality lets us use an interval of fixed length even as the number of sites grows. The formal bound in \cref{app:proof-map-results} retains the commutator norm $\alpha=\sum_{a<b}\|[H_a,H_b]\|$. Specifically, the logarithm in (a) becomes $\ell = \log(\alpha t/(\lambda\varepsilon))$, giving a tighter bound when the generators nearly commute.
\subsection{Proof outline}
\label{sec:proof-overview}

The proof of~\cref{thm:main-informal} combines the two ingredients sketched above: polygon cancellation and the approximation circuits $Q(\delta)$. We then bound the error and the number of exponentials, first at fixed $q$ and then for time $t$ and precision $\varepsilon$. \Cref{fig:construction-overview} gives a schematic and \Cref{app:proof-map} gives more details of the strategy.

\paragraph{Polygon cancellation.} Suppose that approximations $Q(\delta)$ to the same evolution $U$ are available at any time step $\delta>0$, and that they share an even error expansion, $Q(\delta)U^\dagger=I+\delta^2E_2+\delta^4E_4+\cdots$, with coefficients independent of $\delta$. Multiplying $2q-1$ of them, alternately forward and inverted, at  auxiliary time steps $\delta_j=\delta\sin(\pi j/(2q))$ cancels every term below order $\delta^{2q}$ (see \cref{sec:error-cancellation}, \cref{lem:polygon-cancellation}).

\paragraph{A continuous auxiliary time step.} The circuits $Q(\delta)$ consist of Strang stages with nonuniform physical time steps, obtained by mapping equal steps $\delta$ of an auxiliary time onto physical time. A final Strang stage covers any remaining physical time. This produces a finite circuit for each step size required, with the same total physical time for every auxiliary time step (\cref{sec:continuous-family-construction}).

\paragraph{Error and cost.} Combining these circuits gives the polygon circuit for one unit-time evolution. Then~\Cref{thm:order-cost-informal} shows that its error decreases exponentially in $q$ while its number of exponentials grows only polynomially. We then divide the evolution time $t$ into smaller intervals and choose $q$ so that their errors sum to at most $\varepsilon$, giving \cref{thm:main-informal}.
\section{Polygon cancellation}
\label{sec:error-cancellation}

We first show how we can use products of approximations to the same time evolution to cancel their errors, assuming for now that these approximations are available at any small step $\delta$. \Cref{sec:continuous-family-construction} will address how these conditions are met by finite circuits. We use the setting and notation of \cref{sec:setting}. We work at unit time, with target $U=e^{-iH}$; \cref{sec:finite-circuit-cancellation} returns to general $t$.

\subsection{The Strang formula}
\label{sec:strang-formula}
The Strang formula is a symmetric product formula defined as
\begin{equation}
 \begin{aligned}
 S_2(\Delta)={}&e^{-i\Delta H_1/2}\cdots e^{-i\Delta H_{m-1}/2}\,e^{-i\Delta H_m}\\
 &\times e^{-i\Delta H_{m-1}/2}\cdots e^{-i\Delta H_1/2},
 \end{aligned}
 \label{eq:strang-step}
\end{equation}
where $\Delta$ is a real-valued physical time step. Each $S_2(\Delta)$ is one Strang stage. Dividing the unit-time target evolution $e^{-iH}$ into $r$ equal time steps of size $\Delta = 1/r$ produces a second-order product formula
\begin{equation}
 S_2(\Delta)^{r},\qquad \Delta = \frac{1}{r}, \qquad  r=1,2,\ldots.
 \label{eq:uniform-strang}
\end{equation}
An individual Strang stage has error $O(\Delta^{3})$, and so the error of the product is $O(\Delta^{2})$. 

\subsection{Suzuki's recursion}
\label{sec:higher-order}
A product formula $S_p(\Delta)$ has \emph{order} $p$ if $S_p(\Delta)=e^{-i\Delta H}+O(\Delta^{p+1})$. The Strang formula has order two. Suzuki's recursion constructs a formula of every even order from the Strang formula~\cite{S90,S91},
\begin{equation}
 \begin{aligned}
 S_p(\Delta)&=S_{p-2}(u_p\Delta)^2\,S_{p-2}\bigl((1-4u_p)\Delta\bigr)\,S_{p-2}(u_p\Delta)^2,\\
 u_p&=\frac{1}{4-4^{1/(p-1)}},\qquad p=4,6,\ldots.
 \end{aligned}
 \label{eq:suzuki-recursion}
\end{equation}
Each increase of the order by two multiplies the number of Strang stages by five, so that $S_p(\Delta)$ consists of $5^{p/2-1}$ Strang stages. Repeating it $r=1/\Delta$ times gives a product formula for the unit-time evolution. We write $D^{\mathrm{Suz}}_p(\Delta)$ for the number of exponentials it uses. Then
\begin{equation}
 \begin{aligned}
 \|S_p(\Delta)^{r}-U\|&\le C_p\Delta^{p},\\
 D^{\mathrm{Suz}}_p(\Delta)&=O\!\left(\frac{5^{p/2-1}}{\Delta}\right),
 \end{aligned}
 \label{eq:suzuki-order-cost}
\end{equation}
where the constant $C_p$ in the first line depends on $p$ and on norms of nested commutators of the $H_a$~\cite{CSTWZ21}. A higher order thus improves the dependence of the error on the physical time step, but the number of exponentials at a given time step grows exponentially with the order. For a target precision $\varepsilon$, it suffices to take $\Delta\le(\varepsilon/C_p)^{1/p}$ in \cref{eq:suzuki-order-cost}, giving $D^{\mathrm{Suz}}_p=O(5^{p/2}C_p^{1/p}\varepsilon^{-1/p})$. Optimizing $D^{\mathrm{Suz}}_p$ over $p$, with the bounds on $C_p$ of Refs.~\cite{BACS07,WBHS10}, gives $D^{\mathrm{Suz}}_p=e^{O(\sqrt{\log(1/\varepsilon)})}$ for $p$ proportional to $\sqrt{\log(1/\varepsilon)}$. This optimized-order estimate still has a superpolylogarithmic dependence on the inverse precision.

\subsection{Cancellation of the leading error}
\label{sec:general-cancellation}
Suzuki's recursion constructs a high-order product formula by iteratively combining lower-order ones. As \cref{eq:suzuki-order-cost} shows, however, it requires a number of Strang stages that is exponential in the order. Our construction also starts from the Strang formula as a base, but achieves high orders without exponentially many Strang stages. The second-order product formula of \cref{eq:uniform-strang} has a useful expansion whose coefficients are shared by any choice of $\Delta$. Since $S_2(-\Delta)=S_2(\Delta)^{-1}$, the logarithm of $S_2(\Delta)$ contains only odd powers of $\Delta$. Multiplying it by $r=1/\Delta$ and exponentiating gives, for small $\Delta$,
\begin{equation}
 S_2(\Delta)^{r}U^\dagger
 =I+\Delta^{2}\widehat E_2+\Delta^{4}\widehat E_4+\cdots,
 \label{eq:uniform-shared-expansion}
\end{equation}
where the error coefficients $\widehat E_{2k}$ depend on the generators but not on $\Delta$. This independence is important since it allows us to change the physical time step without changing the error coefficients.

The cancellation uses only the form of this expansion. Suppose for the moment that, as an analogue of $S_2(\Delta)^{r}$, a unitary approximation $Q(\delta)$ to $U$ is available at every sufficiently small real $\delta>0$, with an expansion of the same form as \cref{eq:uniform-shared-expansion}:
\begin{equation}
 Q(\delta)U^\dagger=F(\delta^2),\qquad F(z)=I+zE_2+z^2E_4+\cdots,
 \label{eq:shared-error-family}
\end{equation}
with coefficients $E_{2k}$ independent of $\delta$. We call $\delta$ the \emph{auxiliary time step}. For $S_2(\Delta)^{r}$, it is simply the physical time step $\Delta=1/r$, whereas in the circuits of \cref{sec:continuous-family-construction} the physical time steps are no longer equal to $\delta$. We call $F$ the \textit{error factor}, and write $F_j=F(\delta_j^2)=Q(\delta_j)U^\dagger$ for auxiliary time steps $\delta_j$.

We now want to cancel these errors by combining approximations at different auxiliary time steps. Unlike Richardson extrapolation~\cite{R11,RG27}, which takes linear combinations of $Q(\delta)$, we cancel them by multiplying approximations, so that the circuit remains a product formula. If we arrange the factors in an alternating forward and inverse pattern, adjacent factors of $U$ and $U^\dagger$ cancel exactly, so the error factors multiply directly rather than being conjugated by $U$. Since one copy of $U$ must be left over, the product has an odd number of factors.

To see how the construction works, we first consider an example that can cancel the $\delta^2$ term with three factors leaving an error $O(\delta^4)$. Consider \(Q(\delta_1)Q(\delta_2)^{-1}Q(\delta_3)=F_1F_2^{-1}F_3U\). We take the auxiliary time steps to form a palindrome, $\delta_1=\delta_3=a\delta$ and $\delta_2=\delta$, which makes the product symmetric. The coefficient of $\delta^2$ in the product is then $(2a^2-1)E_2$, and choosing $a=1/\sqrt2$ gives
\begin{equation}
 Q(\delta/\sqrt2)Q(\delta)^{-1}Q(\delta/\sqrt2)
 =U+O(\delta^4).
 \label{eq:three-factor-cancellation}
\end{equation}
So three approximations, combined in a single product of unitaries, can cancel the leading $O(\delta^2)$ error purely by a suitable choice of their auxiliary time steps.

However, the second-order product formula of~\cref{eq:uniform-strang} is limited to physical time steps $\Delta = 1/r$ for integer $r$. With $\delta=\Delta=1/r$, the required value $\delta/\sqrt2$ would need $\sqrt2\,r$ stages, which is not an integer. The issue is that $r$ specifies both the width of a stage and how many there are. The immediate idea is rounding, but this perturbs the ratio of the steps, so the leading error is no longer canceled exactly. In multiproduct formulae~\cite{LKW19}, the coefficients of a linear combination can be reweighted to compensate; the factors of a product cannot. One of the main technical contributions of our work (\cref{sec:continuous-family-construction}) is therefore to find approximations $Q(\delta)$ in which $\delta$ can vary continuously, while they still approximate the same target evolution with a modest number of Strang stages.

Moreover, we want to cancel errors to arbitrary order, not just at $\delta^2$. The next subsection does this by a geometric construction, with a number of approximations that is directly proportional to the order.

\subsection{Cancellation to arbitrary order}
\label{sec:polygon-cancellation}
We now cancel errors to higher order with the \emph{polygon product}
\begin{equation}
 \begin{aligned}
 V_q(\delta)&=\prod_{j=1}^{2q-1}Q(\delta_j)^{(-1)^{j+1}}\\
 &=Q(\delta_1)\,Q(\delta_2)^{-1}\,Q(\delta_3)\cdots Q(\delta_{2q-1}),
 \end{aligned}
 \label{eq:polygon-product-def}
\end{equation}
and show that suitable auxiliary time steps $\delta_j$, taking $q$ distinct values, can produce an error of order $\delta^{2q}$. We call $q$ the cancellation order; it counts orders in $z=\delta^2$, and we reserve $p$ for orders in a physical time step. As in the three-factor example, adjacent factors of $U$ and $U^\dagger$ cancel, so that $V_q(\delta)U^\dagger=F_1F_2^{-1}F_3\cdots F_{2q-1}$.

\begin{figure}[!htbp]\centering\resizebox{\columnwidth}{!}{\begin{tikzpicture}\foreach \shift/\angle/\jx/\jy/\ttl/\fa/\fb/\fc/\fd/\fe/\ff in {0/0/1.65/-0.2/{$\theta=0$}/{$F(0)^{-1}$}/{$F(z/4)$}/{$F(3z/4)^{-1}$}/{$F(z)$}/{$F(3z/4)^{-1}$}/{$F(z/4)$},6.4/30/1.45/0.74/{$\theta=\pi/6$}/{$F(c_-z)^{-1}$}/{$F(z/2)$}/{$F(c_+z)^{-1}$}/{$F(c_+z)$}/{$F(z/2)^{-1}$}/{$F(c_-z)$}} {\begin{scope}[xshift=\shift cm]\draw[gray!55] (0,0) circle (1.35);\draw[dashed,gray] (-1.6,0)--(1.6,0);\draw[gray] (\angle:1.35)--(\angle+60:1.35)--(\angle+120:1.35)--(\angle+180:1.35)--(\angle+240:1.35)--(\angle+300:1.35)--cycle;\draw[->,gray!80!black] (50:2.6) arc (50:130:2.6);\ifnum\angle>0\draw[densely dotted,gray] (0,0)--(\angle:1.13);\draw[->,gray!80!black] (0:0.7) arc (0:\angle:0.7);\node[font=\small] at ({\angle/2}:0.98) {$\theta$};\fi\foreach \j/\fac in {0/\fa,1/\fb,2/\fc,3/\fd,4/\fe,5/\ff} {\pgfmathtruncatemacro{\parity}{mod(\j,2)}\ifnum\parity=0\node[circle,fill=red!12,draw=red!65!black,inner sep=2.5pt] at (\angle+60*\j:1.35) {$-$};\else\node[circle,fill=blue!12,draw=blue!65!black,inner sep=2.5pt] at (\angle+60*\j:1.35) {$+$};\fi\node[font=\small,anchor={\angle+60*\j+180}] at (\angle+60*\j:1.6) {\fac};}\node[font=\scriptsize,anchor=north west,inner sep=1pt] at (\jx,\jy) {$j=0$};\node[fill=white,inner sep=1.5pt] at (0,-0.38) {\ttl};\end{scope}}\end{tikzpicture}}
\caption{\label{fig:polygon}
The rotated product $W_3(\theta,z)$ for $q=3$ (hexagon). \Cref{fig:construction-overview}(c) shows the same polygon at $\theta=0$, with vertex $j=0$ at the bottom. Each vertex stands for one factor $F(c_j(\theta)z)^{\pm1}$ of the rotated product and is labeled with that factor. At $\theta=0$, the factors with $j\ge1$ come from the approximations $Q(\delta_j)$ or their inverses. The outer arrows indicate the ordering of factors, running counterclockwise starting at vertex $j=0$ on the right. The sign $+$ marks $F$ and $-$ marks its inverse $F^{-1}$. Each node is obtained from the vertex angle $\phi$ as $(1-\cos\phi)/2$. At $\theta=0$ (left), the nodes are $0,\tfrac{1}{4},\tfrac{3}{4},1,\tfrac{3}{4},\tfrac{1}{4}$. The first factor is $F(0)^{-1}=I$, leaving $W_3(0,z)=F(z/4)F(3z/4)^{-1}F(z)F(3z/4)^{-1}F(z/4)$. After rotation by half a vertex spacing $\theta=\pi/6$ (right, inner arrow), the nodes become $c_-,\tfrac12,c_+,c_+,\tfrac12,c_-$, where $c_\pm=(2\pm\sqrt3)/4$. Vertices reflected across the dashed axis now have equal node values but opposite signs, so adjacent pairs cancel from the middle outward, leaving the identity overall.}
\end{figure}
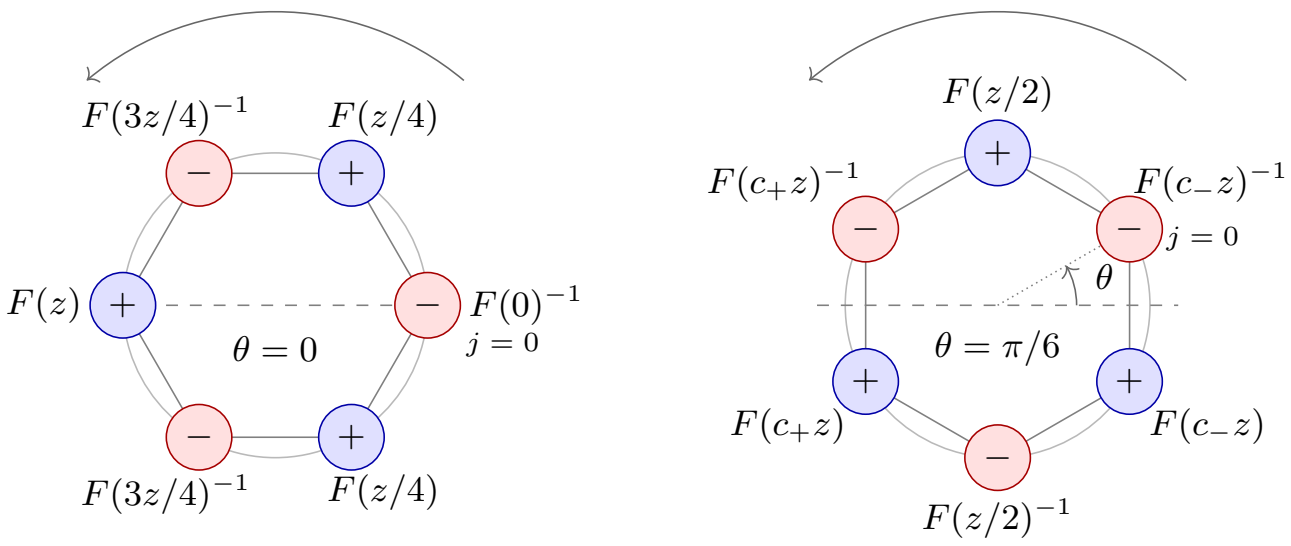

Write the squared auxiliary time steps as $\delta_j^2=c_j\delta^2$. In the three-factor example \cref{eq:three-factor-cancellation}, canceling the leading order required $c_j=\tfrac12,1,\tfrac12$. For $q=3$, canceling the next order requires a five-factor product with $c_j=\tfrac14,\tfrac34,1,\tfrac34,\tfrac14$ (see~\cref{fig:polygon}). Checking the next few orders (\cref{app:polygon-cancellation}) reveals a pattern that matches the squared-sine nodes
\begin{equation}
\label{eq:squared-sine-nodes}
 c_j=\sin^2\frac{\pi j}{2q},\qquad j=1,\ldots,2q-1.
\end{equation}
In $V_qU^\dagger$, a single coefficient $E_{2k}$ appears with the scalar factor $\sum_j(-1)^{j+1}c_j^k$, and this choice of $c_j$ makes these factors vanish for $1\le k<q$. Since $c_{2q-j}=c_j$, the nodes form a palindrome.

However, canceling the terms proportional to a single $E_{2k}$ is not enough, since $V_qU^\dagger$ also contains cross terms, products of several coefficients such as $E_2E_4$ and $E_4E_2$, which need not commute. The choice~\eqref{eq:squared-sine-nodes} turns out to cancel these terms too. To see why, note that the nodes have a geometric meaning. Place the vertices of a regular $2q$-gon on the unit circle, rotate it by $\theta$, and map each horizontal coordinate $x$ to $(1-x)/2$. This gives the rotated nodes $c_j(\theta)$ and the rotated product of error factors $W_q(\theta, z)$,
\begin{equation}
 \begin{aligned}
 c_j(\theta)&=\frac{1-\cos(\theta+\pi j/q)}2,\\
 W_q(\theta,z)&=\prod_{j=0}^{2q-1}
 F(c_j(\theta)z)^{(-1)^{j+1}},
 \end{aligned}
 \label{eq:rotated-polygon}
\end{equation}
with factors written from left to right in increasing $j$. At $\theta=0$, the first factor is $F(0)^{-1}=I$, and $W_q(0,\delta^2)=V_q(\delta)U^\dagger$ is the error factor of the whole polygon product. \Cref{fig:polygon} shows this geometry for $q=3$, embedded in a regular hexagon. 

 We therefore want a guarantee that every coefficient of $W_q(0,z)$ below order $z^q$ to vanish, including the cross terms. The following lemma states this identity.

\begin{lemma}[Polygon cancellation]
 \label{lem:polygon-cancellation}
 Fix an integer $q\ge2$. Let $F(z)=I+\sum_{k\ge1}z^kE_{2k}$ be a formal power series with arbitrary coefficients $E_{2k}$, for example the error factor in \cref{eq:shared-error-family}. Then
 \begin{equation}
 W_q(0,z)=\prod_{j=1}^{2q-1}F\!\left(z\sin^2\frac{\pi j}{2q}\right)^{(-1)^{j+1}}
 =I+O(z^{q}).
 \label{eq:polygon-word}
 \end{equation}
 That is, every coefficient of $z^n$ with $1\le n<q$ vanishes.
\end{lemma}
\begin{proof}[Proof sketch]
After rotating the $2q$-gon by half a vertex spacing, $\theta=\pi/(2q)$, equal-value nodes (\cref{eq:rotated-polygon}) are paired with opposite signs in reverse ordering, so adjacent inverse pairs cancel from the middle outward, giving $W_q(\pi/(2q),z)=I$. The same cancellation occurs at each of the $2q$ angles $\theta_k=(2k+1)\pi/(2q)$, so every coefficient of $z^n$ in $W_q(\theta,z)$ vanishes at these angles.  Each contribution to the coefficient of $z^n$ contains $n$ node factors, counted with multiplicity. By \cref{eq:rotated-polygon}, each node is a trigonometric polynomial of degree one in $\theta$. The coefficient is therefore a trigonometric polynomial in $\theta$ of degree at most $n$, and so cannot have more than $2n$ zeros unless it vanishes identically. For $n<q$ it has $2q>2n$ zeros, and so it vanishes for all $\theta$, including $\theta=0$. \Cref{app:polygon-cancellation} gives the full argument.
\end{proof}

\Cref{lem:polygon-cancellation} is a general algebraic result concerning ordered products of a single power series and its inverses, and may be of standalone interest. With only $2q-1$ factors, it cancels all terms below $z^q$, whereas the expansion itself contains exponentially many ordered products of the error coefficients below this order (\cref{app:polygon-cancellation}), and canceling them by direct matching leads to a blow-up of order conditions. Here a single symmetric choice of auxiliary time steps cancels all of them at once.

We remark that the alternating signs $(-1)^{j+1}$ also appear in the coefficients of well-conditioned multiproduct formulae~\cite{LKW19}, where they multiply coefficients of varying magnitude in a linear combination, and that the squared-sine nodes of~\cref{eq:squared-sine-nodes} are a shifted version of the Chebyshev--Lobatto nodes~\cite{BT04}. The nodes also occur in Uhrig dynamical decoupling~\cite{U07,YL08,PU10}, where a similar cancellation idea suppresses the coupling of a system to its environment. Here the nodes are used to cancel the discretization error between complete approximations to the same evolution.

For approximations of $U$, \cref{lem:polygon-cancellation} gives the error of the polygon product. If $Q(\delta)U^\dagger=F(\delta^2)$ as in \cref{eq:shared-error-family}, then $V_q(\delta)U^\dagger=W_q(0,\delta^2)$ has no terms below $\delta^{2q}$, so that
\begin{equation}
 \bigl\|V_q(\delta)-U\bigr\|=O(\delta^{2q}),\qquad\delta\to0^+.
 \label{eq:polygon-block}
\end{equation}
For $q=2$ this is the three-factor example, and for $q=3$ the five-factor product of \cref{fig:polygon} has error $O(\delta^6)$. Crucially, the bound~\eqref{eq:polygon-block} does not require the expansion~\eqref{eq:shared-error-family} to match exactly to all orders. Suppose that $Q(\delta)U^\dagger=P(\delta^2)+O(\delta^{2q})$, where $P(z)=I+zE_2+\cdots+z^{q-1}E_{2q-2}$. Replacing each factor $Q(\delta_j)U^\dagger$ of the polygon product by $P(c_j\delta^2)$, and each inverse by $P(c_j\delta^2)^{-1}$, changes the product by $O(\delta^{2q})$ at fixed $q$, since these factors and their inverses are bounded for small $\delta$. The remaining product is $W_q(0,\delta^2)$ for $F=P$, which equals $I+O(\delta^{2q})$ by \cref{lem:polygon-cancellation}. Hence \cref{eq:polygon-block} holds whenever, for small $\delta$, the expansion holds up to order $\delta^{2q}$ with coefficients $E_{2k}$ independent of $\delta$:
\begin{equation}
 Q(\delta)U^\dagger=I+\delta^2E_2+\cdots+\delta^{2q-2}E_{2q-2}+O(\delta^{2q}).
 \label{eq:approximate-expansion}
\end{equation}
We call this the \textit{expansion condition}. Our aim is to construct, for every $\delta>0$, approximations $Q(\delta)$ that satisfy the condition of~\cref{eq:approximate-expansion}.

Two questions remain. Can such approximations be realized by finite circuits at all the required steps $\delta_j$? And how do the coefficient of the remainder in \cref{eq:polygon-block} and the number of exponentials of each factor depend on $q$? The next section answers both questions for product formulae constructed only from Strang stages, with a number of exponentials polynomial in $q$. This is where the polylogarithmic overhead of \cref{thm:main-informal} comes from. The formal statements in \cref{app:proof-map-results} also provide the dependence on the Hamiltonian parameters.
\Needspace{8\baselineskip}
\section{Finite circuits with a continuous auxiliary time step}
\label{sec:continuous-family-construction}

We now answer the two questions raised at the end of \cref{sec:error-cancellation}. The product $S_2(\Delta)^r$ of \cref{eq:uniform-strang} allows only the steps $\Delta=1/r$, whereas polygon cancellation needs the $q$ distinct auxiliary time steps $\delta_j$. We first construct, for every $\delta>0$, a finite circuit of Strang stages by dividing auxiliary time into equal intervals and mapping them onto physical time. We then show that these circuits satisfy \cref{eq:approximate-expansion} when they contain enough stages, state the error and the number of exponentials of the resulting polygon circuit (\cref{thm:order-cost-informal}), and prove \cref{thm:main-informal}.

\subsection{Auxiliary time and physical time steps}

\begin{figure*}[t]
 \centering
 \includegraphics[width=0.8\textwidth]{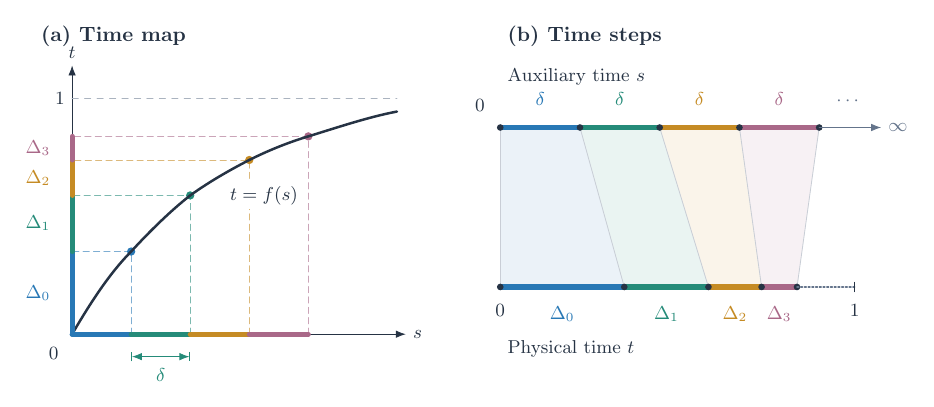}
 \caption{A uniform partition of auxiliary time produces unequal physical time steps. (a) The time map $f(s)$ approaches physical time one as $s\to\infty$. (b) Matching colors identify corresponding intervals. The auxiliary time step $\delta$ can take any positive real value, while the physical time steps are the differences $f((k+1)\delta)-f(k\delta)$.}
 \label{fig:auxiliary-time-map}
\end{figure*}

Let $s\in[0,\infty)$ denote auxiliary time and let the \emph{time map} $f$ be a smooth, strictly increasing map with $f(0)=0$ and $f(s)\to1$ as $s\to\infty$. As auxiliary time increases, the physical time $f(s)$ approaches the target unit time simulation, as illustrated in \cref{fig:auxiliary-time-map}(a). We take equal auxiliary intervals of width $\delta>0$. The lengths of their images under $f$ are the physical time steps
\[
 \Delta_k(\delta)
 =f((k+1)\delta)-f(k\delta),\qquad k=0,1,\ldots.
\]
 In \cref{fig:auxiliary-time-map}(b), matching colors identify an auxiliary interval of width $\delta$ and its physical image of width $\Delta_k$. The endpoint conditions on $f$ make the infinite sum of these physical time steps equal to one for every $\delta>0$. Changing $\delta$ therefore changes the sequence of physical time steps of the same target evolution, so that $\delta$ can take any positive real value, as required for cancellation.

Applying one Strang stage $S_2(\Delta_k)$ on each physical time step gives an infinite ordered product. A circuit, however, can contain only finitely many stages. The first $r$ stages cover the physical time \(\sum_{k=0}^{r-1}\Delta_k(\delta)=f(r\delta)\). The remaining stages, which we call the \emph{tail}, have the physical time $\tau_r=1-f(r\delta)$. Taking $r\delta$ sufficiently large makes the physical time of the tail small. To keep the total physical time equal to one for every finite circuit, we replace all the stages of the tail by one final Strang stage with the physical time of the tail. With the rightmost operator acting first, the finite circuit is
\begin{equation}
 Q_r(\delta)=S_2(\tau_r)\,S_2(\Delta_{r-1})\cdots S_2(\Delta_0),
 \label{eq:finite-normalized-circuit}
\end{equation}
which consists of $r+1$ Strang stages. We call $r$ the \emph{cutoff}. The circuit uses the same types of exponentials of the $H_a$ as the second-order product formula of \cref{eq:uniform-strang}, now at unequal physical time steps.

\subsection{The infinite Strang product and the cutoff}
\label{sec:infinite-strang-product}
We now show that the finite circuits satisfy \cref{eq:approximate-expansion} when the cutoff is large enough. At fixed $\delta>0$, they converge in operator norm as $r\to\infty$ to a unitary $Q_\infty(\delta)=\lim_{r\to\infty}Q_r(\delta)$, the ordered product of all the stages $S_2(\Delta_k)$, $k\ge0$, which we call the \emph{infinite Strang product} (\cref{lem:cutoff-selection} in \cref{app:cutoff-selection}). We first show that the infinite Strang product satisfies the expansion condition of \cref{eq:approximate-expansion}, and then choose the cutoff so that the finite circuits inherit this expansion. Although the physical time steps differ, every stage spans an auxiliary interval of the same width $\delta$. As for \cref{eq:uniform-shared-expansion}, the symmetry of the Strang stage makes the logarithm of one stage odd in $\delta$. Divided by $\delta$, it contains only even powers of $\delta$. For each $q$, \cref{app:finite-circuit-expansion} truncates it at order $\delta^{2q-2}$ and bounds the difference between the resulting evolution and $Q_\infty(\delta)$. This gives the shared even expansion required by the condition of \cref{eq:approximate-expansion}, with control of the remainder as $q$ grows.

The finite circuit differs from the infinite Strang product only in the tail. The final stage of $Q_r(\delta)$ and the stages that it replaces approximate the same evolution $e^{-i\tau_rH}$, each with an error of third order that is proportional to nested commutators of the $H_a$. Hence, for a fixed Hamiltonian, the cutoff error is $\|Q_r(\delta)-Q_\infty(\delta)\|=O(\tau_r^3)$ (\cref{lem:cutoff-selection}), and $Q_r(\delta)$ inherits the expansion of~\cref{eq:approximate-expansion} when the physical time of the tail is small enough. The number of stages that this requires is set by how fast $\tau_r=1-f(r\delta)$ decreases with $r\delta$. We take a time map that approaches one double exponentially, as in the change of variables of tanh--sinh quadrature~\cite{TM74quadrature,M05doubleexponential},
\begin{equation}
 f(s)=1-\exp(1-e^s),\qquad \tau_r=\exp(1-e^{r\delta}).
 \label{eq:chosen-time-map}
\end{equation}
The physical time of the tail is then at most a prescribed \emph{tail time} $0<\eta<1$ for
\begin{equation}
 r=\left\lceil\frac1\delta\log\!\left(1+\log\frac1\eta\right)\right\rceil .
 \label{eq:main-cutoff-stages}
\end{equation}
Choosing $\eta$ proportional to $\delta^{2q}$ makes the cutoff error at most of order $\delta^{2q}$, as required by \cref{eq:approximate-expansion}, and increases the number of stages beyond $1/\delta$ only by a factor $O(\log\log(1/\delta))$ at fixed $q$. Finite circuits of Strang stages are therefore available at every auxiliary time step $\delta>0$ with the expansion~\eqref{eq:approximate-expansion}.

\subsection{Error and number of exponentials at order \texorpdfstring{$q$}{q}}
\label{sec:finite-circuit-cancellation}

We now apply polygon cancellation to the finite circuits. Taking each $Q(\delta_j)$ in the polygon product~\eqref{eq:polygon-product-def} to be the finite circuit $Q_{r_j}(\delta_j)$ gives the \emph{polygon circuit}
\begin{equation}
 V_{q}(\delta)=\prod_{j=1}^{2q-1}Q_{r_j}(\delta_j)^{(-1)^{j+1}},\qquad
 \delta_j=\delta\sin\frac{\pi j}{2q}.
 \label{eq:finite-cancellation-word}
\end{equation}
Each inverse $Q_{r_j}(\delta_j)^{-1}$ uses as many exponentials as $Q_{r_j}(\delta_j)$. We write $D_q(\delta)$ for the number of exponentials of $V_q(\delta)$. At fixed $q$, the cutoffs of \cref{sec:infinite-strang-product} give error $O(\delta^{2q})$ with a cost proportional to $1/\delta$ up to a double logarithm. We now ask how small an auxiliary time step is needed as $q$ grows and how many exponentials it costs. Let $\rho_q$ be a constant depending on $q$ that bounds the auxiliary time step (\cref{app:proof-map-results}).

\begin{informaltheorem}[Error and number of exponentials at order $q$, informal]
\label{thm:order-cost-informal}
For $H=\sum_aH_a$ small enough at unit time (\cref{app:proof-map-results}), let $q\ge2$ be an integer and $0<\delta\le\rho_q$. With a suitable tail time $\eta$ in \cref{eq:main-cutoff-stages}, the polygon circuit approximates the unit-time evolution $U=e^{-iH}$ with error and a number of exponentials given by
\begin{equation}
\begin{aligned}
\|V_q(\delta)-U\| &= O\!\left(q^{-2}(\delta/\rho_q)^{2q}\right),\\
D_q(\delta)&=O\!\left(\frac{q^2\log q}{\delta/\rho_q}\log\!\bigl[q\log(e\rho_q/\delta)\bigr]\right).
\end{aligned}
\end{equation}
For any fixed ratio $0<\delta/\rho_q<1$, the error bound decreases exponentially in $q$. For example, taking $\delta=\rho_q/2$ gives
\begin{equation}
\|V_q(\delta)-U\| = O(4^{-q}/q^2), \qquad D_q(\delta)=O(q^2\log^2 q).
\label{eq:main-order-cost}
\end{equation}
\end{informaltheorem}

\begin{proof}[Proof sketch]
\Cref{app:proof-map} gives a map of the proof. Polygon cancellation removes all terms of the infinite Strang products below order $\delta^{2q}$, but it does not bound the terms that remain. These terms are built from nested commutators of the $H_a$ of degree growing with the order, so they can be large when $H$ is large. The condition that $H$ is small at unit time keeps them under control, so that for every $\delta\le\rho_q$ they sum to $O(q^{-2}(\delta/\rho_q)^{2q})$ (\cref{app:intrinsic-cost}). Each factor $Q_{r_j}(\delta_j)$ of the polygon circuit sits at vertex $j$ of the polygon (\cref{fig:polygon}). Replacing its tail by one Strang stage adds an error proportional to $\eta^3$ (\cref{lem:cutoff-selection}), and our choice of $\eta$ keeps the sum of these errors over the vertices $j=1,\ldots,2q-1$ below this remainder.

The number of exponentials is proportional to the number of Strang stages at the vertices $j=1,\ldots,2q-1$. At vertex $j$ the cutoff is $r_j\approx s_{\rm cut}/\delta_j$, where $s_{\rm cut}=\log(1+\log(1/\eta))$ is the auxiliary time at which the physical time of the tail falls to $\eta$. Although $\log(1/\eta)$ grows linearly in $q$, the double-exponential decay of the tail gives $s_{\rm cut}=O(\log[q\log(e\rho_q/\delta)])$, the last factor of $D_q$. The remaining factor is $\sum_j1/\delta_j$, where the auxiliary time steps $\delta_j$ are set by the positions of the polygon vertices through \cref{eq:squared-sine-nodes}. This sum is $O(q\log q/\delta)$, which gives the factor $q^2\log q/(\delta/\rho_q)$ of \cref{thm:order-cost-informal} (\cref{app:statements}).
\end{proof}

For comparison, \cref{thm:order-cost-informal} has the same form as the bound~\eqref{eq:suzuki-order-cost} for a unit-time Suzuki circuit of the same order $p=2q$, with $\delta/\rho_q$ in place of the physical time step $\Delta$. In the error, the $q$-dependent constant $C_{2q}$ becomes $O(q^{-2})$ for a fixed Hamiltonian. In the number of exponentials, the factor $5^{q-1}$ becomes $q^2$ up to logarithms, so the cost grows polynomially rather than exponentially with the order.

\textit{Proof sketch of~\cref{thm:main-informal}.} \Cref{thm:order-cost-informal} simulates unit-time evolution and requires $H$ to be small enough at unit time. There are two cases. For general Hamiltonians, the sum of the generator norms must lie below a threshold that grows as $q^{2/3}$. For local Hamiltonians, a similar threshold applies but for the growth scale of nested commutators instead. Since evolution for general time $t$ effectively rescales the Hamiltonian, to satisfy these conditions, we divide time $t$ into smaller intervals. Evolution over one interval is unit-time evolution with each generator multiplied by the interval length, with the duration chosen so that the conditions of \cref{thm:order-cost-informal} are met. The total error is bounded by the sum of these interval errors. Choosing $q$ so that this sum is at most $\varepsilon$ and counting the exponentials over all intervals gives \cref{thm:main-informal}. \Cref{app:time-precision-cost} gives the full argument.

\section{Discussion}
\label{sec:discussion}
Product formulae are ubiquitous in classical and quantum simulation, but despite their widespread adoption, known high-order constructions need a number of exponentials that grows exponentially with the order. To our knowledge, POLY is the first product formula whose error decreases exponentially in a cancellation order $q$ while its number of exponentials grows only as $\widetilde O(q^2)$. As a consequence, it simulates Hamiltonian evolution for time $t$ with $\widetilde O(t)$ exponentials, and local Hamiltonians on $N$ sites with $\widetilde O(Nt)$ gates. This scaling is nearly optimal~\cite{BACS07,BCK15,HHKL18}. It had previously been reached only by ancilla-assisted methods, and we reach it without ancilla qubits.

Because POLY is an ordered product of the exponentials $e^{-i\tau H_a}$, it is a single unitary circuit acting only on the system qubits, and so it can serve as a coherent subroutine, for example of phase estimation. Quantum singular value transformation without block encodings builds polynomial transformations of $H$ from Hamiltonian evolutions, using one ancilla qubit for the transformation itself~\cite{CHLSWZ25}. With our circuits for these evolutions, the whole transformation becomes a single unitary circuit whose cost is optimal in every parameter up to polylogarithmic factors. In composite simulations, our circuits can take the place of the product formula for the large terms while qDRIFT treats the small terms probabilistically~\cite{C19,HW23}, which may reduce the cost further for some Hamiltonians. We also expect that probabilistic Trotter error reversal~\cite{MKRST26} applies to our circuits and removes their remaining algorithmic error without bias.
The same circuits can be used directly in classical simulations, for example in tensor-network simulations of local Hamiltonians.

An immediate question is whether the asymptotic improvement already appears at circuit sizes of practical interest. Since we have not optimized the constant factors, our bounds do not establish an advantage at the times and precisions relevant to applications. For standard product formulae, the constant factors have been reduced substantially by optimizing compositions, refining the error analysis and exploiting problem structure (\cref{sec:introduction}), and their actual errors are often far smaller than their worst-case bounds. We expect that improvements of the same kind, for example tuning the time map, the auxiliary time step and the cutoffs to a particular Hamiltonian, can reduce the constant factors of our construction considerably, and that its actual errors are likewise smaller than our bounds.

It remains open how far the polylogarithmic factors in our cost can be reduced. Ancilla-assisted methods reach $O(t+\log(1/\varepsilon))$, in which the dependence on the precision is additive rather than multiplicative (\cref{tab:intro-comparison}). It is also open what lower bounds hold for arbitrary product formulae. Our result shows that such bounds cannot follow from counting order conditions alone, because our circuits reach high precision without relying on a high order in physical time.

Our construction shows that product formulae need not be built by repeating a single step of high order, and that multiplying approximations of the same evolution with different time steps can be more efficient. Polygon cancellation (\cref{lem:polygon-cancellation}) holds for any formal power series $F(z)$ with $F(0)=I$, and so may also be useful beyond the present setting, for example for time-dependent Hamiltonians or for open quantum systems. Our result suggests that product formulae of this more general form deserve further study.

\section*{Author contributions}
K.M.\ formulated the problem and developed the algorithm. J.W.D. formalized and generalized polygon cancellation. K.M. and J.W.D. led the writing of the manuscript. B.K. contributed to the writing and supervised the research.

\begin{acknowledgments}
The authors thank Miho Kobatake, Chusei Kiumi, Po Wei (George) Huang, Yukun Zhang, and Simon C. Benjamin for helpful discussions. K.M.\ was supported by JST ASPIRE Japan, Grant Number JPMJAP2319, the Center of Innovation for Sustainable Quantum AI (SQAI), JST Grant Number JPMJPF2221, and JSPS KAKENHI Grant Number JP24KJ0892.  J.W.D. thanks the Clarendon Fund Scholarship, University of Oxford.
B.K.\ thanks UKRI for the Future Leaders Fellowship Theory to Enable Practical Quantum Advantage (MR/Y015843/1).
For the purpose of Open Access, the authors have applied a CC BY public copyright licence to any Author Accepted Manuscript version arising from this submission.
\end{acknowledgments}

\section*{AI Use Disclosure}
K.M.\ used Claude Opus 5.5 to explore possible constructions during the development of the POLY algorithm, and the authors used various LLMs (Claude Opus 5.5, GPT-6 Astra and GPT-6.2 Astra) to draft the appendices, check mathematical expressions and polish the main text for clarity.

\bibliography{ref}
\clearpage
\input{appendix}

\end{document}

%% file: header.tex
\usepackage{graphicx}  %
\usepackage{bm}        %
\usepackage[colorlinks=true, allcolors=black]{hyperref} %
\usepackage{xcolor}    %
\usepackage{amsthm, amssymb, amsmath}
\usepackage{mathtools} %
\usepackage{mathrsfs}
\usepackage{enumitem}
\usepackage{booktabs}
\usepackage{capt-of}
\usepackage[T1]{fontenc}

\usepackage[capitalize]{cleveref}
\crefname{section}{Sec.}{Secs.}
\Crefname{section}{Section}{Sections}
\usepackage[linesnumbered,algoruled,boxed,lined]{algorithm2e}
\usepackage{mdframed}
\usepackage{tikz}
\usepackage{standalone}
\usetikzlibrary{arrows.meta} %

\newtheorem{theorem}{Theorem}
\newtheorem{informaltheorem}{Theorem}
\crefalias{informaltheorem}{theorem}

\newtheorem{lemma}{Lemma}

\newtheorem{corollary}{Corollary}

\newtheorem{definition}{Definition}

\theoremstyle{definition}

\theoremstyle{definition}

\Crefname{problem}{Problem}{Problems}

\newcommand{\norm}[1]{\left\lVert#1\right\rVert}

\DeclarePairedDelimiterX\braket[2]{\langle}{\rangle}{#1\,\delimsize\vert\,#2}

\DeclareMathOperator{\polylog}{polylog}

\newtheorem{proposition}{Proposition}

\newcounter{proofmappart}
\crefname{proofmappart}{Part}{Parts}
\Crefname{proofmappart}{Part}{Parts}
\newcommand{\proofmappart}[2]{\par\addvspace{.6\baselineskip}\noindent\refstepcounter{proofmappart}\label{#2}\textbf{Part~\theproofmappart. #1.}\enspace\ignorespaces}

%% file: appendix.tex
\onecolumngrid
\appendix
\crefalias{section}{appendix}
\crefalias{subsection}{appendix}
\crefalias{subsubsection}{appendix}

\input{appendices/related-work}

\clearpage
\begin{center}
\textbf{\large Technical details}
\end{center}
\noindent The remaining appendices prove \cref{thm:main,thm:order-cost} and can be read independently of the main text. \Cref{app:proof-map} collects the notation, gives the formal statements (\cref{app:proof-map-results}) and maps out the proof. \Cref{app:polygon-cancellation,app:intrinsic-scales,app:finite-circuit-expansion,app:statements,app:time-precision-cost} contain the proofs.
\bigskip

\input{appendices/proof-map}
\input{appendices/appendix_polygon}
\input{appendices/commutator-growth}
\input{appendices/sweep-comparison}
\input{appendices/polygon-cost}
\input{appendices/simulation-cost}

%% file: appendices/related-work.tex
\section{Related work}
\label{app:related-work}

\Cref{tab:intro-comparison} compares circuits for simulating $e^{-itH}$. Here we examine how the closest constructions cancel error, how the local gate counts in that table are obtained, and what changes when the output is an observable estimate or a random channel. We use the normalization of \cref{tab:intro-comparison} and quote asymptotic costs for $t\ge1$ and small $\varepsilon$, with a fixed number of generators. For local systems we also fix the interaction range, locality, and lattice dimension.

\subsection{Cancellation across resolutions}
\label{app:related-mpf}

Multiproduct formulae run a symmetric product formula $S_2$ at several resolutions over the same physical interval $\Delta$. At fixed $\Delta$, these approximations share an error expansion in even powers of the step size. Both Richardson extrapolation~\cite{R11,RG27,G65,BS66,BCR99,C10b} and POLY use this expansion to cancel the leading errors, as shown in \cref{tab:resolution-comparison}.

\begin{table}[!ht]
\caption{\label{tab:resolution-comparison}Two ways to combine approximations to the same evolution. $Q(\delta_j)$ is one sweep of our circuit at auxiliary step $\delta_j$.}
\begin{ruledtabular}
\begin{tabular}{@{}p{0.20\linewidth}p{0.37\linewidth}p{0.37\linewidth}@{}}
 & Coherent multiproduct formula~\cite{CW12,LKW19} & Polygon formula (this work) \\
\hline
Combination & $\displaystyle\sum_j a_j S_2(\Delta/k_j)^{k_j}$ & $\displaystyle\prod_{j=1}^{2q-1}Q(\delta_j)^{(-1)^{j+1}}$ \\[3pt]
Choice of resolutions & Integer step counts $k_j$, with weights $a_j$ chosen at specific cancellation values & Auxiliary steps $\delta_j=\delta\sin(\pi j/2q)$, realized by varying the physical step widths \\[3pt]
Circuit & An index register selects the controlled evolutions used in the linear combination & Sweeps and inverse sweeps act directly on the system register \\
\end{tabular}
\end{ruledtabular}
\end{table}

In the sum, the entire matrix coefficient of degree $2r$ is multiplied by the scalar moment $\sum_j a_j k_j^{-2r}$. Setting that moment to zero cancels the coefficient. In the product, coefficients from different factors also multiply. For example, if the error factor begins $I+\delta^2 E_2+\delta^4 E_4+\cdots$, both $E_2E_4$ and $E_4E_2$ can appear at degree six. The polygon identity cancels each ordered contribution below the target degree. The continuous family of sweeps matters here because rounding the prescribed step ratios would change the coefficients of these words (\cref{sec:continuous-family-construction}).

The choice of resolutions plays a different role in the multiproduct formula. With $k_j=j$, the sum of the absolute extrapolation weights grows exponentially with the order, making a coherent linear combination expensive to implement. Low, Kliuchnikov, and Wiebe choose integer step counts from a Chebyshev grid and recompute the weights at those integers. For an order-$2n$ formula this gives $\sum_j|a_j|=O(\log n)$. The index register uses $O(\log n)$ qubits, and the factorial decay of the step error allows $n=O(\log(t/\varepsilon)/\log\log(t/\varepsilon))$. They obtain $O(t\log^2(t/\varepsilon))$ controlled Strang stages, including amplitude amplification. Their amplified circuit succeeds with probability $1-O(\varepsilon)$~\cite[Theorem~2]{LKW19}. For fixed $m$, each stage costs a constant number of controlled exponentials. For our formula, writing $\ell=\log(t/\varepsilon)$ under the table normalization, the full bound is $O((t\ell^{4/3}+\ell^2)\log^2\ell)$. An order-$q$ circuit can cover a time proportional to $q^{2/3}$, giving the $4/3$ exponent in the term linear in $t$. The $\ell^2$ term covers the cost of one circuit when the entire evolution fits in a single interval. \Cref{app:related-noncoherent} discusses Richardson extrapolation applied to measured expectation values.

\paragraph{Product formulae and oracle algorithms.}
The fixed-order and order-optimized Suzuki bounds in \cref{tab:intro-comparison} count the same native exponentials as our formula. A formula of order $p$ has an error bound governed by nested commutators of degree $p+1$~\cite{CSTWZ21}. In Suzuki's recursion each increase of two in the order multiplies the number of Strang stages by five~\cite{S90,S91}. Balancing this cost against the number of time steps gives the optimized bound $t\exp[O(\sqrt{\log(t/\varepsilon)})]$~\cite{BACS07,WBHS10,PZ12}.

Several oracle models admit polylogarithmic precision dependence, including fractional-query simulation~\cite{CGMSY09,BCCKS14}, linear combinations of unitaries and truncated Taylor series~\cite{CW12,BCCKS15}, and quantum walks~\cite{C10,BC12,BCK15}. With a block encoding of fixed normalization, quantum signal processing and qubitization give an additive query bound $O(t+\log(1/\varepsilon))$~\cite{LC17a,LC19,GSLW19,MRTC21}. Linear scaling in time is necessary in the worst case~\cite{BACS07,AA17}. Extensions exploit spectral norms and sparsity~\cite{LC17b,L19,BRSL26} or the structure of time-dependent evolution~\cite{LW18,BCSWW20,CGWZ26}. Other work develops the polynomial decompositions and circuit implementations used in signal processing~\cite{H19,MW24,BMPW24,A25}.

Converting a query bound to a gate count requires a circuit for the oracle. Resource estimates for chemistry and lattice models illustrate how this choice and the available workspace affect the comparison~\cite{CMNRS18,C22,KGCJ24}. Approaches that economize on workspace include single-ancilla linear combinations of unitaries~\cite{C24}, low-ancilla block encodings and singular value transformations~\cite{ZS26,CHLSWZ25}, and energy estimation using controlled time evolutions~\cite{LT22,DLT22}.

\paragraph{Earlier cancellation sequences.}
\label{app:related-cancellation}
Multiplicative extrapolation in splitting methods composes approximations at different physical time steps~\cite{QZ94}. Here each factor covers the same physical interval and its auxiliary resolution varies. Products of scaled formulae and their inverses also occur in simulations of commutator exponentials, where the composition isolates the desired commutator~\cite{CW13,CCHJKX22}. In quantum control, composite pulses and dynamical decoupling use ordered operations to suppress errors~\cite{L86,VKL99}. Concatenated decoupling raises the suppression order recursively~\cite{KL05}. The squared-sine nodes also occur in Uhrig dynamical decoupling~\cite{U07,YL08}. There the nodes specify when pulses reverse the sign of an interaction. In the polygon formula they specify the resolutions at which whole sweeps are applied or inverted. Uhrig timing has also been applied to higher-order Zeno sequences~\cite{DKMA26}. Composite pulses for a single qubit likewise admit both recursive constructions~\cite{BHC04} and sequences of length linear in the cancellation order~\cite{LYC14}, with a subsequent connection to quantum signal processing~\cite{LYC16,LC17a}.

\Needspace{0.24\textheight}
\subsection{Gate counts for local Hamiltonians}
\label{app:related-local}

For local Hamiltonians the spacetime volume $Nt$ provides a common scale. \Cref{tab:local-work} gives the overhead over $Nt$ in total gates. For our circuit, each generator exponential is a parallel layer of at most $N$ local gates, so the number of layers also bounds the depth. For Ref.~\cite{HHKL18}, \cref{tab:intro-comparison} gives the stated circuit depth. For the other methods it gives the total gate count divided by $N$, which is a lower bound on the circuit depth.

\begin{table}[!ht]
\caption{\label{tab:local-work}Gate-count overhead for finite-range Hamiltonians on $N$ sites. Multiply the last column by $Nt$ to obtain the total gate-count bound. Here $\ell_N=\log(Nt/\varepsilon)$ and $p$ is a fixed even base order.}
\begin{ruledtabular}
\begin{tabular}{lll}
Method & Extra qubits & Gate count divided by $Nt$ \\
\hline
Fixed-order Suzuki--Trotter~\cite{CS19,CSTWZ21} & $0$ & $O((Nt/\varepsilon)^{1/p})$ \\
Coherent multiproduct formula~\cite{LKW19,M26} & $O(\log\ell_N)$ & $O((N^{1/(p+1)}+\ell_N^2)\ell_N^2\log^2\ell_N)$ \\
Lieb--Robinson blocks with qubitization~\cite{HHKL18} & $O(N)$ & $O(\ell_N^d)$ in $d$ dimensions \\
Lieb--Robinson blocks with coherent corrections~\cite{SWYYZ26} & $O(N)$ & $O(\ell_N)$ \\
Polygon formula (this work) & $0$ & $O(\ell_N^2\log^2\ell_N)$ \\
\end{tabular}
\end{ruledtabular}
\end{table}

Two entries require a choice in translating the published results. Commutator estimates for multiproduct formulae extend to both time-independent and time-dependent generators~\cite{AAT24,MIF26,M26}. For the time-independent case in the table, the proof of Theorem~6 of Ref.~\cite{M26} takes order parameter $n=\Theta(\ell_N)$ and $O(t(N^{1/(p+1)}+\ell_N^2))$ intervals. The well-conditioned weights of Ref.~\cite{LKW19} cost $O(n^2\log^2 n)$ controlled base stages per interval, each using $O(N)$ gates at fixed $p$. Multiplication gives the second row. The shorter entry in \cref{tab:intro-comparison} applies when $N^{1/(p+1)}\ge\ell_N^2$. For the block decomposition, the Lieb--Robinson bound limits how rapidly an interaction can influence distant sites~\cite{LR72,NS06,O06}. Haah, Hastings, Kothari, and Low use this bound to simulate overlapping regions of a lattice separately, obtaining $O(Nt\,\polylog(Nt/\varepsilon))$ gates~\cite{HHKL18}. The displayed logarithmic power uses qubitization on blocks of $O(\ell_N^d)$ sites. Correcting a product formula inside each block gives the fourth row~\cite{SWYYZ26}.

Local product formulae exploit the fact that terms with disjoint support commute, both in lattice field simulations~\cite{JLP12} and in general gate-count bounds~\cite{CS19,CSTWZ21}. In our construction, locality enters through the growth of nested commutators. The local interaction bound of Refs.~\cite{K16information,KMS16floquet} controls this growth independently of $N$ (\cref{app:intrinsic-scales}). This permits an interval of fixed physical duration as the system grows. A separate line of work uses variational optimization, Cartan decomposition, or circuit compression to obtain shorter circuits for particular models~\cite{CHICCS20,KSWFDK22,THL23,ML23}.

\subsection{Observable estimates and randomized simulation}
\label{app:related-noncoherent}

Extrapolation and stochastic compensation also use circuits built from term evolutions. Their costs depend on what must be accurate at the end of the computation. \Cref{tab:sampled-work} distinguishes the quantity controlled by each method from the cost of obtaining it.

\begin{table}[!ht]
\caption{\label{tab:sampled-work}Precision guarantees for sampled product-formula circuits. For an observable estimate, the cost of a circuit and the number of samples are separate resources.}
\begin{ruledtabular}
\begin{tabular}{@{}p{0.32\linewidth}p{0.21\linewidth}p{0.41\linewidth}@{}}
Method & Guarantee & Where precision enters \\
\hline
Richardson extrapolation~\cite{RWW24,WW25,W24} & Observable estimate & Polylogarithmic circuit depth in $1/\varepsilon$ at fixed $t$; $\widetilde O(\varepsilon^{-2})$ circuit repetitions. \\[3pt]
Trotter-error compensation~\cite{WZWZSZGL26,MKRST26} & Observable estimate & Compensation can suppress the bias or remove it exactly. In the unbiased construction the expected rotation count per circuit is independent of $\varepsilon$. \\[3pt]
Randomized simulation (qDRIFT)~\cite{C19} & Average channel & At fixed sum of term norms, $O(t^2/\varepsilon)$ exponentials suffice for diamond-norm error $\varepsilon$. \\
\end{tabular}
\end{ruledtabular}
\end{table}

For observable extrapolation, one measures the same observable at several resolutions and combines the estimates classically~\cite{EZLBY19,CEOW23,FSKKE22,RWW24,WW25}. Related approaches fit the weights of dynamic multiproduct formulae~\cite{ZRB24} or apply extrapolation to eigenvalue transformations~\cite{MWW26}. Another family of methods represents corrections to a product formula by linear combinations of unitaries or samples the error evolution~\cite{ZSJZ25,MCS25,WZWZSZGL26}. Stochastic Zassenhaus expansions sample higher-order commutator terms~\cite{PN25}. Other schemes recover exact dynamics in an ensemble of weighted circuits~\cite{GD24,KK25}. In unbiased compensation, random corrections remove the Trotter bias in the mean, and the sampling variance determines the number of circuits needed for a given additive precision~\cite{MKRST26}. Classical variance reduction can further reduce the number of sampled circuits needed for a given statistical precision~\cite{DHK26}.

Randomization can also improve the accuracy of a product formula, for example by averaging over orderings of its terms~\cite{COS19} or by canceling higher-order errors in the average channel~\cite{CBH24,KG26}. Other randomized compilers use sparsification, higher-order corrections, or a combination of deterministic and sampled terms~\cite{OWC20,NBA24,HW23}. Concentration bounds also quantify the accuracy of a typical sampled circuit~\cite{CHKT21}. For qDRIFT, diamond-norm error controls the error of the averaged simulation inside a larger circuit, including phase estimation with controlled term evolutions~\cite{C19}. For POLY, the operator-norm bound holds for the single circuit returned by the construction, uniformly over input states.

%% file: appendices/proof-map.tex
\section{Map of the proof}
\label{app:proof-map}
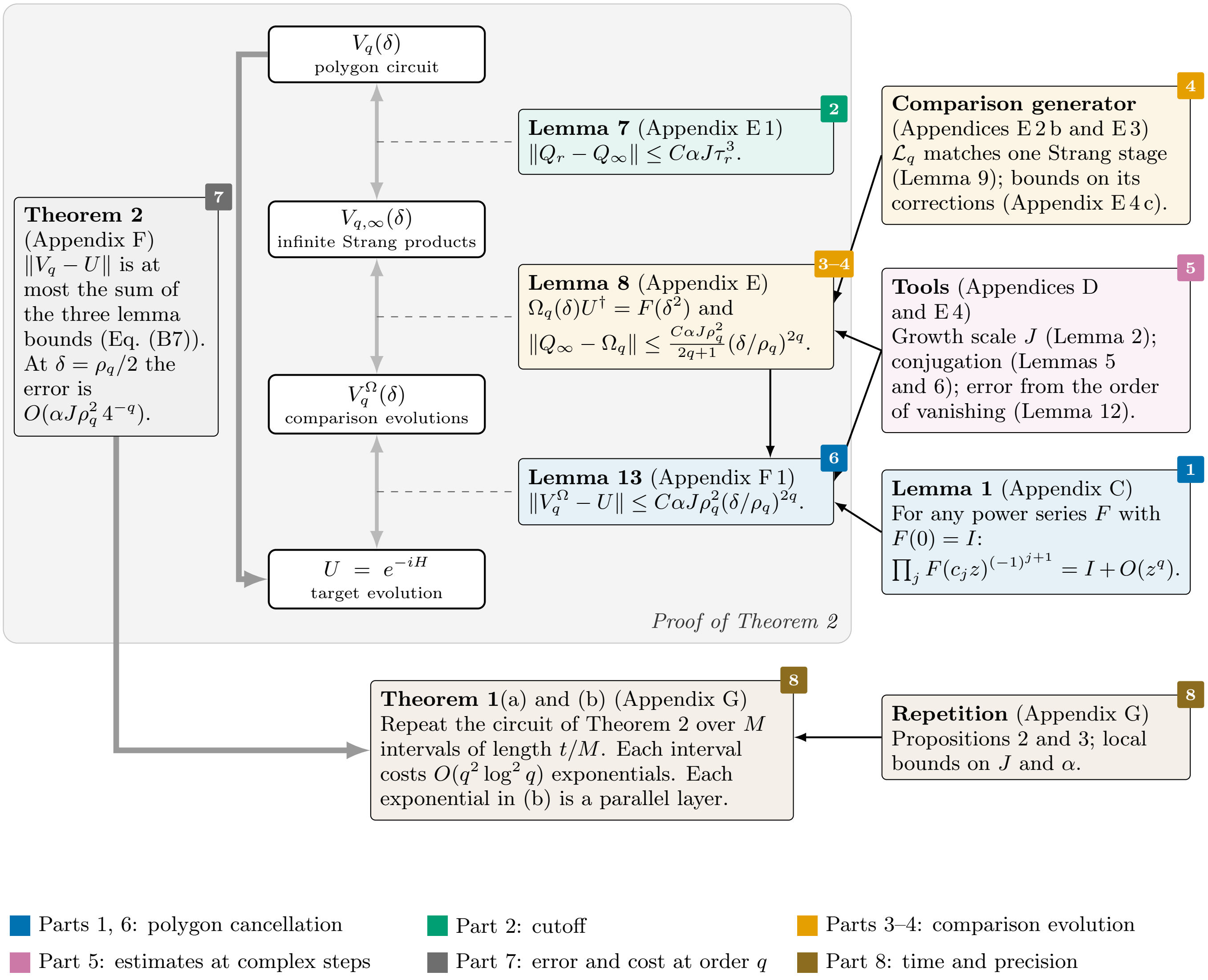
\begin{figure}[htbp]
\centering
\definecolor{pmCancel}{HTML}{0072B2}
\definecolor{pmCutoff}{HTML}{009E73}
\definecolor{pmCompare}{HTML}{E69F00}
\definecolor{pmTools}{HTML}{CC79A7}
\definecolor{pmOrder}{HTML}{6E6E6E}
\definecolor{pmTime}{HTML}{8C6D1F}
\begin{tikzpicture}[
  >=latex,
  box/.style={draw, rounded corners=2pt, align=flush left, text width=4.4cm, inner sep=4pt, font=\small, anchor=west, fill=#1!10},
  side/.style={box=#1, text width=4.3cm},
  op/.style={draw, thick, rounded corners=3pt, align=center, text width=3.0cm, inner sep=3pt, fill=white, font=\small},
  tag/.style={fill=#1, text=white, font=\scriptsize\bfseries, inner sep=1.5pt, minimum height=4mm, minimum width=4mm, rounded corners=1pt},
  key/.style={fill=#1, minimum size=3mm, inner sep=0pt},
  arr/.style={->, thick},
  link/.style={dashed, gray!70!black},
  chain/.style={line width=1.6pt, gray!55},
  thmarrow/.style={line width=2.4pt, #1!70, ->}]
 \fill[gray!9, rounded corners=6pt] (-5.55,0.75) rectangle (7.05,-8.75);
 \draw[gray!45, rounded corners=6pt] (-5.55,0.75) rectangle (7.05,-8.75);
 \node[anchor=south east, font=\small\itshape, text=gray!40!black] at (6.95,-8.7) {Proof of \cref{thm:order-cost}};
 \node[op] (V)  at (0,0)    {$V_q(\delta)$\\[-1pt]{\scriptsize polygon circuit}};
 \node[op] (Vi) at (0,-2.6) {$V_{q,\infty}(\delta)$\\[-1pt]{\scriptsize infinite Strang products}};
 \node[op] (VO) at (0,-5.2) {$V_q^\Omega(\delta)$\\[-1pt]{\scriptsize comparison evolutions}};
 \node[op] (U)  at (0,-7.8) {$U=e^{-iH}$\\[-1pt]{\scriptsize target evolution}};
 \draw[chain,<->] (V)  -- (Vi);
 \draw[chain,<->] (Vi) -- (VO);
 \draw[chain,<->] (VO) -- (U);
 \node[box=pmCutoff]  (cut)  at (2.1,-1.3) {\textbf{\Cref{lem:cutoff-selection}} (\cref{app:cutoff-selection})\\ $\|Q_r-Q_\infty\|\le C\alpha J\tau_r^3$.};
 \node[box=pmCompare] (comp) at (2.1,-3.9) {\textbf{\Cref{lem:finite-circuit-expansion}} (\cref{app:finite-circuit-expansion})\\ $\Omega_q(\delta)U^\dagger=F(\delta^2)$ and\\ $\|Q_\infty-\Omega_q\|\le\frac{C\alpha J\rho_q^2}{2q+1}(\delta/\rho_q)^{2q}$.};
 \node[box=pmCancel]  (ref)  at (2.1,-6.5) {\textbf{\Cref{lem:reference-polygon-error}} (\cref{app:intrinsic-cost})\\ $\|V_q^\Omega-U\|\le C\alpha J\rho_q^2(\delta/\rho_q)^{2q}$.};
 \draw[link] (0,-1.3) -- (cut.west);
 \draw[link] (0,-3.9) -- (comp.west);
 \draw[link] (0,-6.5) -- (ref.west);
 \draw[arr] ([xshift=1.4cm]comp.south) -- ([xshift=1.4cm]comp.south |- ref.north);
 \node[side=pmCompare] (gen)   at (7.5,-1.5) {\textbf{Comparison generator} (\cref{app:comparison-construction,app:comparison-coefficients})\\ $\mathcal L_q$ matches one Strang stage (\cref{lem:comparison-cell-matching}); bounds on its corrections (\cref{app:quantitative-coefficients}).};
 \node[side=pmTools]   (tools) at (7.5,-4.4) {\textbf{Tools} (\cref{app:intrinsic-scales,app:comparison-global})\\ Growth scale $J$ (\cref{lem:commutator-scale-properties}); conjugation (\cref{lem:comparison-conjugation,lem:intrinsic-ideal}); error from the order of vanishing (\cref{lem:real-step-product}).};
 \node[side=pmCancel]  (pc)    at (7.5,-7.1) {\textbf{\Cref{lem:polygon-cancellation}} (\cref{app:polygon-cancellation})\\ For any power series $F$ with $F(0)=I$: $\prod_jF(c_jz)^{(-1)^{j+1}}=I+O(z^q)$.};
 \draw[arr] (gen.west)   -- ([yshift=2mm]comp.east);
 \draw[arr] (tools.west) -- ([yshift=-2mm]comp.east);
 \draw[arr] (tools.west) -- ([yshift=1.5mm]ref.east);
 \draw[arr] (pc.west)    -- ([yshift=-1.5mm]ref.east);
 \draw[thmarrow=pmOrder] (V.west) -- (-2.05,0) -- (-2.05,-7.8) -- (U.west);
 \node[box=pmOrder, anchor=east, text width=2.75cm] (order) at (-2.35,-3.9) {\textbf{\Cref{thm:order-cost}} (\cref{app:statements})\\ $\|V_q-U\|$ is at most the sum of the three lemma bounds (\cref{eq:proof-map-three-errors}). At $\delta=\rho_q/2$ the error is $O(\alpha J\rho_q^2\,4^{-q})$.};
 \node[box=pmTime, anchor=north, text width=6.0cm] (main) at (3.05,-9.3) {\textbf{\Cref{thm:main}}(a) and (b) (\cref{app:time-precision-cost})\\ Repeat the circuit of \cref{thm:order-cost} over $M$ intervals of length $t/M$. Each interval costs $O(q^2\log^2q)$ exponentials. Each exponential in (b) is a parallel layer.};
 \node[side=pmTime] (rep) at (7.5,-10.15) {\textbf{Repetition} (\cref{app:time-precision-cost})\\ \Cref{prop:time-subdivision,prop:cost-after-repetition}; local bounds on $J$ and $\alpha$.};
 \draw[thmarrow=pmOrder] (order.south) |- (main.west);
 \draw[arr] (rep.west) -- (main.east |- rep.west);
 \node[tag=pmCutoff]  at (cut.north east)   {2};
 \node[tag=pmCompare] at (comp.north east)  {3--4};
 \node[tag=pmCancel]  at (ref.north east)   {6};
 \node[tag=pmOrder]   at (order.north east) {7};
 \node[tag=pmTime]    at (main.north east)  {8};
 \node[tag=pmCompare] at (gen.north east)   {4};
 \node[tag=pmTools]   at (tools.north east) {5};
 \node[tag=pmCancel]  at (pc.north east)    {1};
 \node[tag=pmTime]    at (rep.north east)   {8};
 \begin{scope}[yshift=-12.95cm, every node/.style={font=\small}]
  \node[key=pmCancel]  (k1) at (-5.3,0)     {}; \node[anchor=west] at (k1.east) {Parts 1, 6: polygon cancellation};
  \node[key=pmCutoff]  (k2) at (0.9,0)      {}; \node[anchor=west] at (k2.east) {Part 2: cutoff};
  \node[key=pmCompare] (k3) at (6.4,0)      {}; \node[anchor=west] at (k3.east) {Parts 3--4: comparison evolution};
  \node[key=pmTools]   (k4) at (-5.3,-0.55) {}; \node[anchor=west] at (k4.east) {Part 5: estimates at complex steps};
  \node[key=pmOrder]   (k5) at (0.9,-0.55)  {}; \node[anchor=west] at (k5.east) {Part 7: error and cost at order $q$};
  \node[key=pmTime]    (k6) at (6.4,-0.55)  {}; \node[anchor=west] at (k6.east) {Part 8: time and precision};
 \end{scope}
\end{tikzpicture}
\caption{\label{fig:proof-map}Map of the proof of \cref{thm:main}. The center column is the chain of \cref{eq:proof-map-chain}: the polygon circuit $V_q(\delta)$ is compared with $U$ through two intermediate polygon products, and each link is bounded by the lemma attached to it by a dashed line. \Cref{thm:order-cost} (left) bounds $V_q(\delta)-U$ directly by adding these three bounds in \cref{eq:proof-map-three-errors}; the shaded region contains its proof. Together with repetition of the circuit, it gives \cref{thm:main} (bottom). The supporting results on the right feed into the lemmas; an arrow points from a result to a result whose proof uses it. The color and the number in the corner of each box give the part of this outline (\crefrange{app:proof-map-polygon-identity}{app:proof-map-time}) in which the result is discussed. The symbols are defined in \cref{app:proof-map-notation}.}
\end{figure}

This section gives an overview of the appendices that follow and of how they combine to prove \cref{thm:main,thm:order-cost}. It gives the formal statements of both theorems in \cref{app:proof-map-results}, collects the objects that the proof uses, and explains why each intermediate estimate is needed. Each part names the result and the appendix that contain the details. Together with the appendices that follow, it can be read without the main text. \Cref{fig:proof-map} shows how the results depend on each other, and its colors indicate the part of this appendix in which each result is discussed.

We use the radius $\rho_q$ of \cref{thm:order-cost} and treat the number of generators as fixed.

\subsection{Objects and notation}
\label{app:proof-map-notation}

\emph{Hamiltonian and parameters.} We simulate $H=\sum_{a=1}^mH_a$, where each generator $H_a$ is a Hermitian operator of finite norm whose evolutions $e^{-i\tau H_a}$ are available for every real $\tau$. We work at unit time, with target $U=e^{-iH}$, until \cref{app:proof-map-time}. Write
\[
 \lambda=\sum_a\|H_a\|,\qquad \alpha=\sum_{a<b}\|[H_a,H_b]\|.
\]
A product formula is an ordered product of such exponentials, and its cost is its number of exponentials (\cref{def:native-exponential-count}). We write $D(tH,\varepsilon)$ for the cost of a product formula $V$ with $\|V-e^{-itH}\|\le\varepsilon$.

\emph{Nested commutators and the growth scale $J$.} A nested commutator of degree $d$ is $[H_{a_d},[\cdots,[H_{a_2},H_{a_1}]\cdots]]$. The corrections to the comparison generator and the defect in the logarithmic derivative of a Strang stage are sums of nested commutators of degree at least three. The commutator growth scale $J$ (\cref{app:scale-definition}) is the smallest number such that commutation with a nested commutator of degree $d$ multiplies the norm of an argument by at most $\nu\,d!\,(2J)^d$, whenever the argument is a combination of the identity and nested commutators of degree at most $\nu$. Roughly, each further commutation multiplies a norm by at most $2J$ times the current degree. The proof uses four consequences (\cref{lem:commutator-scale-properties} and \cref{eq:intrinsic-root-bound}):
\begin{enumerate}[nosep]
 \item A nested commutator of degree $d\ge2$ has norm at most $\alpha(2J)^{d-2}(d-1)!$. In particular, $\|[H_c,[H_a,H_b]]\|\le4J\|[H_a,H_b]\|\le4\alpha J$. The innermost pair supplies the factor $\alpha$, and the third generator supplies $J$.
 \item $J\le\max_a\|H_a\|\le\lambda$, and $J(\tau H_1,\ldots,\tau H_m)=|\tau|J(H_1,\ldots,H_m)$.
 \item If each $H_a$ is a sum of terms acting on at most $k$ sites, with interaction strength per site at most $g_{\rm loc}$, then $J\le3k^2g_{\rm loc}$, independently of the number of sites. In particular, $J\le3k^2$ for the layer decomposition in \cref{def:local-hamiltonian}.
 \item $J=0$ if and only if the generators commute.
\end{enumerate}
For these layer decompositions $\lambda$ can grow with the number of sites, while the upper bound on $J$ is independent of system size. This is why the estimates are written in terms of $\alpha$ and $J$ wherever possible.

\emph{Strang stage.} The Strang stage
\[
 S_2(\tau)=e^{-i\tau H_1/2}\cdots e^{-i\tau H_{m-1}/2}\,e^{-i\tau H_m}\,e^{-i\tau H_{m-1}/2}\cdots e^{-i\tau H_1/2}
\]
(\cref{eq:strang-step}) uses $2m-1$ exponentials, satisfies $S_2(-\tau)=S_2(\tau)^{-1}$, and has error (\cref{eq:cutoff-strang-error})
\begin{equation}
 \|S_2(\tau)-e^{-i\tau H}\|\le C\alpha J|\tau|^3 .
 \label{eq:proof-map-strang}
\end{equation}

\emph{Auxiliary time and finite circuits.} The time map $f(s)=1-\exp(1-e^s)$ sends the auxiliary time $s\in[0,\infty)$ to the physical time $f(s)\in[0,1)$; write $g=f'$. For an auxiliary time step $\delta>0$, the physical time steps are $\Delta_k(\delta)=f((k+1)\delta)-f(k\delta)$, and they sum to one. The finite circuit with cutoff $r$ is
\[
 Q_r(\delta)=S_2(\tau_r)\,S_2(\Delta_{r-1})\cdots S_2(\Delta_0),\qquad
 \tau_r=1-f(r\delta)=\exp(1-e^{r\delta}),
\]
so that its last stage covers the physical time $\tau_r$ of the tail. The \emph{infinite Strang product} $Q_\infty(\delta)=\lim_{r\to\infty}Q_r(\delta)$ applies all the stages $S_2(\Delta_k)$, $k\ge0$. Both are defined for every real $\delta>0$.

\emph{Polygon products.} For the cancellation order $q\ge2$, put
\[
 c_j=\sin^2\frac{\pi j}{2q},\qquad \delta_j=\delta\sqrt{c_j}=\delta\sin\frac{\pi j}{2q},\qquad j=1,\ldots,2q-1 .
\]
Given approximations $X(\delta)$ to $U$, the polygon product multiplies them alternately forward and inverted, $\prod_{j=1}^{2q-1}X(\delta_j)^{(-1)^{j+1}}$. With $X=Q_{r_j}$ at vertex $j$ this is the polygon circuit $V_q(\delta)$ of \cref{eq:finite-cancellation-word}; an inverse $Q_r(\delta)^{-1}$ is implemented by reversing the stages and negating their physical time steps, so it uses the same number of exponentials. We write $D_q(\delta)$ for the number of exponentials of $V_q(\delta)$. With $X=Q_\infty$ we write $V_{q,\infty}(\delta)$, and with the comparison evolution $X=\Omega_q$ below we write $V_q^\Omega(\delta)$. Only $V_q(\delta)$ is a circuit.

\emph{Comparison evolution.} The comparison generator $\mathcal L_q(s,\delta)=-ig(s)H+\mathcal K_q(s,\delta)$ adds to the exact generator $-ig(s)H$ a correction $\mathcal K_q$ that is a polynomial in $\delta^2$ (\cref{app:proof-map-comparison}). The comparison evolution $\Omega_q(\delta)$ is the evolution under $\mathcal L_q$ from $s=0$ to $s=\infty$. Its error factor is $F(z)=\Omega_q(\sqrt z)U^\dagger$.

\emph{Radius and decay.} The estimates hold for auxiliary time steps up to $\rho_q=\rho_0/q$ under either hypothesis, where $\rho_0$ is the radius constant in \cref{thm:order-cost}. The function $\mu(s)=\exp(-e^s/4)$ is integrable on $[0,\infty)$ and bounds $g$ and its derivatives near $s$, up to constants. It measures how fast the contribution of a stage decreases with its auxiliary starting time.

\subsection{Formal statements of Theorems 1 and 2}
\label{app:proof-map-results}

Let $H=\sum_{a=1}^m H_a$ have bounded Hermitian generators, with each $e^{-isH_a}$ available for every real $s$. Write $\lambda=\sum_a\|H_a\|$, $\alpha=\sum_{a<b}\|[H_a,H_b]\|$, and let $J$ be the commutator growth scale of \cref{eq:commutator-growth-scale}. We fix $m$ throughout. The constants in the bounds below may depend on $m$, and in \cref{thm:main}(b) also on $k$. They are otherwise independent of the generators, system size, time, order, and precision.

\begin{theorem}[Product formulae with polylogarithmic overhead, formal version of Theorem~\ref{thm:main-informal}]
\label{thm:main}
For $t>0$ and $0<\varepsilon<1$, there is a product formula $V$ with $\|V-e^{-itH}\|\le\varepsilon$ and the following bounds on its number of exponentials.

\noindent(a) \emph{General Hamiltonians.} For $\alpha>0$, put $\ell_\lambda=\max\{2,\log(\alpha t/(\lambda\varepsilon))\}$. The number of exponentials satisfies
\begin{equation}
D(tH,\varepsilon)=O\!\left((t\lambda\,\ell_\lambda^{4/3}+\ell_\lambda^2)\log^2\ell_\lambda\right).
\label{eq:time-cost-general}
\end{equation}
If $\alpha=0$, the commuting generators give an exact circuit with at most $m$ exponentials.

\noindent(b) \emph{Local Hamiltonians.} Suppose $H$ on $N$ sites is given as $m$ layers, each a sum of Hermitian terms of norm at most one with pairwise disjoint supports within the layer. Each term acts on at most $k$ sites. For $t\ge1$, put $\ell_N=\max\{2,\log(Nt/\varepsilon)\}$. The number of layer exponentials satisfies
\begin{equation}
D(tH,\varepsilon)=O\!\left(t\,\ell_N^2\log^2\ell_N\right).
\label{eq:local-implementation-cost}
\end{equation}
The circuit has depth at most $D(tH,\varepsilon)$ and uses at most $N D(tH,\varepsilon)$ gates on at most $k$ sites.
\end{theorem}

Both parts follow from a bound on a single polygon circuit at unit time.

\begin{theorem}[Error and number of exponentials at order $q$, formal version of Theorem~\ref{thm:order-cost-informal}]
\label{thm:order-cost}
There are constants $j_0,\rho_0,\lambda_0,C>0$ with the following property. Let $U=e^{-iH}$, let $q\ge2$ be an integer, and put $\rho_q=\rho_0/q$. For $0<\delta\le\rho_q$, form the polygon circuit $V_q(\delta)$ with $\delta_j=\delta\sin(\pi j/(2q))$ and cutoffs $r_j=\lceil\log(1+\log(1/\eta))/\delta_j\rceil$, where $\eta=(\delta/\rho_q)^{2q}/(4q)$. If $J\le j_0$ or $\lambda\le\lambda_0 q^{2/3}$, the error and the number of exponentials satisfy
\begin{equation}\label{eq:appendix-order-cost}
\|V_q(\delta)-U\|\le C\alpha J\rho_q^2(\delta/\rho_q)^{2q},\qquad D_q(\delta)\le C\frac{q\log q}{\delta}\log\!\bigl[q\log(e\rho_q/\delta)\bigr].
\end{equation}
In particular, at $\delta=\rho_q/2$,
\begin{equation}\label{eq:order-cost-fixed-hamiltonian}
\|V_q(\rho_q/2)-U\|\le C\alpha J\,4^{-q}/q^2,\qquad D_q(\rho_q/2)\le Cq^2\log^2q .
\end{equation}
\end{theorem}

Parts~1--7 below prove \cref{thm:order-cost}, and Part~8 deduces \cref{thm:main}.

\subsection{Outline}
\label{app:proof-map-outline}

The proof follows the three comparisons in \cref{fig:proof-map}. Starting from the finite circuit, we first restore the omitted Strang stages at each vertex, then replace each infinite Strang product by a comparison evolution. All three products use the same polygon vertices. In the last product the vertices share an analytic error expansion, so polygon cancellation applies.
\begin{equation}
 V_q(\delta)\ \xleftrightarrow{\ \text{cutoff}\ }\ V_{q,\infty}(\delta)\ \xleftrightarrow{\ \text{comparison evolution}\ }\ V_q^\Omega(\delta)\ \xleftrightarrow{\ \text{polygon cancellation}\ }\ U .
 \label{eq:proof-map-chain}
\end{equation}
For real steps the factors are unitary. Replacing them one at a time, including at inverse vertices, gives
\begin{equation}
 \|V_q-U\|\le\sum_{j=1}^{2q-1}\|Q_{r_j}(\delta_j)-Q_\infty(\delta_j)\|+\sum_{j=1}^{2q-1}\|Q_\infty(\delta_j)-\Omega_q(\delta_j)\|+\|V_q^\Omega-U\| .
 \label{eq:proof-map-three-errors}
\end{equation}
To transfer the cancellation to a finite circuit, both replacement errors must be as small as the polygon remainder, even as $q$ grows.

\proofmappart{Polygon cancellation}{app:proof-map-polygon-identity}
Suppose the approximations at all vertices share an even expansion $X(\delta)U^\dagger=F(\delta^2)$ with $F(0)=I$. In their alternating product, adjacent copies of $U$ and $U^\dagger$ cancel. At the squared-sine nodes $c_j$, \cref{lem:polygon-cancellation} gives
\[
 \prod_{j=1}^{2q-1}F(c_jz)^{(-1)^{j+1}}=I+O(z^q),\qquad z=\delta^2 .
\]
The coefficients can be noncommuting. Rotating the polygon gives $2q$ angles at which the factors cancel in inverse pairs. The coefficient of $z^n$ has trigonometric degree at most $n$, so for $n<q$ these zeros force it to vanish identically (\cref{app:polygon-proof-details}). This fixes the order of the first possible error term. Part~6 bounds its size and the rest of the remainder as $q$ grows.

\proofmappart{Finite circuits and the cutoff}{app:proof-map-cutoff}
Each finite circuit $Q_r(\delta)$ replaces the tail of $Q_\infty(\delta)$ by one Strang stage covering the same physical time $\tau_r$. Both approximations to that tail have error of order $\tau_r^3$ (\cref{lem:cutoff-selection}). Since $\tau_r$ falls double exponentially with the auxiliary cutoff time $r\delta$, we can make the first sum in \cref{eq:proof-map-three-errors} small at every polygon resolution. Part~7 counts the stages this requires.

\proofmappart{Why a comparison evolution is needed}{app:proof-map-why}
One might try to expand the finite circuits in $\delta$ and pass to the infinite product. The order of the limits matters. At fixed $r$, taking $\delta\to0$ puts almost all the physical time into the final stage, so $Q_r(\delta)\to S_2(1)$. Taking $r\to\infty$ first and then refining $\delta$ gives $U$ (\cref{app:comparison-generator}). Analyticity at each finite cutoff therefore does not justify a common convergent expansion for the limit. For each $q$ we construct one comparison evolution for all the vertices. Its error factor has the shared expansion required by Part~1, and its distance from $Q_\infty$ can be bounded at the order of the polygon remainder.

\proofmappart{The comparison evolution}{app:proof-map-comparison}
Including the advance of the auxiliary clock makes every stage a copy of the same symmetric Strang step. Its formal logarithm is odd in $\delta$, so dividing by the step gives a generator with only even powers. Keeping the terms through $\delta^{2q-2}$ gives a comparison generator that matches one stage through degree $2q$ (\cref{lem:comparison-cell-matching}). Evolving it over the fixed auxiliary interval $[0,\infty)$ gives $\Omega_q(\delta)U^\dagger=F(\delta^2)$ with $F$ analytic and $F(0)=I$ (\cref{lem:finite-circuit-expansion}).

Summing the stage errors gives the estimate at one vertex. Under either hypothesis of \cref{thm:order-cost}, \cref{lem:finite-circuit-expansion} gives
\begin{equation}
 \|Q_\infty(\delta)-\Omega_q(\delta)\|\le\frac{C\alpha J\rho_q^2}{2q+1}(\delta/\rho_q)^{2q},\qquad 0<\delta\le\rho_q .
 \label{eq:proof-map-comparison}
\end{equation}
The factor $1/(2q+1)$ lets us sum over all $2q-1$ vertices without losing a factor of $q$.

\proofmappart{Bounds that do not depend on the system size}{app:proof-map-tools}
Parts~4 and~6 use complex auxiliary steps to bound an error on the real interval. Bounding an analytic function on a disk controls the remainder once its order of vanishing is known. Applying this to a derivative and integrating back along the real interval produces the factor $1/(2q+1)$ for a stage, or $1/q$ for the polygon (\cref{lem:real-step-product}).

The difficulty is that evolution at a complex step can amplify errors. Under the norm hypothesis the proof bounds this amplification directly. For local layers, a bound in terms of $\lambda$ alone can grow exponentially with the number of sites. Here the proof follows how conjugation acts on the nested commutators in the error. Their growth is controlled by $J$, which stays bounded for layers of fixed locality as the system grows, and the estimate retains the pairwise commutator norm $\alpha$ (\cref{lem:comparison-conjugation,lem:intrinsic-ideal}).

\proofmappart{The polygon of comparison evolutions}{app:proof-map-polygon}
Part~4 supplies the same $F$ at every vertex, so Part~1 gives a zero of order at least $q$ in $z=\delta^2$ for $V_q^\Omega(\delta)U^\dagger-I$. To use that zero quantitatively under the norm hypothesis, the proof bounds the product on a larger disk. The condition $\lambda\le\lambda_0 q^{2/3}$ keeps the integrated correction at one vertex bounded, since $\lambda^3/q^2=O(1)$. Shrinking the disk by a fixed factor then absorbs the growth from multiplying the vertices. Under the commutator hypothesis, removing the exact evolution leaves only the corrections to be transported. The derivative estimate from Part~5 compensates for the sum over the vertices. Both arguments give the third term in \cref{eq:proof-map-three-errors} (\cref{lem:reference-polygon-error}),
\begin{equation}
 \|V_q^\Omega(\delta)-U\|\le C\alpha J\rho_q^2(\delta/\rho_q)^{2q}.
 \label{eq:proof-map-polygon}
\end{equation}

\proofmappart{Error and number of exponentials at fixed order}{app:proof-map-assembly}
Choose the cutoffs from Part~2 so that their combined error is at most the bound in \cref{eq:proof-map-polygon}. Summing \cref{eq:proof-map-comparison} over the vertices gives the same bound for the comparison error. Adding the three terms now gives \cref{thm:order-cost} for the finite circuit (\cref{app:alpha-cutoff}).

At $\delta=\rho_q/2$ the error is $O(\alpha J\,4^{-q}/q^2)$. Two logarithms enter the cost. Each vertex needs $O(\log q)$ auxiliary time to reach its cutoff, and summing the inverse sine-spaced steps gives $\sum_j1/\delta_j=O(q\log q/\delta)$ (\cref{eq:polygon-sine-cost}). Since $\delta$ is proportional to $1/q$ and each Strang stage uses $2m-1$ exponentials, one interval costs $O(q^2\log^2q)$ exponentials.

\proofmappart{From one circuit to time and precision}{app:proof-map-time}
To simulate time $t$, apply Part~7 to the scaled generators $(t/M)H_a$ and repeat the circuit $M$ times. Unitarity bounds the total error by $M$ times the error on one interval (\cref{prop:time-subdivision}). The two hypotheses allow different interval lengths.

\emph{General Hamiltonians.} The norm hypothesis permits intervals of length proportional to $q^{2/3}/\lambda$. Thus $O(1+\lambda t/q^{2/3})$ intervals suffice, and multiplying by the cost of one interval gives the power $q^{4/3}$ in the term linear in $t$. The total error is $O((\alpha t/\lambda)4^{-q})$, so $q=\Theta(\ell_\lambda)$ suffices for $\alpha>0$ (\cref{app:general-time-precision}). For commuting generators a single Strang stage is exact.

\emph{Local Hamiltonians.} The layer decomposition gives $J\le3k^2$, independent of $N$. Enforcing the commutator hypothesis on each interval takes $O(k^2t)$ intervals for $t\ge1$. Each costs $O(q^2\log^2q)$ layer exponentials, giving the factor $k^2t$ in the depth bound. Counting overlapping terms gives $\alpha=O(kN)$, so the system size enters only through the choice $q=O(\ell_N)$ (\cref{app:many-native-generators}).

%% file: appendices/appendix_polygon.tex
\section{{Proof of Lemma~\ref{lem:polygon-cancellation} (Polygon cancellation)}}\label{app:polygon-cancellation}

This appendix provides the details behind \cref{sec:polygon-cancellation}. \Cref{app:polygon-proof-details} restates \cref{lem:polygon-cancellation} and gives a self-contained proof, together with an analytic version. \Cref{app:polygon-degree-two} is more expository: it describes how the nodes $c_j=\sin^2(\pi j/2q)$ can be found as a pattern in low-order calculations, writes out the $q=3$ cancellation in full, and discusses how many conditions the cancellation actually has to satisfy. Throughout, $q\ge2$ is an integer, and every asymptotic estimate is at fixed cancellation order $q$ unless stated otherwise.

\subsection{{Proof for formal power series}}\label{app:polygon-proof-details}

For convenience, we restate~\cref{lem:polygon-cancellation} from the main text. We provide a self-contained proof of it in this subsection.

\medskip\noindent\textbf{\Cref{lem:polygon-cancellation} (restated).} \emph{Let $F(z)=I+\sum_{k\ge1}z^kE_{2k}$ be a formal power series with arbitrary coefficients $E_{2k}$. For every integer $q\ge2$,}
\begin{equation}
 \prod_{j=1}^{2q-1}F\!\left(z\sin^2\frac{\pi j}{2q}\right)^{(-1)^{j+1}}=I+O(z^q),
 \label{eq:appendix-restated}
\end{equation}
\emph{that is, every coefficient of $z^n$ with $1\le n<q$ vanishes.}
\medskip

Here, as in the main text, $O(z^q)$ denotes a formal power series whose coefficients below $z^q$ vanish. The analytic statement is given separately in \cref{prop:analytic-polygon} below.

 Formal power series whose coefficients are matrices can be multiplied by collecting terms with the same total power of $z$. The order of the matrix factors is preserved. Since the constant term of $F$ is $I$, it has a unique formal inverse, which we can obtain by expanding $F^{-1}=\sum_{m\ge0}(I-F)^m$. Crucially, only finitely many terms contribute to each power of $z$. Write $\sigma_j=(-1)^{j+1}$ for the alternating signs, and recall the rotated nodes and the rotated polygon product from \cref{eq:rotated-polygon},
\begin{equation}
 c_j(\theta)=\frac{1-\cos(\theta+\pi j/q)}{2},\qquad
 W_q(\theta,z)=\prod_{j=0}^{2q-1}F\bigl(c_j(\theta)z\bigr)^{\sigma_j},
 \label{eq:appendix-rotated}
\end{equation}
with factors written from left to right in increasing $j$. Indices on the nodes and signs are taken modulo $2q$. This is because $c_{j+2q}(\theta)=c_j(\theta)$ and, since $2q$ is even, $\sigma_{j+2q}=\sigma_j$. At $\theta=0$ we have $c_0(0)=0$, so the first factor is $F(0)^{-1}=I$, and $c_j(0)=\sin^2(\pi j/2q)$ for the remaining factors. Hence $W_q(0,z)$ is exactly the product on the left-hand side of \cref{eq:appendix-restated}. The proof has four steps: exact cancellation at one rotation, exact cancellation at $2q$ rotations, a bound on how the coefficients depend on $\theta$, and root counting to finish.

\paragraph{{Step 1. Cancellation at a half-step rotation.}} Let $\theta_*=\pi/(2q)$ be half of the spacing between vertices. Vertex $j$ then sits at angle $(2j+1)\pi/(2q)$, and vertex $2q-1-j$ at angle $2\pi-(2j+1)\pi/(2q)$. These are reflections of each other in the horizontal axis, and since $\cos(2\pi-\phi)=\cos\phi$, we have
\begin{equation}
 c_{2q-1-j}(\theta_*)=c_j(\theta_*),\qquad \sigma_{2q-1-j}=-\sigma_j,
\end{equation}
where the second identity holds because $j$ and $2q-1-j$ have opposite parity (see~\cref{fig:polygon} for a visualization). For convenience, abbreviate $Y_m=F(c_m(\theta_*)z)^{\sigma_m}$ for $m=0,\dots,2q-1$, and $X_j=Y_j$ for $0\le j<q$. The two identities above say that $Y_{2q-1-j}=X_j^{-1}$, so the factors around the rotated polygon occur in the following order.
\begin{equation}\label{eq:appendix-pairing}
 W_q(\theta_*,z)=Y_0Y_1\cdots Y_{2q-1}=X_0\cdots X_{q-1}X_{q-1}^{-1}\cdots X_0^{-1}=I.
\end{equation}
The last equality follows by canceling each factor against the adjacent inverse, from the middle outward. Note that the ordering as we traverse the polygon is important. 

\paragraph{\texorpdfstring{{Step 2. The $2q$ zeros.}}{Step 2. The 2q zeros.}} Then observe that this cancellation actually occurs at several different angles. Let $\theta_k=\theta_*+k\pi/q=(2k+1)\pi/(2q)$ for $k=0,\dots,2q-1$. These are $2q$ distinct angles in $[0,2\pi)$. Rotating by $k$ vertex spacings shifts the nodes to $c_j(\theta_k)=c_{j+k}(\theta_*)$, and since $2q$ is even, the parity becomes $\sigma_j=(-1)^k\sigma_{j+k}$. The factor at position $j$ of $W_q(\theta_k,z)$ is therefore $F(c_{j+k}(\theta_*)z)^{(-1)^k\sigma_{j+k}}=Y_{j+k}^{(-1)^k}$, and
\begin{equation}
 W_q(\theta_k,z)=Y_k^{\eta}Y_{k+1}^{\eta}\cdots Y_{2q-1}^{\eta}\,Y_0^{\eta}\cdots Y_{k-1}^{\eta},\qquad \eta=(-1)^k.
\end{equation}
This is a cyclic shift of the product $Y_0^{\eta}Y_1^{\eta}\cdots Y_{2q-1}^{\eta}$. For $\eta=+1$, this product equals $I$ by \cref{eq:appendix-pairing}. For $\eta=-1$, it equals $X_0^{-1}\cdots X_{q-1}^{-1}X_{q-1}\cdots X_0$, which also equals $I$ by the same middle-outward cancellation. Finally, a cyclic shift of a product equal to $I$ is again $I$: if $Z_0Z_1\cdots Z_m=I$, then
\begin{equation}
 Z_k\cdots Z_mZ_0\cdots Z_{k-1}=(Z_0\cdots Z_{k-1})^{-1}(Z_0Z_1\cdots Z_m)(Z_0\cdots Z_{k-1})=I.
\end{equation}
We have therefore shown that
\begin{equation}\label{eq:appendix-polygon-zeros}
 W_q(\theta_k,z)=I,\qquad \theta_k=\frac{(2k+1)\pi}{2q},\qquad k=0,\ldots,2q-1,
\end{equation}
as an identity of formal power series in $z$. Equivalently, every coefficient of $z^n$ with $n\ge1$ in $W_q(\theta_k,z)$ vanishes, for each of these $2q$ angles.

\paragraph{{Step 3. Angular degree of the coefficients.}} We now track how the coefficients of $W_q(\theta,z)$ depend on $\theta$. For $\sigma\in\{+1,-1\}$, write the expansion of $F$ (or its inverse) as
\begin{equation}
 F(y)^{\sigma}=\sum_{r\ge0}y^rB_r^{(\sigma)},\qquad B_0^{(\sigma)}=I,
\end{equation}
where the matrices $B_r^{(\sigma)}$ are fixed polynomials in the $E_{2k}$ and do not depend on $\theta$. Substituting $y=c_j(\theta)z$ into each factor and collecting powers of $z$, the coefficient of $z^n$ in $W_q(\theta,z)$ is
\begin{equation}\label{eq:appendix-angular-coefficient}
 M_n(\theta)=\sum_{r_0+\cdots+r_{2q-1}=n}\left(\prod_{j=0}^{2q-1}c_j(\theta)^{r_j}\right)B_{r_0}^{(\sigma_0)}B_{r_1}^{(\sigma_1)}\cdots B_{r_{2q-1}}^{(\sigma_{2q-1})},
\end{equation}
where the sum runs over nonnegative integers. Each summand is a scalar function of $\theta$ times a matrix product that keeps its original ordering but does not depend on $\theta$. To see how the scalar factor depends on $\theta$, put $w=e^{i\theta}$ and $\omega=e^{i\pi/q}$. Then
\begin{equation}
 c_j(\theta)=\frac14\left(2-\omega^jw-\omega^{-j}w^{-1}\right)
\end{equation}
contains only the powers $w^{-1}$, $w^0$ and $w^1$. A product of $n$ such factors, as in each summand of \cref{eq:appendix-angular-coefficient}, therefore contains only powers $w^m$ with $-n\le m\le n$. Collecting terms,
\begin{equation}\label{eq:appendix-laurent}
 M_n(\theta)=\sum_{m=-n}^{n}M_{n,m}\,w^m,\qquad w=e^{i\theta},
\end{equation}
for some matrices $M_{n,m}$ that do not depend on $\theta$. In other words, $M_n$ is a matrix-valued trigonometric polynomial in $\theta$ of degree at most $n$. The noncommutativity of the $E_{2k}$ affects only the matrices $M_{n,m}$, which we never need to compute.

\paragraph{{Step 4. Root counting.}} Multiplying \cref{eq:appendix-laurent} by $w^n$ removes the negative powers:
\begin{equation}
 P_n(w)=w^nM_n(\theta)=\sum_{m=-n}^{n}M_{n,m}\,w^{m+n}.
\end{equation}
Each matrix entry of $P_n$ is an ordinary polynomial in $w$ of degree at most $2n$. By \cref{eq:appendix-polygon-zeros}, $M_n(\theta_k)=0$ for every $k$, and since $w\ne0$, each entry of $P_n$ vanishes at the $2q$ distinct points $w_k=e^{i\theta_k}$ on the unit circle. A nonzero polynomial of degree at most $2n$ has at most $2n$ roots. If $n<q$, then $2q>2n$, so every entry of $P_n$ vanishes identically. Hence all $M_{n,m}=0$ and $M_n(\theta)=0$ for every $\theta$. In particular, $M_n(0)=0$ for $1\le n<q$, which is \cref{eq:appendix-restated}. \hfill$\square$

At $n=q$, each entry of $P_q(w)$ has degree at most $2q$ and vanishes at every root of $w^{2q}+1$, so it is a constant multiple of that polynomial. Dividing by $w^q$ gives $M_q(\theta)=M_q(0)\cos(q\theta)$. Thus $z^q$ is the first power at which an error can survive.

The proof uses only linearity, associativity, and the existence of formal inverses, and never uses the error coefficients $E_{2k}$ themselves. This means that the identity holds in any unital associative algebra, not only for square matrices. Since a finite-dimensional unital associative algebra can always be represented faithfully by square matrices (by letting each element act on the algebra by left multiplication) the matrix version above already covers this case.

\paragraph{Analytic version.} When $F$ is an analytic function rather than a formal series, the following analytic version also holds.
\begin{proposition}\label{prop:analytic-polygon}
Suppose that $F$ is analytic on a disk $|z|<\rho$ with $F(0)=I$. Then there are $\rho'\in(0,\rho]$ and $C>0$, depending on $F$ and $q$, such that $W_q(0,z)$ is analytic for $|z|<\rho'$ and
\begin{equation}
 \|W_q(0,z)-I\|\le C|z|^q,\qquad |z|\le\rho'/2.
\end{equation}
\end{proposition}
\begin{proof}
Since $F(0)=I$ and $F$ is continuous, $\|F(z)-I\|\le\tfrac12$ on some disk $|z|<\rho'$, where $F(z)$ is therefore invertible and $F(z)^{-1}=\sum_{m\ge0}(I-F(z))^m$ is analytic. The nodes $c_j=\sin^2(\pi j/2q)$ lie in $[0,1]$, so every factor $F(c_jz)^{\pm1}$ is analytic on the same disk, and so is their product $W_q(0,z)$. Its Taylor coefficients are the formal coefficients computed above, so those of $z^1,\dots,z^{q-1}$ vanish and $W_q(0,z)-I=z^qG(z)$ with $G$ analytic for $|z|<\rho'$. Taking $C=\max_{|z|\le\rho'/2}\|G(z)\|$ proves the bound.
\end{proof}

Note that the constant $C$ depends on $q$, and this argument gives no control over how it grows. Bounding that growth is the task of \cref{app:statements}.

\subsection{{Low-order examples and order conditions}}\label{app:polygon-degree-two}\label{app:polygon-words}

The proof in \cref{app:polygon-proof-details} explains why the nodes $c_j$ work, but it is also interesting to discuss how they can be found. In this subsection we go through the low-order calculations that lead to the pattern, write out the $q=3$ cancellation in full, and discuss how many conditions the cancellation really has to satisfy. Nothing here is needed for the proofs; it is meant to make the mechanism concrete.

\paragraph{{Expansion through second order.}} Consider an arbitrary alternating product $\prod_jF(b_jz)^{\sigma_j}$, where the $b_j$ are real nodes and $\sigma_j\in\{+1,-1\}$, and define the signed moments
\begin{equation}
 s_k=\sum_j\sigma_jb_j^k .
\end{equation}
Inverting $F$ term by term, using $(I+X)^{-1}=I-X+X^2-\cdots$, gives
\begin{equation}
 F(y)^{-1}=I-yE_2+y^2(E_2^2-E_4)+y^3\bigl(-E_6+E_2E_4+E_4E_2-E_2^3\bigr)+O(y^4).
\end{equation}
So already at order $y^2$, inversion produces a product of lower-order coefficients. Multiplying out the alternating product through order $z^2$ gives
\begin{equation}\label{eq:appendix-degree-two}
 \prod_jF(b_jz)^{\sigma_j}=I+zs_1E_2+z^2\left[s_2E_4+\frac{s_1^2-s_2}{2}E_2^2\right]+O(z^3).
\end{equation}
The coefficients of $E_2$ and $E_4$ can be read off directly. The coefficient of $E_2^2$ has two sources: the quadratic terms of the exact inverses contribute $\sum_{\sigma_j=-1}b_j^2$, and choosing the linear terms of two different factors $i<j$ contributes $\sum_{i<j}\sigma_i\sigma_jb_ib_j$. Since $\sigma_j^2=1$, we have $s_1^2=\sum_jb_j^2+2\sum_{i<j}\sigma_i\sigma_jb_ib_j$, and the two sources add up to $(s_1^2-s_2)/2$. So through order $z^2$, the coefficient of $E_2^2$ is canceled by $s_1=s_2=0$.

\paragraph{{The three-factor product.}} For the three-factor product of \cref{eq:three-factor-cancellation}, with nodes and signs $(\tfrac12,+),(1,-),(\tfrac12,+)$, we have $s_1=0$ but $s_2=-\tfrac12$, so only the linear term cancels. Explicitly,
\begin{equation}
 F(z/2)F(z)^{-1}F(z/2)=I-\frac{z^2}{2}\left(E_4-\frac12E_2^2\right)+O(z^3).
\end{equation}
To cancel through order $z^2$, we try the next simplest palindrome, $F(az)F(bz)^{-1}F(z)F(bz)^{-1}F(az)$, with two free ratios. After expanding, we find that the moments are $s_1=2a-2b+1$ and $s_2=2a^2-2b^2+1$. The first condition gives $a-b=-\tfrac12$, and factorizing the second as $2(b-a)(b+a)=1$ then gives $a+b=1$. So $a=\tfrac14$ and $b=\tfrac34$. Comparing $\tfrac12$ with $\tfrac14$ and $\tfrac34$, and recalling that $\sin^2(\pi/4)=\tfrac12$, $\sin^2(\pi/6)=\tfrac14$ and $\sin^2(\pi/3)=\tfrac34$, suggests the nodes $\sin^2(\pi j/2q)$. One can check that the same choice for $q=4$, namely $\sin^2(\pi j/8)$, makes $s_1$, $s_2$ and $s_3$ vanish. \Cref{lem:polygon-cancellation} then shows that this continues for all $q$ by using the polygon argument.

\paragraph{{The five-factor product.}} For $q=3$, all terms of order $z$ and $z^2$ cancel. The first surviving term has order $z^3$, and a direct calculation gives
\begin{equation}\label{eq:appendix-q3-cubic}
 W_3(0,z)=I+\frac{3z^3}{16}\left[E_6-\frac12\bigl(E_2E_4+E_4E_2\bigr)+\frac13E_2^3\right]+O(z^4).
\end{equation}
The prefactor is $s_3=2(\tfrac14)^3-2(\tfrac34)^3+1=\tfrac3{16}$. The two ordered products $E_2E_4$ and $E_4E_2$ appear with equal coefficients here, but this does not persist. At order $z^4$, $W_3(0,z)$ contains, among others,
\begin{equation}
 \frac{17}{128}E_2E_2E_4+\frac{7}{64}E_2E_4E_2+\frac{17}{128}E_4E_2E_2,
\end{equation}
so the coefficient of an ordered product of the error coefficients genuinely depends on the ordering of its factors.

\paragraph{{Order conditions for noncommuting coefficients.}} The error coefficient $E_{2k}$ carries the power $z^k$. The ordered products of the error coefficients of order $z^n$, such as $E_2E_4$ and $E_4E_2$ for $n=3$, correspond to compositions $n=k_1+\cdots+k_r$ into positive integers, of which there are $2^{n-1}$. This gives $2^{q-1}-1$ ordered products at positive orders below $z^q$. For generic error coefficients, the scalar multiplying each such product must vanish. Matching these coefficients one by one, as in the calculation above, quickly becomes impractical.

The vanishing of their coefficients does not give independent conditions, however, and it is worth being precise about which ones the lemma is really needed for. Write $F(y)=\exp L(y)$ with $L(y)=\sum_{r\ge1}y^rL_r$, so that each factor is $F(b_jz)^{\sigma_j}=\exp(\sigma_jL(b_jz))$. We write $W(z)=\prod_jF(b_jz)^{\sigma_j}$ for this product. For a palindromic product, the terms of $\log W$ that involve an even number of these exponents vanish: multiplying every exponent by a scalar $t$ gives a product whose value at $-t$ is its inverse, so its logarithm is odd in $t$. Terms involving three exponents first appear at order $z^3$, and there they vanish, because the only linear terms are multiples of $L_1$. Hence, through order $z^3$,
\begin{equation}
 \log W(z)=zs_1L_1+z^2s_2L_2+z^3s_3L_3+O(z^4).
\end{equation}
Through order $z^3$, therefore, only the signed moments matter. In particular, all four ordered products of order $z^3$ in \cref{eq:appendix-q3-cubic} are controlled by the single condition $s_3=0$, and $\exp(s_3z^3L_3)$ reproduces the bracket there.

From order $z^4$ on, this is no longer true. The logarithm of a palindromic product then also contains the nested commutator $[L_1,[L_1,L_2]]$, whose coefficient now also depends on the ordering of the factors. For example, for $q=4$ the polygon product makes $s_1$, $s_2$ and $s_3$ vanish, and the part of order $z^4$ of its logarithm is $s_4z^4L_4$ plus a nonzero multiple of $[L_1,[L_1,L_2]]$. For $q=5$, both vanish. So the low-order examples above, up to $q=4$, could have been found from the conditions $s_k=0$ on the signed moments alone, and the ordering-dependent cancellation guaranteed by \cref{lem:polygon-cancellation} starts doing real work at $q=5$ and up. For larger $q$, the number of such nested commutators grows quickly, and the lemma allows us to cancel all of them simultaneously without computing any of the moment conditions.

%% file: appendices/commutator-growth.tex
\section{Commutator growth for Hamiltonian sums}
\label{app:intrinsic-scales}

This appendix fixes the notation for the Hamiltonian decomposition and the commutator bounds used in the error estimates. \Cref{app:pf-setting} defines the product formulae, the nested commutators of the generators, and their growth scale $J$. \Cref{app:local-definitions} specifies how we measure the strength of local interactions and derives a system-size-independent bound for $J$. \Cref{app:comparison-tools} uses the same growth scale to control conjugation.

\subsection{Parameters and nested commutators}
\label{app:pf-setting}

{Let $H$ be a time-independent Hamiltonian and write $U(t)=e^{-itH}$ for its time evolution. Given a time $t$ and a precision $\varepsilon$, our goal is to implement a circuit $V$ such that $\|V-U(t)\|\le\varepsilon$, where $\|\cdot\|$ denotes the operator norm.

We are given a decomposition $H=\sum_{a=1}^m H_a$ into Hermitian generators and assume that each evolution $e^{-iuH_a}$ can be implemented individually for any real $u$. A product formula approximates $U(t)$ by an ordered product of these exponentials,
\[
 V=e^{-iu_D H_{a_D}}\cdots e^{-iu_2 H_{a_2}}e^{-iu_1 H_{a_1}} .
\]
The rightmost factor acts first. { The product contains $D$ exponentials. For an approximation to $e^{-itH}$ with precision $\varepsilon$, we denote this count by $D(tH,\varepsilon)$, treating $tH$ as the Hamiltonian for unit time as in Sec.~\ref{sec:setting}.}

For the two-generator decomposition $H=A+B$, one Strang stage
\[
 S_2(u)=e^{-iuA/2}e^{-iuB}e^{-iuA/2}
\]
approximates $U(u)$ with three exponentials.}

For the general decomposition write
\begin{equation}
 \lambda=\sum_{a=1}^m\|H_a\|,\qquad
 \alpha=\sum_{1\le a<b\le m}\|[H_a,H_b]\| .
 \label{eq:sum-parameters}
\end{equation}
The commutator estimate $\|[P,Q]\|\le2\|P\|\|Q\|$ gives $\alpha\le 2\sum_{a<b}\|H_a\|\|H_b\|\le\lambda^2$.

\subsubsection{Nested commutators and conjugation}
\label{app:nested-notation}

{For two generators $A$ and $B$, starting with $B$ and commuting twice with $A$ gives $[A,[A,B]]$. To keep track of the order of these operations, write $\operatorname{ad}_T(X)=[T,X]$ and list the generators in the order in which they enter, starting with the innermost operator. For a word $\boldsymbol w=(X_1,\ldots,X_d)$ with $X_j\in\{H_1,\ldots,H_m\}$, set $C_{(X_1)}=X_1$ and define}
{
\begin{equation}
 C_{\boldsymbol w}=[X_d,C_{(X_1,\ldots,X_{d-1})}]
 =\operatorname{ad}_{X_d}\cdots\operatorname{ad}_{X_2}(X_1),\qquad d\ge2 .
 \label{eq:intrinsic-nested-words}
\end{equation}
{Thus $C_{(B,A,A)}=[A,[A,B]]$ has degree three. In general the degree $d=|\boldsymbol w|$ is the number of letters in the word, counted with repetitions. Error terms may contain words of different degrees. We collect words of degree at most $\nu$ and the identity in the space}
\[
 \mathfrak L_\nu=\operatorname{span}_{\mathbb C}
 \bigl(\{I\}\cup\{C_{\boldsymbol w}:1\le|\boldsymbol w|\le\nu\}\bigr).
\]
Thus $\mathfrak L_0=\operatorname{span}_{\mathbb C}\{I\}$. { Including $I$ makes each $\mathfrak L_\nu$ invariant under scalar shifts of the generators.} For the two-generator example, allowing nested commutators of successively higher degree gives
\[
 \begin{aligned}
 \mathfrak L_1 &= \operatorname{span}_{\mathbb C}\{I,A,B\},\\
 \mathfrak L_2 &= \mathfrak L_1+\operatorname{span}_{\mathbb C}\{[A,B]\},\\
 \mathfrak L_3 &= \mathfrak L_2+\operatorname{span}_{\mathbb C}
                      \{[A,[A,B]],\, [B,[A,B]]\}.
 \end{aligned}
\]
At degree two, $[A,A]=[B,B]=0$ and $[B,A]=-[A,B]$, so it suffices to include $[A,B]$. Commuting it once more with either $A$ or $B$ gives the two entries on the last line. Thus $A+[B,[A,B]]$ belongs to $\mathfrak L_3$ even though its two terms have different degrees.

{Commuting $X\in\mathfrak L_\nu$ with any generator adds at most one to its degree bound. If the acting operator is itself a commutator, the Jacobi identity lets us apply the same rule one generator at a time. For instance,}
\[
 [[A,B],X]=[A,[B,X]]-[B,[A,X]].
\]
{Both terms on the right therefore belong to $\mathfrak L_{\nu+2}$.} More generally,
\begin{equation}
 Y\in\mathfrak L_d,\quad X\in\mathfrak L_\nu
 \quad\Longrightarrow\quad [Y,X]\in\mathfrak L_{d+\nu}.
 \label{eq:commutator-degree-addition}
\end{equation}
By linearity it suffices to take $Y$ to be a single nested commutator, since $[I,X]=0$. The degree-one case follows from the definition. For $Y=[T,C']$ with $T$ a generator, write
$[[T,C'],X]=[T,[C',X]]-[C',[T,X]]$.
If commutation with $C'$ raises the degree bound of any input by at most $d-1$, the two terms raise it by at most $(d-1)+1$ and $1+(d-1)$, respectively. This proves the rule by induction.

{ For $H=\sum_{a=1}^m H_a$, each commutation with $H$ gives $[H,X]=\sum_a[H_a,X]$. Repeating this operation gives the Taylor coefficients of $e^{-itH}Xe^{itH}$.} For invertible $U$, write $\operatorname{Ad}_U(X)=UXU^{-1}$. {Differentiating $e^{-itH}Xe^{itH}$ gives $-i[H,e^{-itH}Xe^{itH}]$, so
\[
 \operatorname{Ad}_{e^{-itH}}(X)
 =e^{-it\operatorname{ad}_H}(X)
 =\sum_{j\ge0}\frac{(-it)^j}{j!}\operatorname{ad}_H^{\,j}(X).
\]}
This series converges in operator norm since $\|\operatorname{ad}_H(X)\|\le2\|H\|\|X\|$. Appendix~\ref{app:intrinsic-cost} uses this expansion for the exact evolution.}

\subsubsection{\texorpdfstring{Definition and properties of $J$}{Definition and properties of the growth scale}}
\label{app:scale-definition}

{Let $H_1,\ldots,H_m$ be Hermitian operators of finite norm. To bound an error term built from these generators, we need an estimate of the norm at each commutation. For a single generator $T\in\{H_1,\ldots,H_m\}$ { we seek a number $J$ such that}}
\[
 \|[T,X]\|\le 2J\nu\|X\|,\qquad X\in\mathfrak L_\nu .
\]
{We choose $J$ so that it also controls commutation with a word of any degree.} The \emph{commutator growth scale} $J=J(H_1,\ldots,H_m)$
is the smallest nonnegative number satisfying
\begin{equation}
 \|[C_{\boldsymbol w},X]\|
 \le \nu\,d!\,(2J)^d\|X\|,
 \qquad X\in\mathfrak L_\nu,
 \label{eq:commutator-growth-scale}
\end{equation}
for every $d,\nu\ge1$ and every nested commutator of degree $d$.
Here $d$ is the degree of $C_{\boldsymbol w}$, while $\nu$
limits the degree of its argument~$X$. The argument may be any complex
linear combination in $\mathfrak L_\nu$. {We first prove homogeneity and the bound in terms of the generator norms.} \Cref{app:local-scale-proof} proves the local bound.

\begin{lemma}[Basic properties of the commutator growth scale]
\label{lem:commutator-scale-properties}
With $\lambda=\sum_a\|H_a\|$, for every real $u$,
\begin{equation}
 \begin{aligned}
 J(uH_1,\ldots,uH_m)&=|u|J(H_1,\ldots,H_m),\\
 J(H_1,\ldots,H_m)&\le\max_a\|H_a\|\le\lambda.
 \end{aligned}
 \label{eq:main-global-commutator-scale}
\end{equation}
If each $H_a$ is a sum of terms supported on at most $k$ sites, with interaction strength per site (Definition~\ref{def:locality-strength}) at most $g_{\mathrm{loc}}$, then
\begin{equation}
 {J}\le3k^2g_{\mathrm{loc}}.
 \label{eq:main-local-commutator-scale}
\end{equation}
\end{lemma}
Set $\Lambda=\max_a\|H_a\|$. The elementary estimates
$\|C_{\boldsymbol w}\|\le2^{d-1}\Lambda^d$ and
$\|[C_{\boldsymbol w},X]\|\le(2\Lambda)^d\|X\|$
show that $J\le\Lambda\le\lambda$.
{To see that a smallest admissible scale exists, take one half} of the supremum of
$\bigl(\|[C_{\boldsymbol w},X]\|/(\nu d!\|X\|)\bigr)^{1/d}$
over the nonzero arguments in the defining inequality of $J$, Eq.~\eqref{eq:commutator-growth-scale}.
{This number is finite by the preceding estimate and satisfies every inequality in the definition.}

For $u\ne0$, replacing $(H_1,\ldots,H_m)$ by $(uH_1,\ldots,uH_m)$ leaves each $\mathfrak L_\nu$ unchanged
and multiplies a nested commutator of degree $d$ by $u^d$. {The case $u=0$ is immediate, giving the homogeneity in Eq.~\eqref{eq:main-global-commutator-scale}.}
{ Scalar shifts of the generators also leave $J$ unchanged, since they preserve $\mathfrak L_\nu$ and every commutator in Eq.~\eqref{eq:commutator-growth-scale}.}

{Applying the single-generator bound first to $[H_a,H_b]$ and then to $[H_b,H_a]$ gives $\|[H_a,H_b]\|\le 2J\min\{\|H_a\|,\|H_b\|\}$. In particular, if $J=0$ then the generators commute pairwise. Conversely, if they commute, every word of degree at least two vanishes and Eq.~\eqref{eq:commutator-growth-scale} holds with $J=0$.}

{ We can now bound a word of any degree starting from its innermost commutator. At degree two its norm is at most $\alpha$. At degree three, for instance, $\|[H_c,[H_a,H_b]]\|\le 4J\|[H_a,H_b]\|$ since $[H_a,H_b]\in\mathfrak L_2$. At each subsequent degree $j$, the single-generator bound multiplies the preceding estimate by $2J(j-1)$. Thus}
\begin{equation}
 \|C_{\boldsymbol w}\|\le\|[X_2,X_1]\|(2J)^{d-2}(d-1)!\le\alpha(2J)^{d-2}(d-1)!,\qquad d\ge2.
 \label{eq:intrinsic-root-bound}
\end{equation}

In particular the third letter costs at most $4J$. If the later
commutators are estimated using $\|[H_a,X]\|\le2\lambda\|X\|$, we
instead obtain
\begin{equation}
 \|C_{\boldsymbol w}\|\le 4\alpha J(2\lambda)^{d-3},\qquad d\ge3 .
 \label{eq:intrinsic-mixed-word}
\end{equation}
This keeps the same cubic prefactor when the higher degrees are
controlled using $\lambda$.

\subsection{Local growth bound}
\label{app:local-definitions}

{ We now specify the local interactions that make up a Hamiltonian on $N$ sites. We first define locality and interaction strength for a single Hamiltonian, then apply these definitions to each generator $H_a$ in the product formula.}

\begin{definition}[Interaction strength per site]
\label{def:locality-strength}
{Consider a decomposition of a Hamiltonian into interactions supported on at most $k$ sites each,
\begin{equation}
 H=\sum_{|S|\le k}h_S,\qquad h_S=h_S^\dagger .
 \label{eq:local-decomposition}
\end{equation}
To measure the strength incident on one site, fix $i$ and sum the norms of the interactions whose supports contain it. For the specified decomposition we may use any upper bound $g_H$ satisfying
\begin{equation}
 \max_i\sum_{S\ni i}\|h_S\|\le g_H .
 \label{eq:local-site-strength}
\end{equation}
{ For the decomposition used in the product formula, write $H_a=\sum_S h_{a,S}$, where $h_{a,S}$ acts only on the sites in $S$. Applying the same site sum to each generator gives}
\[
 g_{\mathrm{loc}}=\max_{1\le a\le m}\max_i\sum_{S\ni i}\|h_{a,S}\|.
\]}
\end{definition}
For the decomposition in \cref{def:local-hamiltonian}, disjointness within each layer and the unit norm bound on each term give $g_{\mathrm{loc}}\le1$.

{For example, the open spin-$1/2$ Heisenberg chain has $H=\sum_{i=1}^{N-1}h_{i,i+1}$ with $h_{i,i+1}=J_{\mathrm{ex}}\,\boldsymbol S_i\cdot\boldsymbol S_{i+1}$, so $k=2$. Here $J_{\mathrm{ex}}$ is the exchange coupling. With $S_i^\gamma=\sigma_i^\gamma/2$, the singlet and triplet bond energies are $-3J_{\mathrm{ex}}/4$ and $J_{\mathrm{ex}}/4$, giving $\|h_{i,i+1}\|=3|J_{\mathrm{ex}}|/4$. At most two bonds meet at any site, so we may take $g_H=3|J_{\mathrm{ex}}|/2$. Grouping the odd bonds into $H_1$ and the even bonds into $H_2$ leaves at most one bond per site in each generator, giving $g_{\mathrm{loc}}=3|J_{\mathrm{ex}}|/4$.}

\subsubsection{Commutators of local sums}
\label{app:local-sum-commutators}

{In the odd--even decomposition above, each term of $H_1$ or $H_2$ acts on two neighboring sites. {In the commutator $[H_1,H_2]=H_1H_2-H_2H_1$, an odd bond can contribute only when paired with an adjacent even bond. The resulting term acts on at most three consecutive sites.} { More generally, only terms with overlapping supports contribute to the commutator of two local sums.} { To repeat the calculation for $[H_1,[H_1,H_2]]$, we also need the interaction strength per site of $[H_1,H_2]$. The following lemma bounds the locality and site strength of the commutator of any two local sums.}}

\begin{lemma}[Commutator of local sums]
\label{lem:local-commutator-strength}
{Given local decompositions $P=\sum_S p_S$ and $Q=\sum_T q_T$ with localities $k_P,k_Q$ and interaction strengths per site
$g_P,g_Q$, the expansion $[P,Q]=\sum_{S,T}[p_S,q_T]$ has locality at most}
$k_P+k_Q$ and interaction strength per site satisfying
\begin{equation}
 g_{[P,Q]}\le2(k_P+k_Q)g_Pg_Q.
 \label{eq:local-commutator-strength}
\end{equation}
\end{lemma}

\begin{proof}
Write $[P,Q]=\sum_{S,T}[p_S,q_T]$. Disjoint supports give zero;
each remaining term is supported on $S\cup T$ and has norm at most
$2\|p_S\|\|q_T\|$. {Fix a site $i$ and first consider the pairs with $i\in S$. For a given $S$,}
\[
 \sum_{T:T\cap S\ne\varnothing}\|q_T\|
 \le \sum_{j\in S}\sum_{T\ni j}\|q_T\|\le k_Pg_Q .
\]
{Multiplying by $2\|p_S\|$ and summing over $S\ni i$ gives a contribution at most $2k_Pg_Pg_Q$. The pairs with $i\in T$ contribute at most $2k_Qg_Pg_Q$.}
Counting some pairs twice only enlarges the sum. Adding the two
bounds proves the claim.
\end{proof}

\subsubsection{\texorpdfstring{A bound on $J$ for local Hamiltonians}{A bound on the local growth scale}}
\label{app:local-scale-proof}

{Suppose each generator has locality $k$ and interaction strength per site at most $g_{\mathrm{loc}}$. { To prove $J\le 3k^2g_{\mathrm{loc}}$, we first bound the site strength of a nested commutator $C_{\boldsymbol w}$ and then estimate its action on $X\in\mathfrak L_\nu$.} Expanding a word of length $d$ in local terms gives supports of at most $dk$ sites. Let $g_d$ be the largest site strength of these expansions over all words of length $d$. Adding one generator to a word of length $d-1$ combines supports of sizes at most $k$ and $(d-1)k$. Lemma~\ref{lem:local-commutator-strength} therefore gives $g_d\le 2dk\,g_{\mathrm{loc}}g_{d-1}$. Starting from $g_1\le g_{\mathrm{loc}}$, we obtain}
\begin{equation}
 g_d\le
 g_{\mathrm{loc}}(2kg_{\mathrm{loc}})^{d-1}d!.
 \label{eq:intrinsic-derived-strength}
\end{equation}

{To use this site-strength estimate in Eq.~\eqref{eq:commutator-growth-scale}, we need a bound on $\|[C_{\boldsymbol w},X]\|$ that retains $\|X\|$, including cancellations among the local terms of $X$.} We use Theorem~2.1 of Ref.~\cite{K16information}
(also Lemma~4 of Ref.~\cite{KMS16floquet}).

\begin{lemma}[Local commutator action]
\label{lem:local-commutator-action}
Let $P$ be Hermitian, with locality $k_P$ and interaction strength per site at most $g_P$
for the decomposition $P=\sum_Sp_S$. If $X$ is any complex {$k_X$-local}
operator of finite norm{\footnote{The cited proof allows general $X$. When the local terms of $P$ commute, a term of $X$ on at most $k_X$ sites can connect energies differing by at most $2g_P k_X$. For energy bins wider than this, the only off-diagonal blocks connect neighboring bins. Writing $E_j$ for the bin projections, $\|\sum_j E_{j+1}XE_j\|\le\|X\|$ by orthogonality of the input and output blocks. The same estimate holds in the reverse direction. Neither estimate requires $X$ to be Hermitian, and the decomposition of $P$ into commuting sums in the general case is independent of $X$.}}, then
\begin{equation}
 \|[P,X]\|\le{6g_Pk_Pk_X}\|X\|.
 \label{eq:intrinsic-local-action}
\end{equation}
No bound on the sum of the term norms of $X$ is required.
\end{lemma}

{ For odd $d$, $C_{\boldsymbol w}$ is Hermitian, and for even $d$, $iC_{\boldsymbol w}$ is Hermitian. We may therefore apply the lemma with $k_P=dk$ and $g_P=g_d$ in either case.} Every $X\in\mathfrak L_\nu$ is a sum
of terms on at most $\nu k$ sites. Hence
\begin{equation}
 \|[C_{\boldsymbol w},X]\|
 \le6g_d(dk)(\nu k)\|X\|
 \le\nu d!(6k^2g_{\mathrm{loc}})^d\|X\|,
 \label{eq:intrinsic-local-commutator}
\end{equation}
where the last step uses $d\le(3k)^{d-1}$.
Comparison with the defining inequality of $J$, Eq.~\eqref{eq:commutator-growth-scale}, proves
\begin{equation}
 J\le3k^2g_{\mathrm{loc}}.
 \label{eq:intrinsic-local-scale}
\end{equation}
{For fixed $k$ and $g_{\mathrm{loc}}$, this upper bound is independent of the number of sites.}

\subsection{Conjugation}
\label{app:comparison-tools}

\subsubsection{Sums of nested commutators}
\label{app:scale-sums}

{ In the error expansion in Appendix~\ref{app:finite-circuit-expansion}, each correction to the comparison generator is a sum of nested commutators [Eq.~\eqref{eq:comparison-coefficient-bound}]. For such a sum $Y=\sum_{\boldsymbol w}\gamma_{\boldsymbol w}C_{\boldsymbol w}$, we need estimates for both $[Y,X]$ and $Y$ itself. To bound $[Y,X]$, we apply Eq.~\eqref{eq:commutator-growth-scale} term by term. A word of degree $d=|\boldsymbol w|$ carries the weight $d!(2J)^d$, which leads us to define}
\begin{equation}
 {a(Y)=\sum_{\boldsymbol w}|\boldsymbol w|!\,(2J)^{|\boldsymbol w|}\,
                |\gamma_{\boldsymbol w}| .}
 \label{eq:real-step-action}
\end{equation}
{ The value of $a(Y)$ refers to the chosen expansion of $Y$. Summing the commutator bounds for that expansion gives, for $X\in\mathfrak L_\nu$,}
\begin{equation}
 {\|[Y,X]\|\le\nu\,a(Y)\,\|X\| .}
 \label{eq:scale-sum-commutator}
\end{equation}
For a Hermitian generator $Y$, {the operator $X(t)=e^{itY}Xe^{-itY}$} has initial derivative $i[Y,X]$. Thus $\nu\,a(Y)\|X\|$ bounds the initial rate of change of $X(t)$ in operator norm.

{ To bound $\|Y\|$ itself, suppose that every term has degree at least three, as in each correction to the comparison generator. If $J=0$, all these terms vanish. For $J>0$, Eq.~\eqref{eq:intrinsic-root-bound} and $(2J)^{d-2}(d-1)!\le d!(2J)^d/(12J^2)$ give}
\begin{equation}
 {\|Y\|\le\frac{\alpha}{12J^2}\,a(Y) .}
 \label{eq:scale-sum-norm}
\end{equation}
{Each term of $a(Y)$ in this estimate contains at least three powers of $J$, so at least one remains after division by $J^2$.}

\subsubsection{A commutator bound for conjugation by the remaining evolution}

{To use Lemma~\ref{lem:real-step-product}, we need an upper bound $M$ for $\|\Psi\|$ on the disk. In Appendices~\ref{app:finite-circuit-expansion} and~\ref{app:intrinsic-cost}, the derivative estimates involve terms of the form $\Pi X\Pi^{-1}$, where a change $X$ at one point in the evolution is conjugated by the remaining evolution $\Pi$. To see why, write a product as $W=\Pi T V$, where $T$ is the factor being varied and $V$ acts before it. In $(\partial_zW)W^{-1}$, the contribution from differentiating $T$ is}
\begin{equation}
{\bigl[\Pi(\partial_zT)V\bigr]W^{-1}=\Pi X\Pi^{-1},\qquad X=(\partial_zT)T^{-1}.}
 \label{eq:route-derivative}
\end{equation}
{For a continuous evolution the same product rule gives an integral. A change in the generator at time $x$ contributes its derivative conjugated by the evolution $\Pi$ from $x$ to the endpoint. Appendix~\ref{app:finite-circuit-expansion} uses this identity for the generator derivative within one Strang interval [Eq.~\eqref{eq:comparison-stage-derivative}]. In Appendix~\ref{app:intrinsic-cost} the remaining evolution extends to the end of the polygon product.}

For ordinary unitary time evolution, $\|\Pi X\Pi^{-1}\|=\|X\|$. The bound $M$ must also hold at complex step sizes, where $\Pi$ need not be unitary. How large can the conjugated operator become? Consider first an interval of length $t$ with constant generator $\Lambda$, so $\Pi=e^{t\Lambda}$. Differentiating $Y(t)=e^{t\Lambda}Xe^{-t\Lambda}$ gives $Y'(t)=[\Lambda,Y(t)]$, hence
\[
 e^{t\Lambda}Xe^{-t\Lambda}
 =X+t[\Lambda,X]+\frac{t^2}{2!}[\Lambda,[\Lambda,X]]+\cdots .
\]
{Now fix the step parameter $z$ and let $x$ denote time along the evolution. For a time-dependent generator $\Lambda(x)$ on $[x_0,x_*]$, the conjugated operator satisfies $Y'(x)=[\Lambda(x),Y(x)]$ with $Y(x_0)=X$. Write $\Pi$ for the evolution from $x_0$ to $x_*$. Iterating this equation in integral form gives $Y(x_*)$ as}
\begin{equation}
 {\Pi X\Pi^{-1}
 =\sum_{n\ge0}\ \int_{x_0\le x_1\le\cdots\le x_n\le x_*}
   [\Lambda(x_n),[\cdots,[\Lambda(x_1),X]\cdots]]\,dx_1\cdots dx_n .}
 \label{eq:real-step-conjugation}
\end{equation}
{The $n=0$ term is $X$. For one commutation, Eq.~\eqref{eq:scale-sum-commutator} gives
\[
 \|[\Lambda,X]\|\le \nu\,a(\Lambda)\|X\|,\qquad X\in\mathfrak L_\nu .
\]
By Eq.~\eqref{eq:commutator-degree-addition}, a term $C_{\boldsymbol w}$ of degree $d$ in $\Lambda$ gives $[C_{\boldsymbol w},X]\in\mathfrak L_{\nu+d}$. { For example, in $[C_2,[C_1,X]]$ with $C_1$ of degree $d_1$, the factor $\nu$ in the first estimate becomes $\nu+d_1$ in the second.} To bound the successive degree factors in Eq.~\eqref{eq:real-step-conjugation}, we give each term of degree $d$ an extra weight $e^d$. For a chosen expansion $Y=\sum_{\boldsymbol w}\gamma_{\boldsymbol w}C_{\boldsymbol w}$, define more generally, for $c\ge0$,}
\begin{equation}
 {a_c(Y)=\sum_{\boldsymbol w}e^{cd}\,d!\,(2J)^d\,|\gamma_{\boldsymbol w}| ,\qquad d=|\boldsymbol w|,\quad a_0=a .}
 \label{eq:scale-sum-weighted}
\end{equation}
{{ For example, in $[H_c,[H_a,H_b]]$ the inner bracket has norm at most $\alpha$, $J$ bounds its growth under the outer bracket, and $a_1$ counts the coefficient of this word with the weight for degree three.} The next lemma bounds the term with $n$ commutators in Eq.~\eqref{eq:real-step-conjugation} by a multiple of $L^n$, where
\[
 L=\int_{x_0}^{x_*}a_1(\Lambda(x))\,dx .
\]
For $L<1$ these norm bounds sum to a geometric series.}

\begin{lemma}[Conjugation by an evolution]
\label{lem:comparison-conjugation}
Let $\Pi$ be the evolution over an interval under a generator
$\Lambda(x)=\sum_{\boldsymbol w}\gamma_{\boldsymbol w}(x)C_{\boldsymbol w}$ with
{complex coefficients. If $L<1$, then for every $X\in\mathfrak L_\nu$,}
\begin{equation}
 \|\Pi X\Pi^{-1}\|\le\frac{e^{\nu}}{1-L}\,\|X\| .
 \label{eq:comparison-conjugation}
\end{equation}
In particular, if $X=\sum_{\boldsymbol w}x_{\boldsymbol w}C_{\boldsymbol w}$
contains only nested commutators of degree at least three, then {$X=0$ for $J=0$, and for $J>0$}
\begin{equation}
 \|\Pi X\Pi^{-1}\|\le\frac{\alpha}{12J^2(1-L)}\,a_1(X) .
 \label{eq:comparison-conjugation-size}
\end{equation}
\end{lemma}
\begin{proof}
{Consider first $[C_2,[C_1,X]]$, where $C_i$ is a nested commutator of degree $d_i$ from $\Lambda(x_i)$.
The inner commutator takes $X\in\mathfrak L_\nu$ as its input. The outer one takes $[C_1,X]\in\mathfrak L_{\nu+d_1}$ by Eq.~\eqref{eq:commutator-degree-addition}.
Applying the norm estimate~\eqref{eq:commutator-growth-scale} to these two inputs in turn gives
\[
 \|[C_2,[C_1,X]]\|\le
 \nu(\nu+d_1)\,d_1!d_2!(2J)^{d_1+d_2}\|X\|.
\]

For $n$ commutators the successive degree bounds are $\nu,\nu+d_1,\ldots,\nu+d_1+\cdots+d_{n-1}$.
Writing $\nu'=\nu+\sum_i d_i$, their product is at most
$(\nu')^n\le n!e^{\nu'}=n!e^\nu\prod_i e^{d_i}$, by the exponential series.
At time $x_i$ we sum over the terms in $\Lambda(x_i)$. By definition their weights add to
\[
 \sum_{\boldsymbol w}|\gamma_{\boldsymbol w}(x_i)|
 e^{|\boldsymbol w|}|\boldsymbol w|!\,(2J)^{|\boldsymbol w|}
 =a_1(\Lambda(x_i)).
\]
{After taking norms and summing over the terms of $\Lambda(x_i)$ at each time, the scalar integrand $\prod_{i=1}^n a_1(\Lambda(x_i))$ is symmetric in the $x_i$. The ordered region thus contributes $1/n!$ of its integral over the cube $[x_0,x_*]^n$.} The $n$th term of Eq.~\eqref{eq:real-step-conjugation} thus has norm at most
\[
 n!e^\nu\|X\|
 \int_{x_0\le x_1\le\cdots\le x_n\le x_*}
 \prod_{i=1}^n a_1(\Lambda(x_i))\,dx_1\cdots dx_n
 = e^\nu\|X\|L^n .
\]
Since $L<1$, summing over $n\ge0$ gives Eq.~\eqref{eq:comparison-conjugation}.

For an infinite interval or an absolutely summable expansion the same bounds justify passage to the limit. Expanding $X$, applying Eqs.~\eqref{eq:comparison-conjugation} and~\eqref{eq:scale-sum-norm} to each term, and summing gives Eq.~\eqref{eq:comparison-conjugation-size}.}
\end{proof}

{For $J\le j_0$, Eq.~\eqref{eq:comparison-conjugation-size} bounds the integral in Eq.~\eqref{eq:comparison-stage-derivative}, where the derivative of the correction to the generator is conjugated by the remaining stage evolution. Appendix~\ref{app:intrinsic-cost} uses the same estimate for the derivative of the polygon product. In both cases the inserted operator has only terms of degree at least three, so its contribution retains the factor $\alpha$ after conjugation. The resulting derivative bounds on the disk supply $M$ in Lemma~\ref{lem:real-step-product} and hence bound the single-stage error $\|\Omega-S\|$ in Appendix~\ref{app:finite-circuit-expansion} and the polygon error $\|V_q^\Omega(\delta)-U\|$ in Appendix~\ref{app:intrinsic-cost}.}

\subsubsection{Conjugation by the exact evolution}

Appendix~\ref{app:intrinsic-cost} removes the exact evolution from each factor of the polygon product. The following lemma bounds the resulting $a_1$ weight using the $a_4$ estimate from Appendix~\ref{app:finite-circuit-expansion}.

\begin{lemma}[Conjugation by the exact evolution]
\label{lem:intrinsic-ideal}
For $2emJ\le1/2$, a real number $b$ with $|b|\le1$, and a sum
$Y$ of nested commutators with $a_4(Y)<\infty$, the operator $e^{-ibH}Ye^{ibH}$ has a nested-commutator expansion satisfying
\begin{equation}
 a_1\bigl(e^{-ibH}Ye^{ibH}\bigr)\le 2\,a_4(Y).
 \label{eq:intrinsic-ideal-dressing}
\end{equation}
\end{lemma}
\begin{proof}[Proof of Lemma~\ref{lem:intrinsic-ideal}]
If $J=0$, the claim follows because all generators commute. Otherwise, the conjugation is the series
$e^{-ibH}C_{\boldsymbol w}e^{ibH}
=\sum_{j\ge0}\frac{(-ib)^j}{j!}[H,[\cdots,[H,C_{\boldsymbol w}]\cdots]]$
with $j$ commutators. Since $H=\sum_a H_a$, the term with $j$ commutators is a
sum of $m^j$ nested commutators of degree $d+j$, each with a coefficient
of absolute value $|b|^j/j!$. In
$a_1$ a nested commutator of degree $d+j$
counts $e^{j}(2J)^j(d+j)!/d!$ times more than one of degree $d$. The sum
is therefore multiplied by at most
$\sum_{j\ge0}\binom{d+j}{j}(2emJ|b|)^j=(1-2emJ|b|)^{-d-1}$. Since
$2emJ|b|\le1/2$, this is at most $2^{d+1}\le2e^{3d}$, which
the factor $e^{4d}$ in $a_4$ supplies, since $a_1$ already contains $e^d$. Summing this estimate over the terms of $Y$ gives Eq.~\eqref{eq:intrinsic-ideal-dressing} and completes the proof of Lemma~\ref{lem:intrinsic-ideal}.
\end{proof}

%% file: appendices/sweep-comparison.tex
\section{Comparison of a finite Strang sweep with a continuous evolution}
\label{app:finite-circuit-expansion}

We construct the comparison evolution used in the proof of Theorem~\ref{thm:order-cost} and bound its distance from the finite circuit at one polygon vertex. Throughout this appendix, $C$ denotes a constant that may depend on the fixed number $m$ of generators and may vary between estimates.

\subsection{Finite sweep and cutoff}
\label{app:cutoff-selection}

{Consider one vertex and write $\delta,r$ for its step and cutoff.} The finite circuit $Q_r(\delta)$ of Eq.~\eqref{eq:finite-normalized-circuit} consists of the first $r$ stages of the {infinite Strang product} $Q_\infty(\delta)$ and one final stage with the physical time of the tail.

\begin{lemma}[Cutoff]
 \label{lem:cutoff-selection}
 For bounded Hermitian $H_1,\ldots,H_m$ and any $\delta>0$, the limit
 \[
 Q_\infty(\delta)=\lim_{r\to\infty}Q_r(\delta)
 \]
 exists in operator norm, is unitary, and is the ordered product of all the stages $S_2(\Delta_k)$, $k\ge0$. Write $\tau_r=1-f(r\delta)$ for the physical time of the tail. Then
 \begin{equation}
 \|Q_r(\delta)-Q_\infty(\delta)\|\le C\alpha J\tau_r^3 .
 \label{eq:cutoff-bound}
 \end{equation}
{For $0<\delta\le1$, let $0<\eta\le e^{-2}$ be the allowed physical time remaining in the tail. Then $\tau_r\le\eta$, and hence the cutoff error is at most $C\alpha J\eta^3$, for}
 \begin{equation}
 r=\left\lceil\frac1\delta\log\!\left(1+\log\frac1\eta\right)\right\rceil=O\!\left(\frac1\delta\log\log\frac1\eta\right).
 \label{eq:independent-cutoff}
 \end{equation}
\end{lemma}

\begin{proof}
{Fix the auxiliary time step $\delta>0$. The finite circuit retains the first $r$ Strang stages, whose product is $P_r=S_2(\Delta_{r-1})\cdots S_2(\Delta_0)$, and covers the remaining physical time $\tau_r=1-f(r\delta)$ with one final stage. { Thus the cutoff determines how far the time interval is subdivided. The final stage covers all the time still left, which can be most of the interval when $r$ is fixed and $\delta$ is small. To prove Lemma}~\ref{lem:cutoff-selection}, we estimate how much the circuit changes when this final interval is subdivided. Increasing the cutoff by one retains an additional stage of width $\Delta_r=f((r+1)\delta)-f(r\delta)$. By Eq.~\eqref{eq:finite-normalized-circuit},}
\[
 \begin{aligned}
 Q_r(\delta)&=S_2(\tau_r)P_r,\\
 Q_{r+1}(\delta)&=S_2(\tau_{r+1})S_2(\Delta_r)P_r .
 \end{aligned}
\]
{The common unitary $P_r$ drops out of $\|Q_{r+1}(\delta)-Q_r(\delta)\|$. Since $\Delta_r+\tau_{r+1}=\tau_r$, the two products acting after $P_r$ approximate the same exact evolution $e^{-i\tau_rH}$.}

{For convergence we need a bound that holds however far we extend the cutoff. For any $r'>r$, the stages after $P_r$ in $Q_{r'}(\delta)$ have widths $\Delta_r,\ldots,\Delta_{r'-1},\tau_{r'}$. To compare their product with $e^{-i\tau_rH}$, we use the single-stage error bound~\cite{CSTWZ21}. For $m=2$, $H=A+B$, it is}
\[
 \|S_2(\tau)-e^{-i\tau H}\|
 \le\frac{|\tau|^3}{12}\|[B,[B,A]]\|+\frac{|\tau|^3}{24}\|[A,[A,B]]\|.
\]
For $m$ generators, recursively split off $H_a$ from $H_a+B_a$, where $B_a=\sum_{b>a}H_b$. Replacing the exponential of $B_a$ by the next Strang stage preserves the norm of its error since its two neighbors are unitary. Summing the bounds for the successive splittings gives
\begin{equation}
 \|S_2(\tau)-e^{-i\tau H}\|
 \le |\tau|^3\sum_{a<m}\left(\tfrac1{24}\|[H_a,[H_a,B_a]]\|
                 +\tfrac1{12}\|[B_a,[B_a,H_a]]\|\right)
 \le C\alpha J|\tau|^3 .
 \label{eq:cutoff-strang-error}
\end{equation}
The inner commutator $[H_b,H_a]$ has degree two, so the single-generator case of Eq.~\eqref{eq:commutator-growth-scale} gives
\[
 \|[H_c,[H_b,H_a]]\|\le 4J\|[H_b,H_a]\|.
\]
The sum of the first double-commutator norms is at most $4J\alpha$, and the sum of the second is at most $4(m-1)J\alpha$.

{Replace these stages one at a time by exact evolutions of the same widths. The surrounding unitaries preserve the norm of each replacement error, so the triangle inequality bounds the total error by the sum of the single-stage bounds. The widths are nonnegative and sum to $\tau_r$, hence}
\[
 \sum_{k=r}^{r'-1}\Delta_k^3+\tau_{r'}^3
 \le\left(\sum_{k=r}^{r'-1}\Delta_k+\tau_{r'}\right)^3=\tau_r^3 .
\]
{Thus the product after $P_r$ in $Q_{r'}(\delta)$ and the single stage $S_2(\tau_r)$ in $Q_r(\delta)$ each satisfy the bound in Eq.~\eqref{eq:cutoff-strang-error} with $\tau=\tau_r$. Adding their errors relative to $e^{-i\tau_rH}$ gives}
\[
 \|Q_{r'}(\delta)-Q_r(\delta)\|
 \le C\alpha J\tau_r^3 .
\]

{The right-hand side is independent of $r'$. Since $\tau_r\to0$ as $r\to\infty$, the circuits $Q_r(\delta)$ form a Cauchy sequence in operator norm. Their limit $Q_\infty(\delta)$ is unitary by taking limits in $Q_r^\dagger Q_r=Q_rQ_r^\dagger=I$. The partial products of the infinite Strang product are the $P_r$, which omit the final stage. Since $S_2(\tau_r)\to I$, the identity $P_r=S_2(\tau_r)^{-1}Q_r(\delta)$ shows that these partial products also converge to $Q_\infty(\delta)$.}

Taking $r'\to\infty$ in the bound on $\|Q_{r'}(\delta)-Q_r(\delta)\|$ gives Eq.~\eqref{eq:cutoff-bound}. For two generators the preceding two-term bound gives, in particular, a cutoff error of at most $\tfrac43\alpha J\tau_r^3$.

It remains to choose how many stages to retain. {Let $\eta$ be the allowed upper bound on the remaining physical time. For the time map of Eq.~\eqref{eq:chosen-time-map}, $\tau_r=\exp(1-e^{r\delta})$, so the smallest integer cutoff with $\tau_r\le\eta$ is
\[
 r=\left\lceil\frac{\log(1+\log(1/\eta))}{\delta}\right\rceil .
\]
With these $r$ stages and one final stage for the remaining time, the cutoff error is at most $C\alpha J\eta^3$.}
\end{proof}

\subsection{Modified generator for one cell}
\label{app:comparison-generator}
\label{app:route-comparison}

\begin{figure}[htbp]
 \centering
 \includegraphics[width=0.89\linewidth]{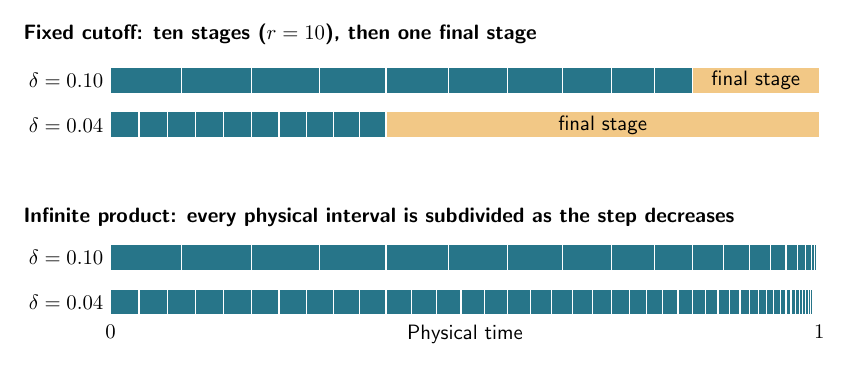}
 \caption{Physical time intervals for the finite circuit $Q_{10}(\delta)$ (top) and the infinite Strang product $Q_\infty(\delta)$ (bottom). Each white divider marks a stage boundary at $f(k\delta)$. In the finite circuit, decreasing $\delta$ from $0.10$ to $0.04$ increases the final stage (orange) from about $18\%$ to $61\%$ of the total time. In the infinite product, the largest stage shrinks as $\delta$ decreases.}
 \label{fig:cutoff-limits}
\end{figure}

The limit defining $Q_\infty$ is taken at fixed $\delta>0$. To see why its dependence on $\delta$ requires care, keep $r$ fixed instead and let $\delta\downarrow0$. The first $r$ physical steps $\Delta_k=f((k+1)\delta)-f(k\delta)$ tend to zero, while the final step $\tau_r=1-f(r\delta)$ tends to one. Thus $Q_r(\delta)$ tends to $S_2(1)$. {Figure~\ref{fig:cutoff-limits} compares the two partitions of physical time for the same values of $\delta$.}

{Taking $r\to\infty$ first gives a different result in general. The physical steps sum to one, and their maximum width tends to zero since $\Delta_k\le\delta\sup_{s\ge0} f'(s)$. Summing the cubic Strang errors of Eq.~\eqref{eq:cutoff-strang-error} gives a vanishing error because
\[
 \sum_k\Delta_k^3\le(\sup_k\Delta_k)^2\sum_k\Delta_k=(\sup_k\Delta_k)^2\longrightarrow0 .
\]
Thus}
\[
 \lim_{r\to\infty}\lim_{\delta\downarrow0} Q_r(\delta)=S_2(1),
 \qquad
 \lim_{\delta\downarrow0}\lim_{r\to\infty} Q_r(\delta)=U .
\]
{Analyticity of the finite circuits therefore gives no direct expansion of $Q_\infty$ at zero. To apply polygon cancellation with a convergent even expansion, we introduce $\Omega_q(\delta)$ in Eq.~\eqref{eq:route-evolution}, an even analytic approximation to $Q_\infty(\delta)$ with error $O(\delta^{2q})$ at fixed $q$. This replacement preserves the order of the error after cancellation.}

{We define $\Omega_q$ through its generator, chosen to reproduce one Strang stage on each auxiliary interval. The exact evolution has elapsed physical time $f(s)$ at auxiliary time $s$, so
\[
 \partial_s e^{-if(s)H}=-ig(s)H\,e^{-if(s)H},\qquad g(s)=f'(s).
\]
Over $[k\delta,(k+1)\delta]$ this generator gives $e^{-i\Delta_kH}$. To match $S_2(\Delta_k)$ through order $\delta^{2q}$, we add corrections to the generator in even powers of $\delta$, obtaining the \emph{comparison generator}}
\begin{equation}
 \mathcal L_q(s,\delta)=-ig(s)H+\delta^2\mathcal K^{[3]}(s)
   +\delta^4\mathcal K^{[5]}(s)+\cdots+\delta^{2q-2}\mathcal K^{[2q-1]}(s),
 \label{eq:route-generator}
\end{equation}
where each correction $\mathcal K^{[\ell]}(s)$ is a sum of nested commutators of the $H_a$ of degree at least three, multiplied by products of $g$ and its derivatives. {Evolving with $\mathcal L_q$ from $s=0$ to $s=\infty$ then gives the} \emph{comparison evolution},
\begin{equation}
 \Omega_q(\delta)=\mathcal T\exp\!\left(\int_0^\infty\mathcal L_q(s,\delta)\,ds\right),
 \label{eq:route-evolution}
\end{equation}
where $\mathcal T$ places later $s$ to the left.

{We chose the corrections by matching a single Strang stage. To replace a whole infinite Strang product at a polygon vertex, we need to control the accumulated error over all stages. The following lemma gives this estimate and shows that the comparison evolution has the even expansion required for cancellation and is unitary for real steps.}

\begin{lemma}[Comparison evolution]
 \label{lem:finite-circuit-expansion}
 {For every integer $q\ge2$,} the comparison evolution $\Omega_q(\delta)$ is defined for every complex $\delta$ and has the following properties.
\begin{enumerate}[nosep]
\item $\Omega_q(\delta)U^\dagger=F(\delta^2)$ has the form of Eq.~\eqref{eq:shared-error-family}, with a power series $F(z)=I+zE_2+z^2E_4+\cdots$ that converges for every $z$.
\item $\Omega_q(\delta)$ is unitary for real $\delta$.
\item {For $0<\delta\le\rho_q=\rho_0/q$, if $J\le j_0$ or $\lambda\le \lambda_0 q^{2/3}$, with the constants for fixed $m$ in \cref{thm:order-cost},}
 \begin{equation}
 \|Q_\infty(\delta)-\Omega_q(\delta)\|
 \le\frac{C\alpha J\rho_q^2}{2q+1}\,(\delta/\rho_q)^{2q}.
 \label{eq:alpha-infinite-comparison}
 \end{equation}
\end{enumerate}
\end{lemma}
{In particular, truncating the power series of property~1 after $z^{q-1}$ and using property~3 gives
$Q_\infty(\delta)U^\dagger=I+\delta^2E_2+\cdots+\delta^{2q-2}E_{2q-2}+O(\delta^{2q})$ with coefficients independent of $\delta$, which is the expansion condition~\eqref{eq:approximate-expansion} stated in \cref{sec:continuous-family-construction}. For a given Hamiltonian, $\lambda\le\lambda_0q^{2/3}$ holds for all sufficiently large $q$. Choosing such an order first and truncating further gives this expansion for every fixed order. The coefficients shared by two such truncations agree by uniqueness of the asymptotic coefficients.}
{The factor $1/(2q+1)$ in property~3 allows us to sum the replacement errors over all $2q-1$ vertices. \Cref{app:quantitative-coefficients} estimates the higher commutators using $J$ in the first case and the operator norms of the generators in the second. To match one Strang stage through order $\delta^{2q}$, we added terms $\delta^{2j}\mathcal K^{[2j+1]}(s)$ to the exact generator $-ig(s)H$ in Eq.~\eqref{eq:route-generator}. Taking $\delta\le \rho_0/q$ controls their sum uniformly in $q$, with the factor $\alpha J\rho_q^2$ shown above.

\Cref{app:comparison-construction} constructs the corrections and \cref{app:comparison-coefficients} bounds their coefficients. \Cref{app:comparison-global} uses those bounds to define the evolution on $[0,\infty)$ and to sum the errors of the individual stages.}

\subsubsection{Derivation of the first generator correction}
\label{app:first-stage-correction}

On the auxiliary interval $[s,s+\delta]$ the physical time step is
\begin{equation}
 \Delta(s,\delta)=f(s+\delta)-f(s).
 \label{eq:interval-physical-time}
\end{equation}
To work out the first correction, take $m=2$ and write $H=A+B$. The first error in a single stage is cubic in its physical time step,
\begin{equation}
 S_2(\tau)e^{i\tau H}=I+\tau^3 C_3+O(\tau^4),\qquad
 C_3=-\tfrac{i}{24}[A,[A,B]]-\tfrac{i}{12}[B,[A,B]] .
 \label{eq:first-strang-local}
\end{equation}
Which correction to the generator reproduces this cubic error? Write $L_0(u)=-ig(u)H$ and add $\delta^2K(u)$, with $K$ smooth. In the time-ordered expansion of the resulting evolution, the term with one integral changes by
\[
 \int_s^{s+\delta}\!\bigl(L_0(u)+\delta^2K(u)\bigr)\,du
 -\int_s^{s+\delta}\!L_0(u)\,du
 =\delta^3K(s)+O(\delta^4).
\]
A term with two or more integrals that contains $K$ is at least fourth order: it carries $\delta^2$ from the correction and at least two more powers of $\delta$ from the integrations. The terms containing only $L_0$ sum to the exact step $e^{-i\Delta H}$, where $\Delta=\Delta(s,\delta)$. Thus the corrected evolution and the Strang stage have the expansions
\[
 \begin{aligned}
 \mathcal T\exp\!\left(\int_s^{s+\delta}[L_0(u)+\delta^2K(u)]\,du\right)
 &=e^{-i\Delta H}+\delta^3 K(s)+O(\delta^4),\\
 S_2(\Delta)&=e^{-i\Delta H}+\delta^3 g(s)^3 C_3+O(\delta^4).
 \end{aligned}
\]
The second line follows from Eq.~\eqref{eq:first-strang-local} since $\Delta=g(s)\delta+O(\delta^2)$ and $e^{-i\Delta H}=I+O(\delta)$. Matching the cubic coefficients gives the first correction,
\[
 \mathcal K^{[3]}(s)=g(s)^3C_3 .
\]
The same matching gives this identity for every $m$ if $C_3$ denotes the cubic coefficient of $S_2(\tau)e^{i\tau H}$.

\subsubsection{Matching the generator to one Strang stage}
\label{app:comparison-construction}

We now seek the higher corrections that reproduce a Strang stage through order $\delta^{2q}$ for every starting time $s$.

\begin{lemma}[Comparison generator]
\label{lem:comparison-cell-matching}
There are operators $\mathcal K^{[\ell]}(s)$, $\ell=3,5,\ldots$, that
depend on $s$ but not on $q$ or $\delta$, such that for every integer
$q\ge2$ the generator
\begin{equation}
 \begin{aligned}
 \mathcal L_q(s,\delta)&=-ig(s)H+\mathcal K_q(s,\delta),\\
 \mathcal K_q(s,\delta)&=\delta^2\mathcal K^{[3]}(s)
   +\delta^4\mathcal K^{[5]}(s)+\cdots
   +\delta^{2q-2}\mathcal K^{[2q-1]}(s),
 \end{aligned}
 \label{eq:comparison-finite-generator}
\end{equation}
has the following properties. The term $\delta^{\ell-1}\mathcal K^{[\ell]}$ in the generator first contributes at order $\delta^\ell$ over one interval.
\begin{enumerate}[nosep]
\item The evolution under $\mathcal L_q(\cdot,\delta)$ from $s$ to
 $s+\delta$,
 \begin{equation}
  \Omega_{q,s}(\delta)=\mathcal T\exp\!\left(\int_s^{s+\delta}
     \mathcal L_q(s',\delta)\,ds'\right),
  \label{eq:comparison-cell-evolution}
 \end{equation}
 where $\mathcal T$ places later $s'$ to the left, agrees with the Strang
 stage of the same interval through order $\delta^{2q}$. At fixed $s$, $q$ and
 $H_1,\ldots,H_m$, as $\delta\to0$,
 \begin{equation}
  S_2\bigl(\Delta(s,\delta)\bigr)-\Omega_{q,s}(\delta)=O(\delta^{2q+1}).
  \label{eq:comparison-cell-matching}
 \end{equation}
\item $\mathcal K^{[3]}(s)=g(s)^3C_3$, where $C_3$ is the cubic coefficient of $S_2(\tau)e^{i\tau H}$. Each $\mathcal K^{[\ell]}(s)$ is a
 sum of nested commutators of the $H_a$ of degree at least three, multiplied by products of $g$ and its derivatives.
 It is anti-Hermitian for real $s$.
\end{enumerate}
\end{lemma}

The first property matches one stage at a fixed starting time $s$.
The second explains why the resulting error can carry a factor of
$\alpha$. Each nonzero nested commutator in a correction contains an innermost pair $[H_a,H_b]$. For example, for $m=2$ Eq.~\eqref{eq:first-strang-local} and $\|C_3\|\le\alpha J/2$ give
\[
 \|\mathcal K^{[3]}(s)\|\le\tfrac12\alpha J\,g(s)^3 .
\]
\Cref{app:comparison-coefficients} estimates the higher coefficients.
Their dependence on $s$ will matter when we sum the errors over all stages.

\begin{proof}
{In \cref{app:first-stage-correction} we found the first correction by matching the cubic term of one stage. { To obtain the higher corrections together, we take the logarithm of a step and read off the successive coefficients of its generator. Consider first a step of fixed physical width $\Delta$.} Near $\Delta=0$ the effective generator is $\Delta^{-1}\log S_2(\Delta)$, since evolving with it for time $\Delta$ gives precisely $S_2(\Delta)$.

Here the physical width $f(s+\delta)-f(s)$ depends on the starting time. Once a stage has acted on the state, the next stage must start at $s+\delta$. We define $\mathbb S_\delta$ to apply the Strang stage at auxiliary time $s$ and assign the resulting state to $s+\delta$. { We take the logarithm of this combined operation.}}

\emph{Step 1. The stage operator $\mathbb S_\delta$.} For a function $\Psi(s)$ with values in the state space, define
\begin{equation}
 (\mathbb S_\delta\Psi)(s+\delta)
 =S_2\bigl(\Delta(s,\delta)\bigr)\,\Psi(s).
 \label{eq:comparison-strang-shift}
\end{equation}
{ Applying the same operator twice gives
\[
 (\mathbb S_\delta^2\Psi)(s+2\delta)=
 S_2(\Delta(s+\delta,\delta))\,S_2(\Delta(s,\delta))\,\Psi(s).
\]
The shift selects the width of the second stage at its own starting time. Iterating from $s=0$ gives
\[
 (\mathbb S_\delta^{\,k}\Psi)(k\delta)
 =S_2(\Delta_{k-1})\cdots S_2(\Delta_0)\,\Psi(0).
\]
}As $k\to\infty$, the right-hand side tends to $Q_\infty(\delta)\Psi(0)$.
Thus a single operation generates the successive stages even though their
physical widths differ.

\emph{Step 2. Strang factorization of $\mathbb S_\delta$.} { To compute the logarithm of $\mathbb S_\delta$, we express it as a symmetric product of exponentials of $m$ operators independent of $\delta$. Define}
\begin{equation}
 \mathsf X_a=b_a\partial_s-ig(s)H_a,\qquad b_a=2(-1)^a\ (a<m),
 \qquad b_m=(-1)^m,\qquad
 \sum_a\mathsf X_a=-\partial_s-ig(s)H .
 \label{eq:comparison-clock-fields}
\end{equation}
Form the symmetric product of Eq.~\eqref{eq:strang-step} with $\mathsf X_a$ in place of $-iH_a$ and step $\delta$. A factor $e^{h\mathsf X_a}$ carries a state from $u$ to $u-b_a h$, multiplying it by
$\exp[-iH_a\int_0^h g(u-b_a v)\,dv]$.
Each half step has $h=\delta/2$ and the middle factor has $h=\delta$. In the order of application the displacements alternate between $+\delta$ and $-\delta$, beginning and ending with $+\delta$. Thus consecutive factors traverse $[s,s+\delta]$ in opposite directions and the final state is at $s+\delta$. In either direction the integral is $(h/\delta)[f(s+\delta)-f(s)]$, so each exponential has the required physical duration. This proves Eq.~\eqref{eq:comparison-strang-shift} for the product of the $\mathsf X_a$.

For example, when $m=2$ we have
\[
 \mathsf A=\mathsf X_1=-2\partial_s-ig(s)A,\qquad
 \mathsf B=\mathsf X_2=\partial_s-ig(s)B,\qquad
 \mathsf A+\mathsf B=-\partial_s-ig(s)H .
\]
{ Write $\Delta=f(s+\delta)-f(s)$. The actions of the two exponentials needed for one stage are
\[
 \begin{aligned}
 (e^{(\delta/2)\mathsf A}\Psi)(s+\delta)&=e^{-i\Delta A/2}\Psi(s),\\
 (e^{\delta\mathsf B}\Psi)(s)&=e^{-i\Delta B}\Psi(s+\delta).
 \end{aligned}
\]
Indeed, during the first evolution the integrated coefficient of $A$ is $\int_0^{\delta/2}g(s+2h)\,dh=\Delta/2$. During the second the coefficient of $B$ integrates to $\int_0^\delta g(s+\delta-h)\,dh=\Delta$. Applying the first identity, then the second, then the first again gives the required action $S_2(\Delta)\Psi(s)$ at $s+\delta$. Thus}
\begin{equation}
 \mathbb S_\delta
 =e^{(\delta/2)\mathsf A}\,e^{\delta\mathsf B}\,e^{(\delta/2)\mathsf A}.
 \label{eq:comparison-strang-formula}
\end{equation}
{ For this example the two constructions can now be compared directly.}
\begin{center}
\renewcommand{\arraystretch}{1.35}
\begin{tabular}{@{}l@{\qquad}l@{\qquad}l@{}}
 & Equal steps (Sec.~\ref{sec:error-cancellation}) & Auxiliary time (this proof)\\
\hline
Two operators & $-iA$, $-iB$ & $\mathsf A$, $\mathsf B$\\
Their sum & $-iH$ & $\mathsf A+\mathsf B$\\
Step & $\Delta=1/r$ & $\delta$\\
One stage & $S_2(\Delta)=e^{-i\Delta A/2}e^{-i\Delta B}e^{-i\Delta A/2}$
 & $\mathbb S_\delta=e^{(\delta/2)\mathsf A}e^{\delta\mathsf B}e^{(\delta/2)\mathsf A}$\\
Whole product & $S_2(1/r)^r$ & repetition of $\mathbb S_\delta$\\
\end{tabular}
\end{center}
\emph{Step 3. The logarithm of $\mathbb S_\delta$.} { Write $V(\delta)$ for the symmetric product of Eq.~\eqref{eq:strang-step} in the noncommuting symbols $x_1,\ldots,x_m$ in place of $-iH_a$. Its formal logarithm has coefficients $p_\ell(x_1,\ldots,x_m)$, with $p_1=\sum_a x_a$. For $m=2$ and $(x_1,x_2)=(x,y)$, for example,}
\begin{equation}
 \log\bigl(e^{\delta x/2}e^{\delta y}e^{\delta x/2}\bigr)
 =\sum_{\ell\ge1}\delta^\ell p_\ell(x,y),\qquad
 p_1=x+y,\qquad
 p_3=-\tfrac1{24}[x,[x,y]]-\tfrac1{12}[y,[x,y]] .
 \label{eq:comparison-free-logarithm}
\end{equation}
{ Since $V(-\delta)=V(\delta)^{-1}$, its formal logarithm is odd, so $p_\ell=0$ for even $\ell$. The Baker--Campbell--Hausdorff formula expresses the remaining $p_\ell$ as real linear combinations of nested commutators for $\ell\ge3$. Substitution of the $\mathsf X_a$ from Step~2 gives, writing $\boldsymbol{\mathsf X}=(\mathsf X_1,\ldots,\mathsf X_m)$,}
\[
 \log\mathbb S_\delta
 =\delta\sum_a\mathsf X_a+\delta^3p_3(\boldsymbol{\mathsf X})
  +\delta^5p_5(\boldsymbol{\mathsf X})+\cdots .
\]
Commutators of the $\mathsf X_a$ contain no $\partial_s$. For
constants $c,c'$ and operators $M_1(s),M_2(s)$ that contain no
$\partial_s$, with a prime for $d/ds$,
\begin{equation}
 [c\,\partial_s+M_1,\ c'\partial_s+M_2]=cM_2'-c'M_1'+[M_1,M_2],
 \label{eq:comparison-clock-bracket}
\end{equation}
and for example
$[\mathsf A,\mathsf B]=ig'(s)(A+2B)-g(s)^2[A,B]$. Hence
$p_\ell(\boldsymbol{\mathsf X})$, $\ell\ge3$, is an operator that depends on
$s$ and contains no $\partial_s$. We denote it by $\mathcal K^{[\ell]}(s)$.
Then
\[
 \log\mathbb S_\delta=\delta\bigl(-\partial_s+\mathcal L(s,\delta)\bigr),
 \qquad
 \mathcal L(s,\delta)=-ig(s)H+\delta^2\mathcal K^{[3]}(s)
   +\delta^4\mathcal K^{[5]}(s)+\cdots .
\]
By
Eq.~\eqref{eq:comparison-clock-bracket}, each commutator either
differentiates a scalar coefficient or adds a commutator with a generator $H_a$.
Each $\mathcal K^{[\ell]}(s)$ is therefore a sum of terms, each a product
of $g$ and its derivatives times a generator or a nested commutator of generators.
We now use these coefficients to match a stage at finite order, then
check which degrees in the $H_a$ survive in each correction.

\emph{Step 4. Matching the stage through order $\delta^{2q}$.}
{ We now check that evolving with the truncated generator $\mathcal L_q$ reproduces $S_2(\Delta(s,\delta))$ to the required order. }For an operator $M(s)$ that contains no $\partial_s$, the operation with logarithm
$\delta(-\partial_s+M)$ carries the value at $s$ to $s+\delta$ and applies
the evolution under $M$ over the interval $[s,s+\delta]$,
\begin{equation}
 (e^{\delta(-\partial_s+M)}\Psi)(s+\delta)
 =\left[\mathcal T\exp\!\int_{s}^{s+\delta}M(s')\,ds'\right]\Psi(s).
 \label{eq:comparison-shift-evolution}
\end{equation}
The identity follows by varying the interval width with $M$ held fixed.
For $M=-ig(s)H$ it gives the exact step $e^{-i\Delta(s,\delta)H}$.
To match a stage through order $\delta^{2q}$ we need only finitely many
coefficients of the formal logarithm. Keeping the terms of $\mathcal L$
through $\delta^{2q-2}$ gives $\mathcal L_q$. At each fixed $\delta$ we
can now take $M(s')=\mathcal L_q(s',\delta)$. By
Eq.~\eqref{eq:comparison-shift-evolution}, the operation with logarithm
$\delta(-\partial_s+\mathcal L_q)$ carries the value at $s$ to $s+\delta$
and applies $\Omega_{q,s}(\delta)$. This logarithm agrees with
$\log\mathbb S_\delta$ through order $\delta^{2q}$, because the difference
starts at $\delta^{2q+1}p_{2q+1}$. { Exponentiating preserves this agreement, so the two operations apply the same stage through order $\delta^{2q}$.}
The map $f$ in Eq.~\eqref{eq:chosen-time-map} is entire, so the stage depends analytically on its endpoints. Here we used that, applied to an analytic $\Psi$, both operations are
analytic in $\delta$ and their Taylor series are the formal exponentials
of their logarithms. Taking $\Psi$ equal to a constant vector gives
Eq.~\eqref{eq:comparison-cell-matching}.

\emph{Step 5. Commutator degree of the corrections.} {Step~4 gives the order of agreement for one stage. To bound the error in terms of $\alpha$ and $J$, we now show that each correction contains only nested commutators of degree at least three.} { For the two-generator example at $\ell=3$, the individual commutators contain terms proportional to $A$, $B$ and $[A,B]$. We can see explicitly how these terms cancel in $\mathcal K^{[3]}$.} By Eq.~\eqref{eq:comparison-clock-bracket},
\[
 \begin{aligned}
 [\mathsf A,\mathsf B]&=ig'(A+2B)-g^2[A,B],\\
 [\mathsf A,[\mathsf A,\mathsf B]]
   &=-2ig''(A+2B)+6gg'[A,B]+ig^3[A,[A,B]],\\
 [\mathsf B,[\mathsf A,\mathsf B]]
   &=ig''(A+2B)-3gg'[A,B]+ig^3[B,[A,B]].
 \end{aligned}
\]
Inserting these in $p_3$ of Eq.~\eqref{eq:comparison-free-logarithm}, the
terms with $g''$ cancel since $\tfrac2{24}-\tfrac1{12}=0$, the terms with
$gg'$ cancel since $-\tfrac6{24}+\tfrac3{12}=0$, and
\[
 \mathcal K^{[3]}(s)
 =g(s)^3\Bigl(-\tfrac i{24}[A,[A,B]]-\tfrac i{12}[B,[A,B]]\Bigr)
 =g(s)^3C_3
\]
remains.

{To see why this cancellation holds for every odd $\ell\ge3$, replace each $H_a$ by $\xi H_a$. The power of $\xi$ counts the physical operators in a term, while the power of $\delta$ counts its order in the auxiliary step. Both the Strang step and the exact step advance auxiliary time from $s$ to $s+\delta$. Write $\Delta=f(s+\delta)-f(s)$. Scaling every generator by $\xi$ multiplies the physical time in each exponential by $\xi$, so the two steps act on the state by $S_2(\xi\Delta)$ and $e^{-i\xi\Delta H}$, respectively. Since the Strang formula is second order, their terms through degree two in $\xi$ agree:
\begin{equation}
 S_2(\xi\Delta)
 = I-i\xi\Delta H-\tfrac12\xi^2\Delta^2 H^2+O((\xi\Delta)^3)
 = e^{-i\xi\Delta H}+O((\xi\Delta)^3).
 \label{eq:comparison-coupling-quotient}
\end{equation}
Both steps advance auxiliary time by $\delta$, and their actions on the state agree through second order in $\xi$. In the formal logarithm in $\delta$, each coefficient is a polynomial in the step coefficients. Terms containing $\xi^3$ or higher powers therefore cannot change the logarithm through order $\xi^2$. Each correction $\mathcal K^{[\ell]}$ comes from the difference of these logarithms, so terms containing only one or two physical generators cancel. Thus the nested commutators in Step~3 all have degree at least three once these terms are collected.}

Finally, Eq.~\eqref{eq:comparison-clock-bracket} maps anti-Hermitian
$M_1,M_2$ and real $c,c'$ to an anti-Hermitian operator, so every
$\mathcal K^{[\ell]}(s)$ is anti-Hermitian for real $s$.
\end{proof}

\subsection{Decay of the correction coefficients in auxiliary time and growth with order}
\label{app:comparison-coefficients}

{Lemma~\ref{lem:comparison-cell-matching} gives an error of $O(\delta^{2q+1})$ on a single interval starting at $s$. The constant in this estimate can depend on $s$. To add the errors over the intervals starting at $0,\delta,2\delta,\ldots$ in \cref{app:comparison-global}, we need to know how the contribution from each interval decays as its starting time increases.

For the first correction, $\mathcal K^{[3]}(s)=g(s)^3C_3$, the commutator combination $C_3$ is the same on every interval. All the dependence on the starting time is in the coefficient $g(s)^3$. The higher corrections are also sums of nested commutators of the $H_a$, but their coefficients contain derivatives of $g$ as well. We bound the decay of these coefficients with $s$ so that the stage errors can be summed. We also bound their growth with $\ell$, since increasing the order $q$ brings higher corrections into the generator.

Recall how the corrections were constructed. We first expanded the logarithm of a symmetric product in $m$ symbols $x_a$, writing $p_\ell$ for the coefficient of $\delta^\ell$. Substituting $\mathsf X_a$ for $x_a$ then gave $\mathcal K^{[\ell]}=p_\ell(\boldsymbol{\mathsf X})$. We estimate the coefficients in the same order. Lemma~\ref{lem:comparison-nested-logarithm} bounds the sum of the absolute coefficients when $p_\ell$ is written in nested commutators. After substitution, the derivatives $\partial_s$ in the $\mathsf X_a$ act on the factors of $g$. Lemma~\ref{lem:comparison-coefficient-bound} estimates the resulting derivatives and uses the first bound to sum their contributions.}

\begin{lemma}[Coefficients of the logarithm of a product]
\label{lem:comparison-nested-logarithm}
There is a constant $C_{\log}$ depending only on $m$ such that $p_\ell$ can be written as a sum of left-nested commutators
$[[\cdots[y_1,y_2],y_3],\ldots,y_\ell]$, with $y_j\in\{x_1,\ldots,x_m\}$,
whose coefficients have absolute values that sum to at most $C_{\log}^\ell/\ell$.
Moreover, $p_1=\sum_a x_a$ and $p_{2j}=0$.
\end{lemma}

\begin{proof}
The proof has two steps. We first estimate the coefficients of
$p_\ell$ when it is written as a sum of products $y_1y_2\cdots y_\ell$
with $y_j\in\{x_1,\ldots,x_m\}$, and then rewrite the products as nested commutators.
\emph{Coefficients in the word expansion.} Write $V=V(z)$ for the symmetric product above.
When $V-I$ is expanded in products of the $x_a$, all coefficients are
positive. Setting $x_a=1$ for all $a$ shows that the sum of the coefficients of
degree $\ell$ equals the coefficient of $z^\ell$ in $e^{mz}-1$. The logarithm is
$\log V=\sum_{k\ge1}(-1)^{k+1}(V-I)^k/k$, and the sum of absolute
coefficients of a product is at most the product of the sums. The sum of
the absolute coefficients of $p_\ell$ is thus at most the coefficient
$m_\ell$ of $z^\ell$ in
\[
 \sum_{k\ge1}\frac{(e^{mz}-1)^k}{k}=-\log(2-e^{mz}).
\]
This function has radius of convergence $(\log 2)/m$ and positive coefficients. Its value at
$z_0=m^{-1}\log\tfrac32$ is $\log2<1$. Hence $m_\ell z_0^\ell<1$, that
is, $m_\ell<z_0^{-\ell}$.

\emph{Expansion in nested commutators.} By the Baker--Campbell--Hausdorff formula,
$p_\ell$ is a Lie polynomial, that is, a combination of commutators of the $x_a$. For a homogeneous Lie polynomial
$P=\sum c_{y_1\cdots y_\ell}\,y_1\cdots y_\ell$ of degree $\ell$, the
Dynkin--Specht--Wever theorem \cite[Theorem~5.1]{EnomotoSatoh2014} states
that replacing each product by the corresponding nested commutator
multiplies $P$ by $\ell$,
\[
 \sum c_{y_1\cdots y_\ell}\,
 [[\cdots[y_1,y_2],y_3],\ldots,y_\ell]=\ell\,P .
\]
For example, $P=[x,y]=xy-yx$ gives $[x,y]-[y,x]=2P$. Dividing by $\ell$
expresses $p_\ell$ as a sum of nested commutators whose coefficients are
$1/\ell$ times the coefficients of the products. Their sum of absolute
values is therefore at most $z_0^{-\ell}/\ell$, which is the statement
with $C_{\log}=1/z_0$.
{ The assertions about $p_1$ and the even coefficients follow from Step~3 of the preceding proof.}
\end{proof}

The form of the estimate in
Lemma~\ref{lem:comparison-coefficient-bound} below follows from
Eq.~\eqref{eq:comparison-clock-bracket}. The operator
$\mathcal K^{[\ell]}$ is a combination of nested commutators of $\ell$
operators from the $\mathsf X_a$, and by
Eq.~\eqref{eq:comparison-clock-bracket} each commutator either
differentiates a scalar coefficient or raises the degree of a nested
commutator of the $H_a$ by one. A nested
commutator of $\ell$ operators therefore consists of terms of the form
\[
 \kappa(s)\,C_{\boldsymbol w},\qquad |\boldsymbol w|=d,\qquad 1\le d\le\ell,
\]
where $C_{\boldsymbol w}$ is the nested commutator of
Eq.~\eqref{eq:intrinsic-nested-words}, of degree $d$, and $\kappa(s)$
is a product of $d$ functions $g$ that carries $\ell-d$ derivatives in
total. For example, $[\mathsf A,\mathsf B]$ contains $ig'(s)(A+2B)$, with
$d=1$ and one derivative, and $-g(s)^2[A,B]$, with $d=2$ and no
derivative. In $\mathcal K^{[\ell]}$ the terms with $d\le2$ cancel by the
second property of Lemma~\ref{lem:comparison-cell-matching}, which leaves
$3\le d\le\ell$. The two kinds of steps affect the size of a term
differently. A nested commutator of degree $d$ is controlled by the scale $J$ of
Appendix~\ref{app:intrinsic-scales}. A derivative of $g$ is not controlled
by $J$.

The coefficients $\kappa(s)$ are products of
$g$ and its derivatives. The derivatives of $g(s)=\exp(1+s-e^s)$ are
\[
 g'(s)=(1-e^s)\,g(s),\qquad g^{(j)}(s)=P_j(e^s)\,g(s),
\]
with a polynomial $P_j$ of degree $j$. They are larger than $g$ by powers of $e^s$, so no multiple
of $g$ bounds them. We therefore use the function
\begin{equation}
 \mu(s)=\exp(-e^s/4),
 \label{eq:comparison-time-weight}
\end{equation}
which decays more slowly than $g$. It has two properties. First, it bounds
$g$ at complex points near $s$,
\[
 |g(u)|\le G\,\mu(s),\qquad |u-s|\le\tfrac14,\qquad G=4e^{1/4},
\]
as shown in Eq.~\eqref{eq:comparison-anchored-scalar} below. Complex
points are included because the proof of
Lemma~\ref{lem:comparison-coefficient-bound} bounds a derivative at a
point by the values of the function on a disk around that point. Second, $\mu$ is
integrable,
\[
 \int_0^\infty \mu(s)\,ds<\infty .
\]
{ Cauchy's inequality on the disk gives the same decay $\mu(s)$ for each fixed derivative of $g$. Products of these derivatives retain this decay since $\mu(s)\le1$. The following lemma also tracks how the bound grows with the order of the correction.}

\begin{lemma}[Coefficients of the corrections]
\label{lem:comparison-coefficient-bound}
There is a constant $C>0$ depending only on $m$ such that, for every odd $\ell\ge3$, the
corrections can be written, also for complex $u$, as
\begin{equation}
 \mathcal K^{[\ell]}(u)=
 \sum_{d=3}^{\ell}\sum_{|\boldsymbol w|=d}
 \kappa_{\ell,\boldsymbol w}(u)C_{\boldsymbol w},
 \label{eq:comparison-physical-representation}
\end{equation}
with coefficients that satisfy, for $3\le d\le\ell$ and all real $s$,
\begin{equation}
 \sum_{|\boldsymbol w|=d}
 \sup_{|u-s|\le1/8}|\kappa_{\ell,\boldsymbol w}(u)|
 \le C^\ell\binom{\ell-1}{d-1}\,(\ell-d)!\,\mu(s).
 \label{eq:comparison-coefficient-bound}
\end{equation}
\end{lemma}

The three quantities on the right have different origins. The function
$\mu(s)$ gives the decay in $s$. The factorial $(\ell-d)!$ comes from the
$\ell-d$ derivatives. This growth determines the scale $\rho_q=\rho_0/q$ used to sum the corrections in \cref{app:quantitative-coefficients}. The binomial coefficient is the number of ways to
choose, among the $\ell-1$ commutators, the $d-1$ that raise the degree. The supremum over complex $u$ near $s$ is used in
\cref{app:quantitative-coefficients}.

\begin{proof}
{ Consider one nested commutator of $\ell$ operators from the $\mathsf X_a$ in $\mathcal K^{[\ell]}$. We follow a term $\kappa(u)C$ in its expansion, where $\kappa$ is a scalar coefficient and $C$ is a nested commutator of the physical generators. Taking another commutator either differentiates $\kappa$ or multiplies it by $g$ and adds one generator to $C$. We bound $g$ and the derivative first, then add the bounds for terms of the same degree.}

\emph{Step 1. A bound on $g$ near the real axis.} For $|u-s|\le\tfrac14$ we have
$\operatorname{Re}e^u\ge e^{s-1/4}\cos\tfrac14\ge e^s/2$. Hence
\begin{equation}
 |g(u)|=\bigl|e^{1+u-e^u}\bigr|\le e^{5/4}\,e^{s}\exp(-e^{s}/2)
 \le G\mu(s),\qquad G=4e^{1/4},
 \label{eq:comparison-anchored-scalar}
\end{equation}
where the last step uses $e^{s}\exp(-e^{s}/4)\le4/e$. Since $\mu(s)\le1$,
this gives in particular
\begin{equation}
 |g(u)|\le G,\qquad |u-s|\le\tfrac14 .
 \label{eq:comparison-g-constant}
\end{equation}

{
\emph{Step 2. The effect of one derivative.} To estimate $\kappa'(u)$,
Cauchy's inequality uses the values of $\kappa$ on a circle about $u$:
\begin{equation}
 |\kappa'(u)|\le \frac{1}{h}\max_{|v-u|=h}|\kappa(v)| .
 \label{eq:comparison-cauchy}
\end{equation}
To use the coefficient bound from Step 1, this circle must lie inside
the disk on which the bound holds. Figure~\ref{fig:comparison-cauchy-disks}
shows how that disk shrinks over two successive derivatives.

Step 1 gives a bound on the disk of radius $1/4$, and the lemma asks for
a bound on the disk of radius $1/8$. A nested commutator of $\ell$
operators requires $\ell-1$ brackets, each of which can differentiate
a coefficient once. Choosing $h=(1/4-1/8)/\ell=1/(8\ell)$ leaves room
for all these derivatives. After the first $n$ operators we use the radius
\[
 r_n=\tfrac14-(n-1)h.
\]
For $2\le n\le\ell$, suppose $|u-s|\le r_n$. Every point $v$ on the
circle of radius $h$ about $u$ then satisfies
\[
 |v-s|\le |v-u|+|u-s|\le h+r_n=r_{n-1}.
\]
Thus the entire circle used to estimate $\kappa'(u)$ lies in the
preceding disk. Applying Eq.~\eqref{eq:comparison-cauchy} with
$h=1/(8\ell)$ and taking the supremum over $u$ gives
\[
 \sup_{|u-s|\le r_n}|\kappa'(u)|
 \le 8\ell\,\sup_{|v-s|\le r_{n-1}}|\kappa(v)|.
\]

\begin{figure}[htbp]
 \centering
 \includegraphics[width=0.96\linewidth]{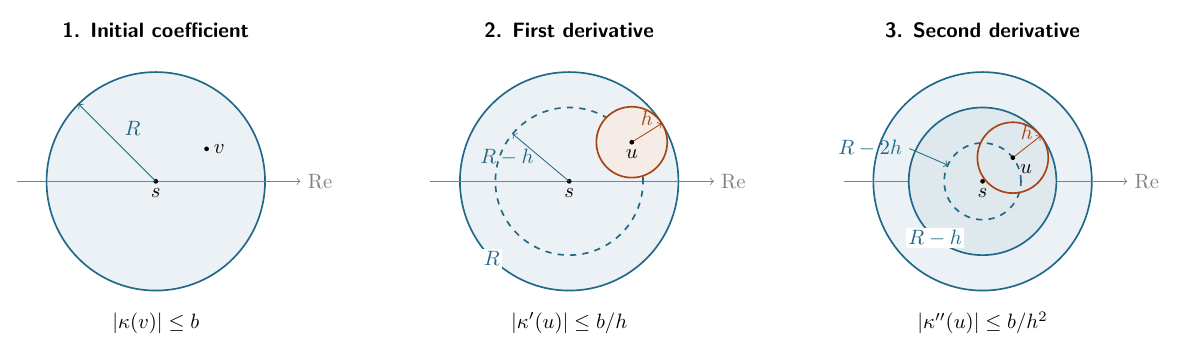}
 \caption{Successive applications of Cauchy's inequality about a real
 starting time $s$. Initially $|\kappa|\le b$ on the disk of radius $R$
 (left). To bound $\kappa'(u)$, the orange circle of radius $h$ about $u$
 must fit inside that disk, allowing $u$ to range over the disk of radius
 $R-h$ (middle). Applying the same argument to $\kappa'$ bounds
 $\kappa''$ by $b/h^2$ on the disk of radius $R-2h$ (right).}
 \label{fig:comparison-cauchy-disks}
\end{figure}

\emph{Step 3. One nested commutator.} Write the successive operators as
$c_i\partial_s+gT_i$, where $|c_i|\le2$ and $T_i$ is one of the $-iH_a$.
The first bracket is
\[
 [c_1\partial_s+gT_1,c_2\partial_s+gT_2]
 =c_1g'T_2-c_2g'T_1+g^2[T_1,T_2].
\]
The derivative operators have canceled. Each remaining term is a scalar
coefficient $\kappa_j(u)$ times a nested commutator $C_j$ of the $T_i$,
where a single $T_i$ has degree one. Taking the next bracket preserves
this form:
\begin{equation}
 [\kappa_j C_j,c\,\partial_s+gT]=-c\,\kappa_j'C_j+g\kappa_j[C_j,T].
 \label{eq:comparison-chain-step}
\end{equation}
The first term retains the degree of $C_j$. By Step 2 and $|c|\le2$,
its coefficient bound grows by at most a factor of $16\ell$.
The second term raises the degree by one, and its coefficient bound
grows by at most $G$ because $|g|\le G$.

After $n$ operators, write the expansion as $\sum_j\kappa_j(u)C_j$.
For each degree $d$, sum the bounds for the terms of that degree:
\begin{equation}
 M_{n,d}=\sum_{j:\,\deg C_j=d}\ \sup_{|u-s|\le r_n}|\kappa_j(u)|.
 \label{eq:comparison-coefficient-size}
\end{equation}
For $n=1$ we count only the multiplication term $gT_1$, so
$M_{1,1}=\sup_{|u-s|\le1/4}|g(u)|\le G\mu(s)$.
Set $M_{n,d}=0$ outside $1\le d\le n$.
The first bracket displayed above has two degree-one terms involving
$g'$ and one degree-two term with coefficient $g^2$.
Thus $M_{2,1}\le32\ell M_{1,1}$ and $M_{2,2}\le G M_{1,1}$. At each subsequent bracket, a term of degree $d$ comes either from differentiating the coefficient of a degree-$d$ term or from commuting a degree-$(d-1)$ term with $T$. Using $32\ell$ for the derivative bound also covers the first bracket and gives
\begin{equation}
 M_{n,d}\le32\ell\,M_{n-1,d}+G M_{n-1,d-1},\qquad 2\le n\le\ell .
 \label{eq:comparison-chain-recurrence}
\end{equation}
To reach degree $d$, exactly $d-1$ of the $n-1$ brackets must raise the degree. The other $n-d$ differentiate a coefficient. { For example, when $n=4$ and $d=2$, exactly one of the three brackets raises the degree, giving three choices for its position. Counting these choices and using} the initial bound $G\mu(s)$ gives
\begin{equation}
 M_{n,d}\le\binom{n-1}{d-1}(32\ell)^{n-d}G^d\mu(s).
 \label{eq:comparison-chain-bound}
\end{equation}

\emph{Step 4. Summation over commutators.} Since $r_\ell\ge1/8$, the bound for $M_{\ell,d}$ holds on the disk in Eq.~\eqref{eq:comparison-coefficient-bound}. By Lemma~\ref{lem:comparison-nested-logarithm}, the coefficients of the nested commutators in $p_\ell$ sum in absolute value to at most $C_{\log}^\ell/\ell$. Multiplying Eq.~\eqref{eq:comparison-chain-bound} by this bound and using $\ell^{\ell-d}\le e^\ell(\ell-d)!$ gives
\[
 \sum_{|\boldsymbol w|=d}\sup_{|u-s|\le1/8}|\kappa_{\ell,\boldsymbol w}(u)|
 \le C^\ell\binom{\ell-1}{d-1}(\ell-d)!\,\mu(s)
\]
with $C=32eGC_{\log}$. The terms of degrees one and two cancel by Lemma~\ref{lem:comparison-cell-matching}, leaving the stated range $3\le d\le\ell$.}
\end{proof}

\subsection{From one cell to the full sweep}
\label{app:comparison-global}

{We now use the generator constructed in \cref{app:comparison-generator} over the full auxiliary interval. We first establish existence, analyticity and unitarity as stated in Lemma~\ref{lem:finite-circuit-expansion}. For real $\delta>0$, we then compare the products over the same intervals, with $s_k=k\delta$ and $\Delta_k=f(s_k+\delta)-f(s_k)$,
\[
 Q_\infty(\delta)=\overleftarrow{\prod}_{k\ge0}S_2(\Delta_k),
 \qquad
 \Omega_q(\delta)=\overleftarrow{\prod}_{k\ge0}\Omega_{q,s_k}(\delta).
\]}

\subsubsection{\texorpdfstring{Proof of properties 1 and 2 (Analyticity and unitarity of $\Omega_q$)}{Proof of properties 1 and 2 (Analyticity and unitarity of Omega q)}}
\label{app:comparison-summary}

{Fix $q$. The correction added to the exact generator $-ig(s)H$ is the sum of $q-1$ terms,
\[
 \mathcal K_q(s,\delta)=\sum_{j=1}^{q-1}\delta^{2j}\mathcal K^{[2j+1]}(s).
\]
By Lemma~\ref{lem:comparison-coefficient-bound}, the norm of each coefficient is bounded by a multiple of $\mu(s)=\exp(-e^s/4)$, which is integrable on $[0,\infty)$. The same bound holds for $g(s)$ by Eq.~\eqref{eq:comparison-anchored-scalar}. Hence for each finite $R$ { there is a constant $C_R$, which may depend on $q,H_1,\ldots,H_m$, such that} $\sup_{|\delta|\le R}\|\mathcal L_q(s,\delta)\|\le C_R\mu(s)$. In the time-ordered expansion of
\begin{equation}
 \Omega_q(\delta)=\mathcal T\exp\!\left(\int_0^\infty\mathcal L_q(s,\delta)\,ds\right),
 \label{eq:comparison-full-evolution}
\end{equation}
the $n$-fold integral is bounded by $(C_R\int_0^\infty\mu(s)\,ds)^n/n!$. The series of polynomial terms therefore converges uniformly on every such disk and defines an entire function of $\delta$.

For one stage the upper endpoint $s+\delta$ moves with $\delta$, so its
evolution can contain odd powers even when its generator is even.
Here the endpoints are $0$ and $\infty$. Replacing $\delta$ by $-\delta$
leaves both the integration region and the generator unchanged in every
term. Thus $\Omega_q$ is even. Since $\Omega_q(0)=e^{-iH}=U$, we obtain $\Omega_q(\delta)U^\dagger=F(\delta^2)$ with $F$ entire and $F(0)=I$. For real $\delta$ the generator is anti-Hermitian, so $\Omega_q(\delta)$ is unitary. This proves properties~1 and~2.}

\subsubsection{Bounding the real-step error from the derivative on a complex disk}

{Two applications require a bound on an error whose order in $\delta$ is already known. On one auxiliary interval starting at $s$, the comparison evolution $\Omega=\Omega_{q,s}(\delta)$ matches the Strang stage $S=S_2(\Delta(s,\delta))$ through order $\delta^{2q}$. For the polygon product $V_q^\Omega$, cancellation removes the powers below $\delta^{2q}$. Thus, at fixed $q$ and Hamiltonian,
\[
 \|\Omega-S\|=O(\delta^{2q+1}),\qquad
 \|V_q^\Omega(\delta)-U\|=O(\delta^{2q}).
\]
To prove Theorem~\ref{thm:order-cost}, Appendices~\ref{app:finite-circuit-expansion} and~\ref{app:intrinsic-cost} must also bound the constants in these estimates in terms of $q$ and the Hamiltonian.

In each case, let $P$ be the relative evolution in the table below. For real step sizes the comparison factors are unitary, so the corresponding error is exactly $\|P-I\|$. The variable $z$ is the auxiliary step $\delta$ for a single stage and $\delta^2$ for the polygon product.}
\begin{center}
\renewcommand{\arraystretch}{1.2}
\begin{tabular}{@{}l@{\qquad}c@{\qquad}c@{\qquad}c@{}}
Application & $P$ & $z$ & {Order of $P(z)-I$} \\
\hline
One stage & {$S^{-1}\Omega_{q,s}(\delta)$} & $\delta$ & $O(z^{2q+1})$ \\
Polygon product & $V_q^\Omega(\delta)U^{-1}$ & $\delta^2$ & $O(z^q)$
\end{tabular}
\end{center}
{Let $k$ be the exponent in the last column. Since the analytic function $P(z)-I$ vanishes to order at least $k$, the maximum principle applied to $(P(z)-I)/z^k$ gives}
\begin{equation}
 \|P(z)-I\|\le\Bigl(\max_{|w|=R}\|P(w)-I\|\Bigr)(|z|/R)^k,
 \qquad |z|\le R .
 \label{eq:direct-maximum-principle}
\end{equation}
{The factor $(|z|/R)^k$ gives the required power of $z$. Bounding the maximum on the circle is harder: at complex step sizes the evolution factors need not be unitary, and bounding their norms separately can give a prefactor exponential in the system size.

On the real interval, $P(0)=I$ gives $\|P(z)-I\|\le\int_0^z\|P'(x)\|\,dx$. We can therefore estimate the error using an analytic operator $\Psi$ whose norm equals $\|P'\|$ on this interval. For the polygon product, for example, $\Psi=P'P^{-1}$ has this property by unitarity. Lemma~\ref{lem:real-step-product} applies the maximum principle to $\Psi$ and integrates the resulting bound along the real interval.}

\begin{lemma}[From the order of the first term to an estimate]
\label{lem:real-step-product}
Let $P(z)$ be differentiable and unitary for $0\le z\le R$, with $P(0)=I$,
and suppose that $\|\partial_zP(z)\|=\|\Psi(z)\|$ there, where $\Psi$ is
analytic for $|z|\le R$, satisfies $\|\Psi(z)\|\le M$ on this disk, and
{vanishes to order at least $k-1$ at zero, for an integer $k\ge1$.} Then
\begin{equation}
 \|P(z)-I\|\le\frac{MR}{k}\,(z/R)^k,\qquad 0\le z\le R .
 \label{eq:real-step-product-bound}
\end{equation}
\end{lemma}
\begin{proof}
By the maximum principle, applied to $\Psi(z)/z^{k-1}$, $\|\Psi(z)\|\le M(|z|/R)^{k-1}$ for
$|z|\le R$, and
$\|P(z)-I\|\le\int_0^z\|\partial_xP(x)\|\,dx\le MR(z/R)^k/k$.
\end{proof}

\subsubsection{Proof of property 3 (Error for the infinite Strang product)}
\label{app:quantitative-coefficients}
\label{app:alpha-comparison}
\label{app:normalization-threshold}

{It remains to prove property~3, the error estimate in Eq.~\eqref{eq:alpha-infinite-comparison}. For real $\delta>0$ the factors in both products are unitary. Replacing one factor at a time and taking the limit of the finite products gives
\[
 \|Q_\infty(\delta)-\Omega_q(\delta)\|
 \le \sum_{k\ge0}\|S_2(\Delta_k)-\Omega_{q,s_k}(\delta)\|.
\]
Lemma~\ref{lem:comparison-cell-matching} gives the order $O(\delta^{2q+1})$ of each term. To sum them we retain the factor $\mu(s_k)$ in the bound for each stage. Since $\sum_k\mu(k\delta)=O(1/\delta)$, this gives the required order $\delta^{2q}$ for the whole product.

To obtain the prefactor, Step~1 differentiates the relative evolution of a single stage. Step~2 bounds this derivative on the real axis by estimating an analytic operator on a complex disk. We estimate conjugation inside the interval using Lemma~\ref{lem:comparison-conjugation} for small $J$, or directly in operator norm when $\lambda\le \lambda_0 q^{2/3}$. The resulting disk bound supplies $M$ in Lemma~\ref{lem:real-step-product}, giving the error of one stage. Step~3 sums the bounds.}
Throughout we assume $q\ge2$ and set $\rho_q=\rho_0/q$, with $\rho_0>0$ chosen sufficiently small below. For the estimate under $J\le j_0$, taking $0<j_0\le\min\{1/(32e^4),1/(4em)\}$ suffices for both the $a_1$ and $a_4$ bounds on higher commutator degrees in Step~2 and the conjugation estimate in Lemma~\ref{lem:intrinsic-ideal} in Appendix~\ref{app:intrinsic-scales}. We take $J>0$, since for $J=0$ the generators commute and $Q_\infty(\delta)=\Omega_q(\delta)=U$.

\emph{Step 1. Derivative of the relative stage evolution.}
Consider the stage that starts at $s$, and write
$S(\delta)=S_2(\Delta(s,\delta))$ for the Strang stage and
$\Omega(\delta)=\Omega_{q,s}(\delta)$ for the comparison evolution over the same
auxiliary interval. The operator
\begin{equation}
 P_s(\delta)=S(\delta)^{-1}\Omega(\delta)
 \label{eq:comparison-stage-quotient}
\end{equation}
equals $I+O(\delta^{2q+1})$, and for real $\delta$, where $S$ and $\Omega$ are
unitary, $\|S(\delta)-\Omega(\delta)\|=\|P_s(\delta)-I\|$. { Differentiating $P_s=S^{-1}\Omega$ gives
\[
 \partial_\delta P_s=S^{-1}\Phi\,\Omega,\qquad
 \Phi=(\partial_\delta\Omega)\Omega^{-1}-(\partial_\delta S)S^{-1}.
\]
Thus $\|\partial_\delta P_s\|=\|\Phi\|$ on the real axis. To use $\Phi$ as the operator $\Psi$ in Lemma~\ref{lem:real-step-product}, we need to bound it on the complex disk. We compute the two derivatives separately.} For the Strang stage, $\partial_\delta\Delta(s,\delta)=g(s+\delta)$ gives
\begin{equation}
 (\partial_\delta S)S^{-1}
 =g(s+\delta)\bigl[-iH+\mathcal D(\Delta(s,\delta))\bigr],
 \qquad \mathcal D(\tau)=S_2'(\tau)S_2(\tau)^{-1}+iH .
 \label{eq:comparison-strang-derivative}
\end{equation}
{ Thus $\mathcal D(\tau)$ is the difference between the logarithmic derivative of the Strang stage and the exact generator $-iH$.} For the comparison
evolution, $\delta$ enters through the end $s+\delta$ of the interval and
through the generator $\mathcal L_q(s',\delta)=-ig(s')H+\mathcal
K_q(s',\delta)$, in which only $\mathcal K_q$ depends on $\delta$. Hence
\begin{equation}
 (\partial_\delta\Omega)\Omega^{-1}
 =\mathcal L_q(s+\delta,\delta)
  +\int_s^{s+\delta}\Omega(s+\delta,s')\,
   (\partial_\delta\mathcal K_q)(s',\delta)\,\Omega(s+\delta,s')^{-1}\,ds' ,
 \label{eq:comparison-evolution-derivative}
\end{equation}
where $\Omega(s+\delta,s')$ is the evolution under $\mathcal L_q(\cdot,\delta)$
from $s'$ to $s+\delta$. In the difference of
Eqs.~\eqref{eq:comparison-evolution-derivative} and
\eqref{eq:comparison-strang-derivative} the terms $-ig(s+\delta)H$, which
are not small, cancel, and
\begin{equation}
 \begin{aligned}
 \Phi(\delta)&=\underbrace{\mathcal K_q(s+\delta,\delta)}_{\Phi_{\rm end}}
  +\underbrace{\int_s^{s+\delta}\Omega(s+\delta,s')\,
   (\partial_\delta\mathcal K_q)(s',\delta)\,\Omega(s+\delta,s')^{-1}\,ds'}_{\Phi_{\rm gen}}
  -\underbrace{g(s+\delta)\,\mathcal D(\Delta(s,\delta))}_{\Phi_{\rm Strang}} .
 \end{aligned}
 \label{eq:comparison-stage-derivative}
\end{equation}
For real $\delta$, $\|\partial_\delta P_s\|=\|\Phi\|$, and therefore
\begin{equation}
 \|S(\delta)-\Omega(\delta)\|=\|P_s(\delta)-I\|
 \le\int_0^\delta\|\Phi(x)\|\,dx .
 \label{eq:comparison-stage-integral}
\end{equation}
Since $P_s(\delta)-I$ starts at order $\delta^{2q+1}$, $\Phi(\delta)$
starts at order $\delta^{2q}$. All operators above are analytic in $\delta$.
For complex $\delta$, $\Omega_{q,s}(\delta)$ is defined as the evolution along
the segment from $s$ to $s+\delta$, and
the same derivative identities hold. By
Lemma~\ref{lem:real-step-product} with $\Psi=\Phi$ and $k=2q+1$, it
remains to estimate $\|\Phi(\delta)\|$ for complex $\delta$ with
$|\delta|\le\rho_q$. The term $\Phi_{\rm end}$ comes from the moving end of the interval,
$\Phi_{\rm gen}$ from the dependence of the generator on $\delta$, and
$\Phi_{\rm Strang}$ from the error of the Strang stage. Each term arises from nested commutators of degree at least three. Their norms supply the factor $\alpha J$, and the two powers of the step size supply $\rho_q^2$ on the disk.

\emph{Step 2. A bound on $\Phi$ for complex step sizes.} {{Stage matching has already shown that the full $\Phi$ vanishes to order at least $2q$ at the origin. We bound its three terms separately to obtain the disk bound $M$ in Lemma~\ref{lem:real-step-product}, which combines $M$ with this order of vanishing.} At fixed $s$, we need
\[
 \sup_{|\delta|\le\rho_q}\|\Phi(\delta)\|\le C\alpha J\rho_q^2\mu(s).
\]
We first estimate the endpoint correction $\Phi_{\rm end}$ and the Strang defect $\Phi_{\rm Strang}$ directly. For $\Phi_{\rm gen}$ we must also bound conjugation by the remaining evolution on the interval.}

\emph{Case $J\le j_0$.} Let $|\delta|\le\rho_q$, and choose $\rho_0$ so that $\rho_q\le\tfrac18$. Then the points
$s+\delta$ and $s'$ lie in the disk of radius $\tfrac18$ around $s$ on which
Eq.~\eqref{eq:comparison-coefficient-bound} estimates the coefficients
$\kappa_{\ell,\boldsymbol w}$. We first compute the size of
$\mathcal K_q(s',\delta)=\sum_\ell\delta^{\ell-1}\mathcal K^{[\ell]}(s')$
for $s'$ in this disk. The correction $\mathcal K^{[\ell]}$ carries
$\delta^{\ell-1}$, which is at most $\rho_q^{\ell-1}$ in absolute value,
and the binomial coefficients in
Eq.~\eqref{eq:comparison-coefficient-bound} sum to at most $2^{\ell-1}$.
{Here we first use the smallness of $J$. Since $2eJ\le 2e j_0\le1$ and $d\ge3$, the commutator weight satisfies $(2eJ)^d=(2eJ)^3(2eJ)^{d-3}\le(2eJ)^3$ for every degree in the correction. For $d\le\ell<2q$ we also have $(\ell-d)!\,d!\le\ell!\le6(2q)^{\ell-3}$.} Summing over $\ell$ gives
\begin{equation}
 a_1\bigl(\mathcal K_q(s',\delta)\bigr)
 \le C J^3\rho_q^2\, \mu(s)\sum_{\ell\ge3}(2Cq\rho_q)^{\ell-3}
 \le C J^3\rho_q^2\, \mu(s),
 \label{eq:comparison-generator-majorants}
\end{equation}
where the last step holds because $2q\rho_q=2\rho_0$ is small. The choice $\rho_q=\rho_0/q$ thus controls the growth of the higher
corrections with $q$. Since all
nested commutators in $\mathcal K_q$ have degree at least three,
$\|C_{\boldsymbol w}\|\le\alpha\,d!(2J)^d/(12J^2)$ turns
Eq.~\eqref{eq:comparison-generator-majorants} into the estimate of the
term $\Phi_{\rm end}$,
\begin{equation}
 \|\mathcal K_q(s',\delta)\|\le C\alpha J\rho_q^2\mu(s) .
 \label{eq:comparison-correction-norm}
\end{equation}
{ Appendix~\ref{app:intrinsic-cost} uses Eq.~\eqref{eq:comparison-generator-majorants} with $a_4$. The same proof applies because $2e^4J<1$. In the variable $z=\delta^2$, write $\widehat{\mathcal K}_q(s',z)=\mathcal K_q(s',\delta)$. Differentiating the powers $z^{(\ell-1)/2}$ multiplies each coefficient by $(\ell-1)/2$, which is summable in the same geometric bound. Consequently, on $|z|\le\rho_q^2$,
\[
 a_4(\widehat{\mathcal K}_q(s',z))\le C J^3|z|\mu(s),\qquad
 a_4(\partial_z\widehat{\mathcal K}_q(s',z))\le C J^3\mu(s).
\]}

For $\Phi_{\rm Strang}$, write
$S_2(\tau)=\prod_{j=1}^{2m-1}e^{\tau\beta_j T_j}$ in the order of Eq.~\eqref{eq:strang-step}, where $T_j$ is the corresponding $-iH_a$ and the positive stage coefficients sum to $m$. For $m=2$,
$(T_1,T_2,T_3)=(-iA,-iB,-iA)$ and
$(\beta_1,\beta_2,\beta_3)=(\tfrac12,1,\tfrac12)$. In
$S_2'(\tau)S_2(\tau)^{-1}$ each $\beta_jT_j$ is conjugated by the exponentials
that precede it, and expanding these conjugations gives nested
commutators. Those of degree one sum to $-iH$, which cancels against
$iH$ in $\mathcal D(\tau)$, and those of degree two cancel because the
Strang stage is of second order. Those of degree $d\ge3$ have coefficients
whose absolute values sum to at most
\begin{equation}
 \sum_j \beta_j\frac{(\sum_{i<j}\beta_i)^{d-1}}{(d-1)!}|\tau|^{d-1}
 \le \frac{m^d |\tau|^{d-1}}{(d-1)!}.
 \label{eq:comparison-strang-generator-size}
\end{equation}
With $\|C_{\boldsymbol w}\|\le\alpha(2J)^{d-2}(d-1)!$ this gives, for
complex $\tau$ with $2mJ|\tau|\le\tfrac12$,
\begin{equation}
 \|\mathcal D(\tau)\|\le\alpha\sum_{d\ge3}m^d(2J)^{d-2}|\tau|^{d-1}
 \le4m^3\alpha J|\tau|^2 .
 \label{eq:comparison-strang-defect}
\end{equation}
By Eq.~\eqref{eq:comparison-anchored-scalar}, $|g(s+\delta)|\le G\mu(s)$
and $|\Delta(s,\delta)|\le\rho_qG\mu(s)$, which also ensures
$2mJ|\Delta(s,\delta)|\le\tfrac12$, so $\Phi_{\rm Strang}$ is at
most $C\alpha J\rho_q^2\mu(s)$ as well.

{{ It remains to bound $\Phi_{\rm gen}$. Differentiating $\delta^{\ell-1}$ in $\mathcal K_q$ gives $(\ell-1)\delta^{\ell-2}$. The extra factor $\ell-1$ is summable in the geometric bound of Eq.~\eqref{eq:comparison-generator-majorants}, giving
\[
 a_1(\partial_\delta\mathcal K_q(s',\delta))\le CJ^3\rho_q\,\mu(s).
\]
Integration along a segment of length at most $\rho_q$ restores the second power of $\rho_q$.} Inside this integral, $\partial_\delta\mathcal K_q$ is conjugated by $\Omega(s+\delta,s')$. This evolution need not be unitary for complex $\delta$, so we use Lemma~\ref{lem:comparison-conjugation}. Parametrize the segment by $s+t\delta$, $0\le t\le1$. Its generator in $t$ is $\delta\mathcal L_q(s+t\delta,\delta)$. By Eqs.~\eqref{eq:comparison-g-constant} and~\eqref{eq:comparison-generator-majorants}, the integrated weight on any remaining part of this segment satisfies $L\le\rho_q[2emJG+CJ^3\rho_q^2]\le\tfrac12$, where $G=4e^{1/4}$. Equation~\eqref{eq:comparison-conjugation-size} therefore gives
\[
 \|\Phi_{\rm gen}\|\le
 \rho_q\,\frac{\alpha}{6J^2}\sup_{s'\in[s,s+\delta]}a_1(\partial_\delta\mathcal K_q(s',\delta))
 \le C\alpha J\rho_q^2\mu(s).
\]}

\emph{Case $\lambda\le \lambda_0 q^{2/3}$.}
A nonzero nested commutator of degree three has norm at most $4\alpha J$ by Eq.~\eqref{eq:commutator-growth-scale}. Each additional commutation with any $H_a$ multiplies this bound by at most $2\lambda$. Hence
\[
 \|C_{\boldsymbol w}\|\le 4\alpha J(2\lambda)^{d-3},\qquad d=|\boldsymbol w|\ge3 .
\]
We estimate the corrections on a disk of radius $r\le R=r_0/q$, where $r_0>0$ depends only on $m$ and $\rho_0\le r_0$. In Eq.~\eqref{eq:comparison-coefficient-bound}, $(\ell-d)!\le(2q)^{\ell-d}$ for $d\le\ell<2q$. Choosing $\lambda_0\le1$ gives $\lambda\le q$, so the commutator and the factorial together contribute at most $4\alpha J(2q)^{\ell-3}$. Summing the binomial coefficients over $d$ costs at most $2^{\ell-1}$. Thus, uniformly for $|u-s|\le1/8$,
\[
 r^{\ell-1}\|\mathcal K^{[\ell]}(u)\|
 \le C\alpha J r^2\mu(s)(Cqr)^{\ell-3}.
\]
Choose $r_0$ so that $Cr_0\le1/2$ and $R\le1/8$. Differentiating $\delta^{\ell-1}$ adds a multiplier $(\ell-1)/r$ to this estimate. Summing the two geometric series gives
\begin{equation}
 \|\mathcal K_q(u,\delta)\|+r\|\partial_\delta\mathcal K_q(u,\delta)\|
 \le C\alpha J r^2\mu(s),\qquad |\delta|\le r,\quad 0<r\le R .
 \label{eq:comparison-generator-norm-disk}
\end{equation}
Appendix~\ref{app:intrinsic-cost} will use this estimate on the full disk of radius $R$ for the polygon product.

For the stage take $r=\rho_q$. On the segment from $s$ to $s+\delta$, the integrated norm of $-ig(u)H+\mathcal K_q(u,\delta)$ is at most $C(\lambda r+\alpha J r^3)$. Since $J\le\lambda$, $\alpha\le\lambda^2$ and $\lambda r\le r_0$, this integral is $O(1)$. The time-ordered series for the evolution on any part of the segment, forward or backward, therefore gives $O(1)$ for the norms of $\Omega(s+\delta,s')$ and its inverse. Equation~\eqref{eq:comparison-generator-norm-disk} now gives $C\alpha J r^2\mu(s)$ for both the endpoint term and the integral in Eq.~\eqref{eq:comparison-stage-derivative}. For the Strang term, $|\Delta(s,\delta)|\le Gr\mu(s)$ and $J|\Delta(s,\delta)|\le G\lambda r$. Taking $r_0$ sufficiently small gives $2mJ|\Delta|\le1/2$, so Eq.~\eqref{eq:comparison-strang-defect} yields the same estimate.

Altogether
$\|\Phi(\delta)\|\le C\alpha J\rho_q^2\mu(s)$ for $|\delta|\le\rho_q$,
and Lemma~\ref{lem:real-step-product} with $M=C\alpha J\rho_q^2\mu(s)$,
$R=\rho_q$ and $k=2q+1$ gives the estimate on one stage,
\begin{equation}
 \bigl\|S_2\bigl(\Delta(s,\delta)\bigr)-\Omega_{q,s}(\delta)\bigr\|
 \le\frac{C\alpha J \rho_q^3}{2q+1}\,\mu(s)\,(\delta/\rho_q)^{2q+1},
 \qquad 0\le\delta\le\rho_q .
 \label{eq:alpha-stage-comparison}
\end{equation}

{\emph{Step 3. Summing over the intervals.} Set $s=k\delta$ in Eq.~\eqref{eq:alpha-stage-comparison} and sum over $k\ge0$. Since $\mu$ is decreasing and integrable,
\[
 \sum_{k=0}^{\infty}\mu(k\delta)
 \le \mu(0)+\frac{1}{\delta}\int_0^\infty\mu(s)\,ds
 \le \frac{C}{\delta}.
\]
{ The factor $1/\delta$ in this sum turns the single-stage order $\delta^{2q+1}$ into the full-product order $\delta^{2q}$. This gives Eq.}~\eqref{eq:alpha-infinite-comparison} and completes the proof of Lemma~\ref{lem:finite-circuit-expansion}.}

%% file: appendices/polygon-cost.tex
\section{Error and cost of one polygon circuit}
\label{app:statements}

{Theorem~\ref{thm:order-cost} bounds the error and the number of exponentials in the polygon circuit at fixed order $q$. We prove it by choosing the cutoffs at the polygon vertices so that the finite circuit approximates $U=e^{-iH}$ with error of order $\delta^{2q}$ in the auxiliary step $\delta$. To obtain Theorem~\ref{thm:main}, we then choose $q$ for the desired precision and repeat the circuit for the desired physical time. Appendix~\ref{app:time-precision-cost} carries out this last step for the general and local Hamiltonians of parts (a) and (b), respectively.}

{Let $H=\sum_{a=1}^m H_a$, where the $H_a$ are Hermitian operators of finite norm. Write $\alpha=\sum_{a<b}\|[H_a,H_b]\|$ and $J=J(H_1,\ldots,H_m)$ for the commutator norm and growth scale of Appendix~\ref{app:intrinsic-scales}. Use $\rho_q$ from \cref{thm:order-cost}. Throughout this appendix $C$ may depend on $m$.}

\paragraph{Decomposition of the circuit error.}
\label{app:route-map}

The proof compares four operators,
\begin{equation}
 V_q(\delta)
 \ \overset{{\text{Step 1}}}{\longleftrightarrow}\
 V_{q,\infty}(\delta)
 \ \overset{{\text{Step 2}}}{\longleftrightarrow}\
 V_{q}^{\Omega}(\delta)
 \ \overset{{\text{Step 3}}}{\longleftrightarrow}\ U .
 \label{eq:polygon-chain}
\end{equation}
{In Step~1 we replace the finite circuit $Q_{r_j}(\delta_j)$ at each vertex by the infinite Strang product $Q_\infty(\delta_j)$, which gives
\begin{equation}
 V_{{q},\infty}(\delta)
 =\prod_{j=1}^{2{q}-1}Q_\infty(\delta_j)^{(-1)^{j+1}}.
 \label{eq:infinite-cancellation-word}
\end{equation}
In Step~2 we replace it by $\Omega_q(\delta_j)$, a comparison evolution whose error factor $\Omega_q(\delta_j)U^\dagger$ has a convergent expansion in $\delta_j^2$. { The first two bounds add the replacement errors of individual vertices. Cancellation in Step~3 takes place only after the vertices have been composed. Both replacements preserve the steps} $\delta_j=\delta\sin(\pi j/(2q))$ and the alternating forward and inverse factors. Step~3 uses polygon cancellation to bound the error of the whole comparison product relative to $U$. The sum of the three bounds controls $\|V_q(\delta)-U\|$.}

\subsection{Error after polygon cancellation}
\label{app:intrinsic-cost}
\label{app:alpha-polygon}

Let $V_q^{\Omega}$ be the polygon product of Eq.~\eqref{eq:infinite-cancellation-word} with every {infinite Strang product} replaced by the comparison evolution,
\begin{equation}
 V_q^{\Omega}(\delta)=\prod_{j=1}^{2q-1}\Omega_q(\delta_j)^{(-1)^{j+1}},
 \qquad\delta_j=\delta\sin\frac{\pi j}{2q}.
 \label{eq:comparison-polygon-product}
\end{equation}
{With $z=\delta^2$, property~1 of Lemma~\ref{lem:finite-circuit-expansion} gives the error factor at every vertex as
\[
 \Omega_q(\delta_j)U^\dagger=F\!\left(z\sin^2\frac{\pi j}{2q}\right).
\]
By \cref{lem:polygon-cancellation}, all powers through $z^{q-1}$ cancel in $V_q^\Omega U^{-1}-I$. Thus, at fixed $q$, the error of the polygon product is $O(\delta^{2q})$.

To choose the circuit order $q$ for a given precision, we also need to know how large the error can be as $q$ varies. The following lemma bounds $\|V_q^\Omega(\delta)-U\|$ in terms of the relative step size $\delta/\rho_q$ and the Hamiltonian scales $\alpha$ and $J$, with a constant independent of $q$.}

\begin{lemma}[Polygon product of the comparison evolutions]
\label{lem:reference-polygon-error}
Under the hypotheses of property~3 of \cref{lem:finite-circuit-expansion},
\begin{equation}
 \|V_{q}^{\Omega}(\delta)-U\|
 \le C\alpha J\rho_q^2(\delta/\rho_q)^{2q}.
 \label{eq:alpha-polygon-remainder}
\end{equation}
\end{lemma}

\begin{proof}
Write $z=\delta^2$ and $F(z)=\Omega_q(\sqrt z)U^{-1}$ for the error factor of one comparison evolution. The $j$th factor uses the squared step $\delta_j^2=c_jz$, where $c_j=\sin^2(\pi j/(2q))$ for $1\le j\le 2q-1$ [Eq.~\eqref{eq:squared-sine-nodes}]. Substituting $\Omega_q(\delta_j)=F(c_jz)U$ in the alternating product cancels adjacent factors of $U$ and $U^{-1}$. This leaves
\begin{equation}
 \Gamma(z)=V_q^\Omega(\delta)U^{-1}
 =\prod_{j=1}^{2q-1} F(c_jz)^{(-1)^{j+1}}
 =I+O(z^q).
 \label{eq:alpha-polygon-error}
\end{equation}
The last equality in Eq.~\eqref{eq:alpha-polygon-error} is polygon cancellation, Lemma~\ref{lem:polygon-cancellation}, applied to $F$, which is entire with $F(0)=I$ by Lemma~\ref{lem:finite-circuit-expansion}.
For $J\le j_0$ we estimate $\Gamma-I$ by integrating its derivative along the real interval. For $\lambda\le \lambda_0 q^{2/3}$ we use the order of the zero in Eq.~\eqref{eq:alpha-polygon-error} to estimate $\Gamma-I$ directly on a smaller disk. For the derivative estimate, unitarity on the real axis gives $\|\Gamma'\Gamma^{-1}\|=\|\Gamma'\|$. The advantage of $\Gamma'\Gamma^{-1}$ on the complex disk is seen by writing $\Gamma=\Pi T V$. The contribution from differentiating $T$ simplifies to
\[
 (\Pi T' V)\Gamma^{-1}=\Pi(T'T^{-1})\Pi^{-1}.
\]
The earlier factors $V$ cancel, and Lemma~\ref{lem:comparison-conjugation} bounds conjugation by the remaining evolution $\Pi$. We can thus use $\Psi=\Gamma'\Gamma^{-1}$ in Lemma~\ref{lem:real-step-product}.

Both estimates use the generator of a single factor $F$. Fix $\delta$. Evolve with the comparison generator up to auxiliary time $s$, then cover the remaining physical time $1-f(s)$ with the exact evolution. Relative to $U$, this gives
\[
 R(s)=e^{-i[1-f(s)]H}\Omega(s,0)U^{-1},\qquad
 R(0)=I,\qquad \lim_{s\to\infty}R(s)=F(\delta^2).
\]
Differentiating in $s$ cancels the ideal generator $-ig(s)H$ and leaves
\begin{equation}
 (\partial_s R)R^{-1}
 =e^{-i[1-f(s)]H}\mathcal K_q(s,\delta)e^{i[1-f(s)]H}.
 \label{eq:alpha-dressed-correction}
\end{equation}
Writing $\mathcal G(s,\delta^2)=(\partial_sR)R^{-1}$ for the generator above, we obtain the forward factor at step $\delta_j$ as
\[
 F(c_jz)=\mathcal T\exp\!\left(\int_0^\infty\mathcal G(s,c_jz)\,ds\right),
 \qquad c_jz=\delta_j^2 .
\]
The inverse factor is the same evolution followed backward. For $J\le j_0$ we apply Lemma~\ref{lem:comparison-conjugation} to the whole product. Its condition is $L=\int a_1(\Lambda(x))\,dx<1$. Counting $-ig(s)H$ in all $2q-1$ factors would give a weight bound of order $qmJ$. With Eq.~\eqref{eq:alpha-dressed-correction} we need only the $a_1$ weight of the conjugated correction, which Lemma~\ref{lem:intrinsic-ideal} bounds using the $a_4$ estimate from Appendix~\ref{app:finite-circuit-expansion}. We take $j_0$ small enough that $2emj_0\le1/2$, as required by that lemma. For $J=0$ every stage is exact, so we take $J>0$ below.

\emph{The case $J\le j_0$.} We first bound $L$ for the full product in Eq.~\eqref{eq:alpha-polygon-error}. Each factor contributes the integral of $a_1(\mathcal G(s,c_jz))$ along its evolution, with measure $|ds|$ in either direction. By Eq.~\eqref{eq:alpha-dressed-correction} the generator to which we apply Lemma~\ref{lem:intrinsic-ideal} is
\begin{equation}
 \mathcal G(s,z)=e^{-i[1-f(s)]H}\,
                 \mathcal K_q(s,\sqrt z)\,e^{i[1-f(s)]H}.
 \label{eq:alpha-polygon-generator}
\end{equation}

For $|z|\le\rho_q^2$, the coefficient bounds in \cref{eq:comparison-generator-majorants}, summed with $|(c_jz)^{(\ell-1)/2}|\le c_j|z|\rho_q^{\ell-3}$, give $a_4(\mathcal K_q(s,\sqrt{c_jz}))\le CJ^3 c_j|z|\mu(s)$. Since $2emJ\le1/2$, \cref{lem:intrinsic-ideal} bounds the weight after conjugation by
\begin{equation}\label{eq:polygon-factor-weight}
a_1(\mathcal G(s,c_jz))\le CJ^3 c_j\rho_q^2\mu(s).
\end{equation}
Summing over the vertices, using $\sum_j c_j=q$, gives integrated weight
\begin{equation}\label{eq:alpha-polygon-transport-budget}
L\le CqJ^3\rho_q^2\int_0^\infty\mu(s)\,ds\le1/2
\end{equation}
for a sufficiently small $\rho_0$ in $\rho_q=\rho_0/q$ since $mJ\le1/(4e)$. The same geometric series remains summable after differentiation adds the factor $(\ell-1)/2$. This gives $a_1(\partial_z[\mathcal G(s,c_jz)])\le Cc_jJ^3\mu(s)$ by the same conjugation lemma. All terms have commutator degree at least three. After transport through the remaining evolutions, \cref{lem:comparison-conjugation} therefore gives
\begin{equation}\label{eq:alpha-polygon-derivative}
\|(\partial_z\Gamma)\Gamma^{-1}\|\le \frac{C\alpha}{J^2}\sum_j c_jJ^3\int_0^\infty\mu(s)\,ds\le C \alpha J q .
\end{equation}
The derivative vanishes through degree $q-2$. Applying \cref{lem:real-step-product} in $z$ with $k=q$ and radius $\rho_q^2$ yields
\begin{equation}\label{eq:alpha-polygon-lemma-applied}
\|V_q^\Omega(\delta)-U\|\le C\alpha J\rho_q^2(\delta/\rho_q)^{2q}.
\end{equation}

\emph{The case $\lambda\le \lambda_0 q^{2/3}$.}
Let $R=r_0/q$ be the radius in Eq.~\eqref{eq:comparison-generator-norm-disk}.
For $|z|\le R^2$ and $s\ge0$, the conjugation in Eq.~\eqref{eq:alpha-polygon-generator}
is unitary because $1-f(s)$ is real. Thus
\[
 \int_0^\infty\|\mathcal G(s,c_jz)\|\,ds
 \le C\alpha J R^2\int_0^\infty\mu(s)\,ds\le C\alpha J R^2 .
\]
The backward evolution gives the inverse factor. Its time-ordered series converges uniformly on this disk by the same integral bound, so the inverse is analytic there as well.
Concatenating the $2q-1$ evolutions therefore gives total integrated generator
norm at most $b=C(2q-1)\alpha J R^2$. In the time-ordered series for
$\Gamma-I$, the term with $n$ generators has norm at most $b^n/n!$. Hence
\[
 \sup_{|z|\le R^2}\|\Gamma(z)-I\|\le e^b-1\le b e^b
 \le \alpha J R^2 e^{Cq}.
\]
Here $\alpha J R^2\le \lambda^3 R^2=O(1)$, which explains the power
$q^{2/3}$ in the hypothesis.

The zero of order at least $q$ in Eq.~\eqref{eq:alpha-polygon-error} now gives,
by Eq.~\eqref{eq:direct-maximum-principle},
\[
 \|\Gamma(z)-I\|\le \alpha J R^2 e^{Cq}(|z|/R^2)^q .
\]
Choose $\rho_0/r_0$ sufficiently small that
$e^{Cq}(\rho_0/r_0)^{2q-2}\le1$ for every $q\ge2$.
Substituting $z=\delta^2$ yields
$\|V_q^\Omega(\delta)-U\|\le\alpha J\rho_q^2(\delta/\rho_q)^{2q}$
for $0<\delta\le\rho_q$.
\end{proof}

\subsection{Transfer to the finite circuit}
\label{app:alpha-cutoff}

{Under the same assumptions as in Lemma~\ref{lem:reference-polygon-error}, we now choose the cutoff at each vertex of the finite circuit $V_q(\delta)$. If at most $\eta$ of physical time remains at each vertex, Lemma~\ref{lem:cutoff-selection} bounds the total truncation error by $(2q-1)C\alpha J\eta^3$. To keep this within the error bound of Step~3, it suffices to choose $\eta=(\delta/\rho_q)^{2q}/(4q)$. Let $s_{\rm cut}$ be the auxiliary time at which $1-f(s_{\rm cut})=\eta$. Each vertex must reach at least this time. Since $1-f(s)=\exp(1-e^s)$, the cutoffs for $1\le j\le q$ are given by}
\begin{equation}
 \begin{aligned}
 s_{\rm cut}&=\log\!\left[1+2q\log(\rho_q/\delta)+\log(4q)\right],\\
 r_j&=\left\lceil s_{\rm cut}/\delta_j\right\rceil .
 \end{aligned}
 \label{eq:alpha-finite-cutoffs}
\end{equation}
{Set $r_{2q-j}=r_j$ for the remaining vertices, which have the same auxiliary steps as their partners. With these cutoffs we will bound the error and count the exponentials to obtain}
\begin{equation}
 \|V_{q}(\delta)-U\|\le C\alpha J\rho_q^2\,(\delta/\rho_{q})^{2{q}},\qquad
 D_{q}(\delta)=O\!\left(\frac{{q}\log {q}}{\delta}{\log\!\bigl[q\log(e\rho_q/\delta)\bigr]}\right).
 \label{eq:order-cost-error}
\end{equation}

For the three steps in Eq.~\eqref{eq:polygon-chain}, unitary telescoping gives
\begin{equation}
 \|V_q-U\|\le
 \underbrace{\sum_{j=1}^{2q-1}\|Q_{r_j}(\delta_j)-Q_\infty(\delta_j)\|}_{\text{cutoff}}
 +\underbrace{\sum_{j=1}^{2q-1}\|Q_\infty(\delta_j)-\Omega_q(\delta_j)\|}_{\text{comparison evolution}}
 +\underbrace{\|V_q^\Omega-U\|}_{\text{polygon remainder}} .
 \label{eq:polygon-three-errors}
\end{equation}
{\emph{Proof of Theorem~\ref{thm:order-cost}.} We first bound $\|V_{q,\infty}(\delta)-U\|$. Lemma~\ref{lem:reference-polygon-error} already bounds $\|V_q^\Omega(\delta)-U\|$, so it remains to replace each comparison evolution by the corresponding infinite Strang product. All factors of the two}
products are unitary, and $\|R^{-1}-S^{-1}\|=\|R-S\|$ for unitary $R$ and
$S$, so replacing the $2q-1$ factors $\Omega_q(\delta_j)$ by
$Q_\infty(\delta_j)$, one at a time, changes the product by at most the
sum of their deviations by property~3 of \cref{lem:finite-circuit-expansion}. Since $\delta_j\le\delta$,
\begin{equation}
 \|V_{q,\infty}(\delta)-V_q^{\Omega}(\delta)\|
 \le(2q-1)\frac{C\alpha J\rho_q^2}{2q+1}(\delta/\rho_q)^{2q}
 \le C\alpha J\rho_q^2(\delta/\rho_q)^{2q}.
 \label{eq:alpha-polygon-replacement}
\end{equation}
Together with Lemma~\ref{lem:reference-polygon-error},
\begin{equation}
 \|V_{q,\infty}(\delta)-U\|
 \le C\alpha J\rho_q^2(\delta/\rho_q)^{2q}.
 \label{eq:alpha-infinite-polygon-bound}
\end{equation}

{To recover the finite circuit, we now truncate at each vertex. By Eq.~\eqref{eq:alpha-finite-cutoffs}, $r_j\delta_j\ge s_{\rm cut}$, so the remaining physical time satisfies}
\begin{equation}
 \tau_{r_j}=e^{1-e^{r_j\delta_j}}
 \le \frac{1}{4q}(\delta/\rho_q)^{2q} .
 \label{eq:alpha-finite-tail}
\end{equation}
{Applying the same sum over vertices to the single-vertex truncation bound of Lemma~\ref{lem:cutoff-selection} gives}
\begin{equation}
 \|V_q(\delta)-V_{q,\infty}(\delta)\|
 \le(2q-1)C\alpha J\Bigl[\frac{1}{4q}(\delta/\rho_q)^{2q}\Bigr]^3
 \le C\alpha J\rho_q^2(\delta/\rho_q)^{2q}.
 \label{eq:alpha-finite-replacement}
\end{equation}
Here $(2q-1)/(4q)^3\le C/q^2$. Since $q^{-2}=O(\rho_q^2)$ and $\delta/\rho_q\le1$, the last inequality follows. Adding \cref{eq:alpha-infinite-polygon-bound}
gives the error bound in Eq.~\eqref{eq:order-cost-error}.

\subsection{Number of exponentials and proof of Theorem~\ref{thm:order-cost}}

{It remains to count the exponentials. At vertex $j$, the circuit $Q_{r_j}(\delta_j)$ uses $r_j+1$ Strang stages, each implemented with $2m-1$ exponentials. Its inverse uses the same number. Each step $\delta_j$ with $j<q$ occurs twice, while $\delta_q$ occurs once. Thus, before merging adjacent exponentials,}
\begin{equation}
 \begin{aligned}
 \frac{D_q(\delta)}{2m-1}&=\sum_{j=1}^{2q-1}(r_j+1)=2\sum_{j=1}^{q-1}(r_j+1)+r_q+1\\
 &\le 4q-2+\frac{s_{\rm cut}}{\delta}
       \left(2\sum_{j=1}^{q-1}\csc\frac{\pi j}{2q}+1\right).
 \end{aligned}
 \label{eq:alpha-exact-stage-count}
\end{equation}
For $1\le j\le q$, concavity of the sine on $[0,\pi/2]$ gives
$\sin(\pi j/(2q))\ge j/q$, and the sum in
\cref{eq:alpha-exact-stage-count} is at most $2q(1+\log q)+1$. More precisely, $\csc u=1/u+O(1)$ on $0<u\le\pi/2$, so
\begin{equation}
 \sum_{j=1}^{2q-1}\delta_j^{-1}
 =\delta^{-1}\bigl[(4q/\pi)\log q+O(q)\bigr].
 \label{eq:polygon-sine-cost}
\end{equation}
 {For $q\ge2$ and $0<\delta\le\rho_q$, the argument of the logarithm defining $s_{\rm cut}$ in Eq.~\eqref{eq:alpha-finite-cutoffs} is at most $4q\log(e\rho_q/\delta)$.} Hence
\begin{equation}
 D_q(\delta)=O\!\left(
  \frac{q\log q}{\delta}{\log\!\bigl[q\log(e\rho_q/\delta)\bigr]}\right).
 \label{eq:alpha-finite-stage-count}
\end{equation}
At $\delta=\rho_q/2$, the cutoff takes auxiliary time $O(\log q)$. Substituting the common radius $\rho_q=\rho_0/q$ and multiplying the stage counts by $2m-1$ gives the exponential counts in \cref{eq:order-cost-fixed-hamiltonian}.

%% file: appendices/simulation-cost.tex
\section{Simulation at prescribed time and precision}
\label{app:complexity-corollaries}
\label{app:time-precision-cost}

The fixed-order circuit of \cref{thm:order-cost} approximates one unit of evolution when $J\le j_0$ or $\lambda\le \lambda_0 q^{2/3}$. We first choose its order $q$ to meet a prescribed error. For an arbitrary time $t$, we use this circuit for each of $M$ intervals of length $t/M$, allowing error $\varepsilon/M$ in each interval. The condition on $J$ allows an interval length independent of system size for local Hamiltonians. For general generators, the condition on $\lambda$ allows the interval length to grow with $q$. The resulting count $D(tH,\varepsilon)$ is the number of exponentials sufficient for the full evolution.

\subsection{Time subdivision}
\label{app:time-subdivision}

\paragraph{Choice of the cancellation order.}

\begin{corollary}[Unit-time cost]
 \label{cor:normalized-cost}
 For $J\le j_0$ and $\alpha>0$, put $\ell=\max\{2,\log(\alpha/\varepsilon)\}$. Then
 \begin{equation}
 D(H,\varepsilon)=O(\ell^2\log^2\ell).
 \label{eq:appendix-unit-cost-stages}
 \end{equation}
 A choice ${q}=\Theta(\ell)$ in \cref{thm:order-cost} suffices.
\end{corollary}

For $J\le j_0$, the factor $J/q^2$ in \cref{eq:order-cost-fixed-hamiltonian} is bounded by $j_0$. At $\delta=\rho_q/2$, \cref{thm:order-cost} gives
\begin{equation}
 \|V_q-U\|\le C_*\alpha\,4^{-q},\qquad
 D_q\le C_*q^2\log^2q,
 \label{eq:alpha-order-precision}
\end{equation}
after increasing a constant $C_*\ge1$ depending only on $m$.
Choose the smallest integer $q$ such that
\begin{equation}
 q\ge\max\left\{2,
       \frac{\log(C_*\alpha/\varepsilon)}{\log4}\right\}.
 \label{eq:alpha-precision-order}
\end{equation}
Since $C_*$ is fixed, this choice gives $q=\Theta(\ell)$. Substitution in \cref{eq:alpha-order-precision} proves \cref{cor:normalized-cost}.
If $\alpha=0$, one Strang stage is exact for any evolution
time, so order selection is unnecessary.

\paragraph{Cost for \texorpdfstring{$M$}{M} repetitions.}

\begin{proposition}[Repetition]
 \label{prop:time-subdivision}
 For every positive integer $M$, repeating a product formula for an interval of length $t/M$ and precision $\varepsilon/M$ gives a full-time formula whose cost satisfies
 \begin{equation}
 D(tH,\varepsilon)\le M D\!\left(\frac tM H,\frac\varepsilon M\right).
 \label{eq:time-cost-subdivision}
 \end{equation}
\end{proposition}

To prove \cref{prop:time-subdivision}, let $W$ approximate
$U_M=e^{-i(t/M)H}$ to norm error at most $\varepsilon/M$.
Both operators are unitary, and
\begin{equation}
 W^M-U_M^M=\sum_{a=0}^{M-1}
    W^{M-1-a}(W-U_M)U_M^a.
 \label{eq:alpha-unitary-repetition}
\end{equation}
Hence $\|W^M-e^{-itH}\|\le\varepsilon$. Repeating its circuit
$M$ times multiplies the number of exponentials by $M$, proving the
stated inequality for $D$.

With $M$ repetitions, the allowed error is $\varepsilon/M$ per interval and the commutator norm on that interval is $(t/M)^2\alpha$. The ratio of the commutator norm to the allowed error determines the required order $q$. We evaluate the cost at fixed $M$ before choosing the interval length.

\begin{proposition}[Cost after repetition]
\label{prop:cost-after-repetition}
Let $t>0$ and $\alpha>0$. For every positive integer
$M$ satisfying $tJ/M\le j_0$, the repeated polygon circuit gives
\begin{equation}
 \begin{aligned}
 D(tH,\varepsilon)&=O(M L_M^2\log^2L_M),\\
 L_M&=\max\left\{2,
    \log\frac{C_*\alpha t^2}{M\varepsilon}\right\}.
 \end{aligned}
 \label{eq:alpha-all-time-cost}
\end{equation}
The implied constant depends only on $m$.
\end{proposition}

\begin{proof}
Put $\tau=t/M$. The operators in one repetition are $\tau H_1,\ldots,\tau H_m$, with
\begin{equation}
 \lambda_\tau=\tau\lambda,\qquad J_\tau=\tau J\le j_0,\qquad \alpha_\tau=\tau^2\alpha.
 \label{eq:alpha-time-scaled-parameters}
\end{equation}
Allocate error $\varepsilon/M$ to each repetition. The ratio entering
the order selection is exactly
\[
 \frac{C_*\alpha_\tau}{\varepsilon/M}
 =\frac{C_*\alpha(t/M)^2}{\varepsilon/M}
 =\frac{C_*\alpha t^2}{M\varepsilon}.
\]
Applying \cref{cor:normalized-cost} to $\tau H$ at precision $\varepsilon/M$ gives
$O(L_M^2\log^2L_M)$ exponentials per repetition. The repetition inequality
then gives \cref{eq:alpha-all-time-cost}.
\end{proof}

\subsection{General Hamiltonian sums}
\label{app:general-time-precision}

\paragraph{Choice of \texorpdfstring{$M$}{M}.}

We first give the simulation cost in terms of $J$ for a fixed number of generators.

\begin{theorem}[Hamiltonian-dependent simulation cost]
 \label{thm:time-precision-cost}
 Let $t{J}\ge1$. Put
 \begin{equation}
 \ell_J=\log\!\left(\frac{\alpha t}{J\varepsilon}\right).
 \label{eq:time-cost-accuracy-scale}
 \end{equation}
 For $\ell_J\ge2$,
 \begin{equation}
 D(tH,\varepsilon)=O\!\left(t{J}\ell_J^2\log^2\ell_J\right).
 \label{eq:appendix-time-cost-stages}
 \end{equation}
\end{theorem}

Let $K\ge J$ be a positive upper bound on $J$ and choose
\begin{equation}
 M=\lceil tK/j_0\rceil.
 \label{eq:alpha-time-partition}
\end{equation}
This ensures $tJ/M\le j_0$, so
\cref{prop:cost-after-repetition} applies for every $t>0$.
When $tK\ge1$, the ceiling gives
\[
 \frac{tK}{j_0}\le M\le\left(\frac1{j_0}+1\right)tK.
\]
Consequently $\log(C_*\alpha t^2/(M\varepsilon))
=\ell_K+O(1)$ for $\ell_K=\log(\alpha t/(K\varepsilon))$.
For $\ell_K\ge2$, this implies $L_M=\Theta(\ell_K)$ and
\begin{equation}
 D(tH,\varepsilon)=O(tK\ell_K^2\log^2\ell_K).
 \label{eq:alpha-upper-scale-cost}
\end{equation}
Taking $K=J$ proves \cref{thm:time-precision-cost}.

\paragraph{Choice of \texorpdfstring{$M$}{M} using the norm bound.} To obtain \cref{thm:main}(a), use the norm hypothesis of \cref{thm:order-cost} on each interval. For $t,\alpha>0$ and integer $q\ge2$, choose
\begin{equation}
 M=\left\lceil\frac{t\lambda}{\lambda_0 q^{2/3}}\right\rceil,\qquad
 \tau=\frac{t}{M}.
 \label{eq:general-time-partition}
\end{equation}
Then $\tau\lambda\le \lambda_0 q^{2/3}$. Let $W$ be the circuit of \cref{thm:order-cost} for $\tau H_1,\ldots,\tau H_m$ at $\delta=\rho_q/2$. Its error is at most $C\tau^3\alpha J\,4^{-q}/q^2$. Summing the errors by \cref{eq:alpha-unitary-repetition} and using $J\le\lambda$ gives
\begin{equation}\label{eq:general-repetition-error}
\|W^M-e^{-itH}\|\le C\frac{\alpha t}{\lambda}\frac{(\tau\lambda)^2}{q^2}4^{-q}\le C\frac{\alpha t}{\lambda}4^{-q}.
\end{equation}
Thus $q=\max\{2,\lceil\log_4(C\alpha t/(\lambda\varepsilon))\rceil\}$ suffices. Each interval uses $O(q^2\log^2q)$ exponentials, and rounding up the number of intervals gives
\begin{equation}\label{eq:general-rounded-cost}
D(tH,\varepsilon)=O\!\left((q^2+t\lambda q^{4/3})\log^2q\right).
\end{equation}
Since $q=O(\ell_\lambda)$ and \cref{eq:general-rounded-cost} is increasing in $q$, this proves \cref{thm:main}(a).

\subsection{Local implementation}
\label{app:many-native-generators}

For the local decompositions of \cref{app:local-definitions}, \cref{lem:commutator-scale-properties} permits $K=J_{\mathrm{loc}}=3k^2g_{\mathrm{loc}}$, giving
\begin{equation}
 D(tH,\varepsilon)=O(tJ_{\mathrm{loc}}\ell_{J_{\mathrm{loc}}}^2
                       \log^2\ell_{J_{\mathrm{loc}}}),\qquad
 J_{\mathrm{loc}}=3k^2g_{\mathrm{loc}},
 \label{eq:time-cost-local}
\end{equation}
when $tJ_{\mathrm{loc}}\ge1$ and $\ell_{J_{\mathrm{loc}}}\ge2$. For smaller $tK$ or lower precision, use
\cref{eq:alpha-all-time-cost} with the same $M$ of
\cref{eq:alpha-time-partition}; the maximum in $L_M$ already covers
those cases.

For the layer decomposition in \cref{def:local-hamiltonian}, each site meets at most one norm-one or smaller interaction per layer. Thus $J\le3k^2$. A term overlaps at most $k$ terms of each other layer, and a layer contains at most $N$ nonempty supports. Summing $\|[h,h']\|\le2$ over overlapping pairs gives $\alpha\le 2kN\binom m2\le km^2N$.

Take $M=\lceil3k^2t/j_0\rceil$ and $\tau=t/M$. Then $\tau J\le j_0$, $\tau\le C/k^2$ and $M\le Ck^2t$ for $t\ge1$. At half the commutator radius, \cref{eq:order-cost-fixed-hamiltonian} gives total error
\[
M\frac{C(\tau^2\alpha)(\tau J)}{q^2}4^{-q}\le C\frac{\alpha t}{k^2q^2}4^{-q}\le CNt\,4^{-q}.
\]
Take $q=\max\{2,\lceil\log_4(CNt/\varepsilon)\rceil\}$. Then $q=O(\ell_N)$. Each interval uses $O(q^2\log^2q)$ layer exponentials, proving \cref{thm:main}(b). The depth is at most $D$ and the number of gates on at most $k$ sites is at most $ND$.